\RequirePackage{rotating}
\documentclass[fleqn,usenatbib]{mnras}

\usepackage{newtxtext,newtxmath}

\usepackage[T1]{fontenc}

\DeclareRobustCommand{\VAN}[3]{#2}
\let\VANthebibliography\thebibliography
\def\thebibliography{\DeclareRobustCommand{\VAN}[3]{##3}\VANthebibliography}

\usepackage{graphicx}	
\usepackage{amsmath}	

\usepackage{float}
\usepackage{lscape}
\usepackage{rotating}
\usepackage{caption}
\usepackage{subcaption} 
\usepackage{morefloats}
\usepackage[flushleft]{threeparttable} 
\usepackage{multicol} 
\usepackage{hyperref}
\hypersetup{colorlinks=true, linkcolor=blue, urlcolor=blue, citecolor=blue}
\usepackage{pdflscape} 
\usepackage{longtable} 

\usepackage{changepage}
\usepackage{array}

\usepackage{fancyvrb}

\usepackage{wrapfig}
\usepackage{lscape}
\usepackage{rotating}
\usepackage{epstopdf}

\title[3D structure of Galactic molecular cloud complexes]{The three-dimensional structure of Galactic molecular cloud complexes out to 2.5 kpc } 

\author[T. E. Dharmawardena et al.]{
T. E. Dharmawardena$^{1}$\thanks{E-mail: dharmawardena@mpia.de},
C.A.L. Bailer-Jones$^{1}$,
M. Fouesneau $^{1}$,
D. Foreman-Mackey$^{2}$,
P. Coronica$^{3}$, \newauthor
T. Colnaghi$^{3}$, 
T. M\"{u}ller$^{1}$,
and J. Henshaw$^{1}$
\\
$^{1}$Max Plank Institute for Astronomy (MPIA), Königstuhl 17, 69117 Heidelberg, Germany\\
$^{2}$Center for Computational Astrophysics, Flatiron Institute, 162 5th Ave, New York, NY 10010, USA\\
$^{3}$Max Planck Computing and Data Facility, Gießenbachstraße 2, 85748, Garching, Germany
}

\date{Accepted XXX. Received YYY; in original form ZZZ}

\pubyear{2015}

\begin{document}
\label{firstpage}
\pagerange{\pageref{firstpage}--\pageref{lastpage}}
\maketitle


\begin{abstract}
Knowledge of the three-dimensional structure of Galactic molecular clouds is important for understanding how clouds are affected by processes such as turbulence and magnetic fields and how this structure effects star formation within them.
Great progress has been made in this field with the arrival of the Gaia mission, which provides accurate distances to $\sim10^{9}$ stars. Combining these distances with extinctions inferred from optical-IR, we recover the three-dimensional structure of 16 Galactic molecular cloud complexes at $\sim1$pc resolution using our novel three-dimensional dust mapping algorithm \texttt{Dustribution}.
Using \texttt{astrodendro} we derive a catalogue of physical parameters for each complex.
We recover structures with aspect ratios between 1 and 11, i.e.\ everything from near-spherical to very elongated shapes.
We find a large variation in cloud environments that is not apparent when studying them in two-dimensions. 
For example, the nearby California and Orion A clouds look similar on-sky, but we find California to be more sheet-like, and massive, which could explain their different star-formation rates. 
In Carina, our most distant complex, we observe evidence for dust sputtering, which explains its measured low dust mass. 
By calculating the total mass of these individual clouds, we demonstrate that it is necessary to define cloud boundaries in three-dimensions in order to obtain an accurate mass; simply integrating the extinction overestimates masses.
We find that Larson's relationship on mass vs radius holds true whether you assume a spherical shape for the cloud or take their true extents.

\end{abstract}

\begin{keywords}
ISM: clouds < Interstellar Medium (ISM), Nebulae --
ISM: dust, extinction < Interstellar Medium (ISM), Nebulae --
ISM: individual objects: . . . < Interstellar Medium (ISM), Nebulae --
(Galaxy:) local interstellar matter < The Galaxy --
Galaxy: structure < The Galaxy --
methods: numerical < Astronomical instrumentation, methods, and techniques
\end{keywords}



\section{Introduction}
Understanding the three-dimensional structure of galaxies and their constituent parts is a key goal of astronomy. Yet the only information generally available to observers is two-dimensional projections on the sky and a third dimension of frequency. For our Galaxy, however, we can obtain a third spatial dimension using parallaxes, something which has received a major impetus from the Gaia mission \citep{2016Prusti}.
 
Inferring the structure of the interstellar medium is a challenge, because there is no simple way to measure the distance to dust clouds directly.
The main approach is to trace the absorption by the interstellar matter (e.g.\ through extinction or atomic lines) of light from background stars, for which distances can be estimated. 
While every line of sight could be treated individually, in the most complete formulation of the problem the positions (distances) of stars are inferred along with the distribution of absorbing material over all points. 
However, this leads to a very high-dimensional inference problem, scaling with both the number of stars and the size of the grid on which interstellar absorption is discretised.

When doing so, care must also be taken to differentiate between the integrated absorption and the (more interesting) interstellar density. While these are connected by derivatives, inference in density space is more directly interpretable, because it is a local  quantity rather than the result of a heliocentric line-of-sight (LOS).

Pseudo-three-dimensional information on the interstellar medium can be obtained from multi-frequency data: when lines with known rest-frequencies are observed, the line-of-sight velocities as a function of position can be used to model distance. However, this is very dependent on a global kinematic gas model, and if multiple components with the same velocities exist at different locations along the line-of-sight, separating their contributions is impossible without additional information.

In recent years, works by \citep{Sale2018, Green2019_Bayestars19, Marshall2006, Lallement2019, Leike2019, Rezaei2018B, Dharmawardena2022} have exploited large survey data including extinction and distances to map the ISM using dust in 3D where the third dimension is distance. Among these works, Gaussian Processes (GPs) have emerged as a key method, as they produce robust maps while following an intuitive process and remaining easy to interpret physically. With GPs we are able to take into account correlations between points in 3D space and incorporate them explicitly into the GP prior on the density distribution rather than coupling individual lines-of-sight together in the plane of the sky, helping to mitigate artefacts such as the fingers-of-god effect \citep[e.g.,][]{Leike2019, Leike2020, Dharmawardena2022}.

With this new generation of 3D maps, there is great interest in extracting individual structures from them to help understand the ISM and the processes that shape it. A case of particular interest is the structure of Galactic molecular clouds, as these are among the most massive dusty structures in the ISM. This is motivated by the need to understand how processes such as turbulence, magnetic fields, and gravitational collapse affect the star-forming process. The effects of these physical processes are imprinted on these 3D structures: What distinguishes the dense parts of the cloud currently undergoing star formation from other structures in their wider environment that are not forming stars? Understanding the shapes, sizes, masses, densities, and other properties of these clouds and their environs may shed light upon this question.

Recent studies have begun to explore this area. \citet{Hottier2020} mapped the 3D dust structure of the Vela molecular ridge using the FellWalker algorithm to identify the positions and boundaries of clumps and cavities. \citet{Zucker2021} and \citet{Bialy2021} used the 3D dust maps of \citet{Leike2020} to trace the structure of several nearby molecular clouds by decomposing them into skeletons and their envelopes. \citet{Dharmawardena2022} mapped the Perseus, Taurus, and Orion molecular clouds, as well as the more distant Cygnus X region. Curiously, we found evidence for a large free-standing filament in the foreground of the Orion and saw evidence that Cygnus is a monolithic structure with smaller clumps at larger distances along the line-of-sight. We also identified many filamentary structures extending outward from the Perseus and Taurus, and pointed out the importance of using 3D information to compute masses to avoid overestimation, since it does not include excess foreground and background extinction. This is echoed by \citet{Zucker2021} who compared 2D extinction estimates with those derived from the 3D information.

In the present work, we employ our novel algorithm \texttt{Dustribution} \citep{Dharmawardena2022} to map the 3D structure of 16 well known Galactic molecular cloud complexes and their extended environments out to 2.5 kpc. In Sec.~\ref{sec:Method_3D} we summarise our GP-based algorithm which we have now accelerated with GPU computing and lazy evaluation.  
Using a modified version of the dendrogram code \texttt{astrodendro} we extract sub-structures of the molecular clouds and determine their properties such as shapes and sizes, mass, and fill factor (section \ref{sec:Method_Astrodendro}). We present a catalogue of these sub-structures in Sec.~\ref{sec:Results}. We discuss the features observed in each molecular clouds and associate the sub-structures recovered by \texttt{astrodendro} with literature data in the associated region in Sec.~\ref{sec:IndGMCs}. We discuss the overall results in section \ref{sec:Discussion} and conclude in section \ref{sec:Conclusions}.

\section{3D mapping of molecular clouds with \texttt{Dustribution}}
\label{sec:Method_3D}

In \citet{Dharmawardena2022} we introduced our algorithm \texttt{Dustribution}. Here we provide a summary of our code and the improvements made to it. Using arbitrary catalogues of stellar extinction and distances to stars as input, \texttt{Dustribution} infers the 3D dust density and extinction of any user selected region around those stars. It uses latent variable GP models \citep{Rasmussen2006_GPbook} and Variational inference \citep{Bishop2006_VIandELBObook, Blei2017:VI}, which allow it to directly account for the 3D correlations between any density points throughout the mapped region and infer the logarithm of the dust density. 

Our GP prior is defined in the logarithm of density to ensure we only predict positive densities. The predictions from this "latent" model are integrated along all lines-of-sight to which we have observable data in order to compute model extinctions; these are compared to the observed extinctions to compute the (Gaussian) likelihood. The covariance matrix of our GP is described by an Radial basis function (RBF) kernel (i.e., a Gaussian kernel which is not be confused with the GP itself) \citep{Rasmussen2006_GPbook}. The RBF kernel is formed of five hyperparamters: three physical scale lengths in the three Heliocentric Cartesian coordinates ($x$; $y$; $z$) of physical space; one exponential scale factor; and the mean density.

Our model set up -- where our GP prior is defined as a latent variable requiring integration and prediction of the joint density distribution at all points -- results in an extremely large parameter space. To overcome this issue we employ an inducing point method and variational inference. When using inducing points we condition the GP on a subset of points $M < N$ which can represent the density at their locations, 
thereby reducing the dimensionality of the model (e.g.\ \citealt{Quinonero-candela_sparseGP, titsias_sparsevariationalGP, Sale2018}; Figure 1 in \citealt{Uhrenholt2021} demonstrates how this represents the underlying data just as well as a non-sparse GP). On the other hand, the likelihood evaluation still exploits the entire input sample of extinctions. To further reduce the computing time, we use variational inference, which replaces the target posterior with a simpler approximate posterior, by
finding the parameters for this approximation that best reproduces the true posterior. This allows us to directly compute the approximate posterior and its gradient with respect to the free parameters, thus enabling the use of
gradient-descent optimisers in our model. 
We have implemented this using the GpyTorch \citep{Gardner2018_GpyTorch} and Pyro \citep{Bingham2018_Pyro1, Phan2019_Pyro2} packages. 

Since publishing our model in  \citet{Dharmawardena2022},  we have made several improvements to \texttt{Dustribtuion} to accelerate the computations. Several key stages are now computed lazily to eliminate unnecessary operations and memory usage (see Appendix \ref{sec:App:CodeMod} in the supplementary material for more details). We also now exploit pytorch's GPU interface to parallelise the computations. Collectively, these improvements result in a factor of 100 reduction in typical run-times and a factor of 200 reduction in memory usage.

In \citet{Dharmawardena2022} we showed that \texttt{Dustribtuion} effectively recovers the structures of Galactic molecular clouds, including revealing previously unidentified filaments outside the main star-forming clouds. By integrating the densities, we derived cloud masses with $12\%$ statistical uncertainties. However, like all GP methods, the prior introduces some smoothing which tends to average out the sharpest peaks in density. This is compounded by the poor sampling of the highest-density points, both because they are compact (and hence there is a low probability of the required background stars) and because their extinctions are so high that background stars may not be detectable. These two effects tend to reduce the inferred peak densities, so caution is advised when analysing the results derived from GP methods.

Here we use \texttt{Dustribution} to map a sample of 16 molecular cloud complexes. Our sample was selected to cover a wide variety of properties such as distance from the Sun, mass (and hence the presence or absence of massive stars) and age. 
Following \citep[][Sec.4]{Dharmawardena2022} we use as inputs the extinctions and distances from \citet{Fouesneau2022_LBol}, which is based on multiband photometry and Gaia parallaxes. As it also uses ALLWISE photometry, the wide wavelength coverage improves the precision of the extinction estimates, particularly at high extinction, and hence the stellar parameters. 

{The \texttt{Dustribution} code is publicly available in GitHub at \url{https://github.com/Thavisha/Dustribution}. The predicted 3D density and extinction data sets for all 16 molecular cloud complexes are available interactively via the \texttt{Dustribution} website at \url{www.mwdust.com}, where they can also be downloaded from. These data sets are also available to download via Zenodo at \url{https://doi.org/10.5281/zenodo.7061955}. 
}

\section{Deriving properties of the molecular clouds with \texttt{astrodendro}} 
\label{sec:Method_Astrodendro}


To derive properties of our molecular clouds we construct 3D dendrograms from our density data cubes using the \texttt{astrodendro} package\footnote{\url{https://dendrograms.readthedocs.io/en/stable}}. 
\texttt{Astrodendro}, which utilises dendrograms, is a hierarchical structure-finding tool, which perform a top-down segmentation of a dataset starting from the highest data values \citep[e.g.][figure 6.]{Chen2018_astrodendroUsage}. 
A binary tree is used to represent the relationships between (nested) sub-structures (in terms of density, the most dense and compact condensations) and larger parent structures.

\texttt{astrodendro} requires us to specify three control parameters: 1) the minimum data value considered (points with values below this are discarded before computing the tree), i.e., the minimum density above which pixels are considered as contributors to the structure ($min\_value$).; 2) the minimum difference in peak density between two candidate structures below which they will be merged ($min\_delta$), i.e., adjacent peaks must have differences greater than this to be divided into sub-structures; 3) the minimum number of data points (in our case 3D pixels) to form the smallest possible structure, ($npix$), therefore structures cannot be subdivided if either structure would have fewer points than this, even if they are sufficiently different. This results in a set of nested structures, analogous to defining a set of iso-density surfaces which enclose distinct regions. 

In the case of parameter $min\_value$ and $min\_delta$ we begin by setting them based on the final GP hyper-parameters after training is completed: $min\_value$ is set to the final mean density hyper-parameter, while  $min\_delta$ is set to the final scale factor GP hyper-parameter. When testing these parameters we found some tweaking was required in the case of  $min\_value$ to ensure we did not include the diffuse background. Specifically, we decided to set the  $min\_value$ as three times the final mean density hyper-parameter. The $npix$ parameter was set to be the number of pixels within one ellipsoidal volume where the $xyz$ axis radii are equal to the final scale lengths of our trained 3D model. 

The advantages of dendrograms are that they generate a hierarchical segmentation of a dataset, providing the relationships between structures at different density levels, while making minimal assumptions about the underlying shape of the structures. Unlike the classic GaussClumps algorithm\footnote{\url{https://starlink.eao.hawaii.edu/docs/sun255.htx/sun255se2.html\#x3-70002}}, there is no assumption of shape. In contrast to FilFinder \citep{Koch2015_FillFinder}, astrodendro seeks the boundaries of the structure rather than only the structure's core. Although  ClumpFind \citep{Williams1994_ClumpFind} and FellWalker \citep{Berry2015_FellWalker} also make no assumptions about shape, they do not recover nested structures.

In order to handle our 3D data cubes we have modified \texttt{astrodendro} to accommodate 3D spatial data, instead of position-position-velocity data. \texttt{astrodendro}  extracts the structure within regions by decomposing the 3D data cubes and assigning the pixels (the spatial element of the grid of our 3D data cube) to hierarchical structures based on their intensities and the differences in intensities compared to other groups of pixels. This produces a tree-like hierarchy of structures. These structures are bound by the contours of the dendrogram

Each structure recovered from \texttt{astrodendro} can be described as a parent or a child. A parent gives rise to further structures whereas a child is a substructure arising from a parent. The algorithm classifies three types of structures based on their position on the dendrogram: Trunks, which have no parents; Branches where two substructures are merged into a single structure; Leaves, which have no children. Traditionally a trunk could simultaneously be a leaf if it has no parent and can not be further split into any substructure (so no children). However, in this work we consider this type of structure only as a trunk - more specifically an "isolated trunks" - and not count them towards the leaves. "Isolated trunks" allow us to identify dense structures that are independent of the main molecular cloud and are non-substructured density enhancements. Therefore, in this work all leaves arise from a parent structure.

One of the properties we derive for each of the structures is its maximum 3D extent in using our 3D dust density cubes in heliocentric cartesian coordinates ($x,y,z$). From this the resulting ellipsoidal volume with the major axis in the direction of the greatest extent is also derived. This is carried out for each sub-structure identified by \texttt{astrodendro} whether they are trunks, branches or leaves. First we derive the vector that connects the two pixels furthest apart ($v_{max1}$). We then derive the pair of pixels connected by the vector $v^\prime_{max2}$ with the largest projection orthogonal to $v_{max1}$ (i.e. the vector with the largest rejection\footnote{The "vector rejection" is the opposite of the "vector projection" i.e. the projection of a vector orthogonal to a reference vector.} from $v_{max1}$); the projection of $v^\prime_{max2}$ orthogonal to $v_{max1}$ defines $v_{max2}$; we define a 3rd vector $\hat{v}_{3}$ which is the cross product of $v_{max1}$ and $v_{max2}$, and find the pair of pixels connected by the vector with longest projection in the direction of $\hat{v}_{3}$, $v_{max3}$. With this we derive the maximum 3D extent of these structures, which we also convert to Galactic $l,b,d$ coordinates. By observing these structures in the $x,y,z$ Cartesian planes we see they are most similar to an ellipsoidal shape, therefore we derive the volume of an ellipsoid using the above measured maximum extents. 


In addition to this ellipsoid volume, we compute a second volume measure which is simply the volume of all the pixels within the structure. As our 3D density cubes are predicted in Cartesian coordinates all pixels have the same volume and we are able to simply add up the densities in all pixels within the structure. The density of the Edge pixels are considered to fall within the structure. The effect of fully including the edge pixels is negligible as the fraction of edge pixels in all structures is small. By dividing this volume by the ellipsoidal volume derived above from the maximum 3D extents we  derive a filling factor of these structures. 

The mass of the structures are derived by multiplying the structure volume with the density of all pixels within the structure. By doing so we derive mass in mag pc$^{2}$ units. We convert this mass to solar masses using the method described in \citet{Dharmawardena2022}, Sec. 7. We also derive the mass-weighted centroid for each structure giving us the center of mass of the structure. The mass and volume of the parent structures include the mass of their children so that we get an accurate representation of the mass of a given structure. 

{The dendrograms for all 16 complexes are available on Zenodo at \url{https://doi.org/10.5281/zenodo.7061955}. These publicly available dendrograms can be used to directly compare our dataset to others, for example by comparing the structures that are recovered, or determining the mass within the same surfaces (contours). 
}

\subsection{Uncertainty calculations}

To measure the peak density error we use the 16th and 84th percentile maps produced by \texttt{Dustribution} to derive the upper and lower uncertainties. In the case of total mass uncertainties we follow the same prescription as given by \citet{Dharmawardena2022}, Sec.~7. 

To calculate the statistical uncertainties on the $l,b,d$ and $x,y,z$ minimum and maximum extents we begin by extracting $1\%$ of $npix$ of the pixels in each structure at the minimum and maximum extent in each of $l,b,d$ and $x,y,z$ directions. One per cent was chosen to balance the competing requirements for as few pixels as possible (to ensure we sample the extrema) and as many pixels as possible (for a reliable average), and particularly to ensure there are always enough samples for the central limit theorem to apply. Each of these pixels has a density gradient given as 

\begin{equation}
    \frac{\partial \rho}{\partial x},
\end{equation}

for the x direction as an example. We derive this derivative numerically for all pixels and extract the values for the selected pixels. However, since we are interested in the uncertainty on the spatial direction, the gradient required to derive the uncertainty is: 

\begin{equation}
    \frac{\partial x}{\partial \rho} \approx \frac{1}{ \frac{\partial \rho}{\partial x}}. \label{eq:invgrad}
\end{equation}

Therefore to derive the uncertainty on the e.g. x extent $\sigma_{x}$ we follow the standard approach for propagating uncertainties based on gradients and compute:

\begin{equation}
    <\sigma_x> = < \sigma_\rho \frac{\partial x}{\partial \rho} >, \label{eq:sigma_x}
\end{equation}

where $\sigma_\rho$ is the uncertainty on density at each selected pixel. This is repeated for each maximum and minimum $x,y,z$ and $l,b,d$ extents. 

This typically produces an uncertainty of up to a few per cent in each direction, up to approximately 10\%. However, at the edges of the map, the gradients may become unstable, resulting very large spatial gradients. In these cases, this edge effect may corrupt the average uncertainty even if only a few pixels are affected, resulting in outliers. As we cannot trust the uncertainty in these cases, we truncate their uncertainty to 10\% of the extent for these outliers. 

To calculate the uncertainties in Galactic coordinates, we first compute the gradient vector $\nabla \rho = \left(\frac{\partial \rho}{\partial x}, \frac{\partial \rho}{\partial y}, \frac{\partial \rho}{\partial z} \right) $. We then compute the projection of $\nabla \rho$ in the $l, b, d$ directions, respectively, to give $\frac{\partial \rho}{\partial l}$ (or $b$ or $d$). The resulting gradient is then inserted into eqns. \ref{eq:invgrad} \& \ref{eq:sigma_x} as in the case of the cartesian extents to give the required uncertainties.

Using this method we find a median uncertainty of $0.2^{\circ}$ for $l$ and $b$ minimum and maximum extents and $3$ pc for the minimum and maximum distance extents. The full set of these statistical uncertainties for each individual structure is given in the electronic versions of Tables~\ref{tab:region_params}, \ref{tab:maintrunk_params} and \ref{tab:leaf_params}. We stress that all the statistical uncertainties calculated in this work reflect only the formal internal uncertainties of our method; the systematic errors arising from values such as the extinction data, assumptions and conversion factors (not included) are expected to be at least an order of magnitude larger.

\section{Results}
\label{sec:Results}

\begin{sidewaysfigure*}
    \centering
    \includegraphics[width=\textwidth]{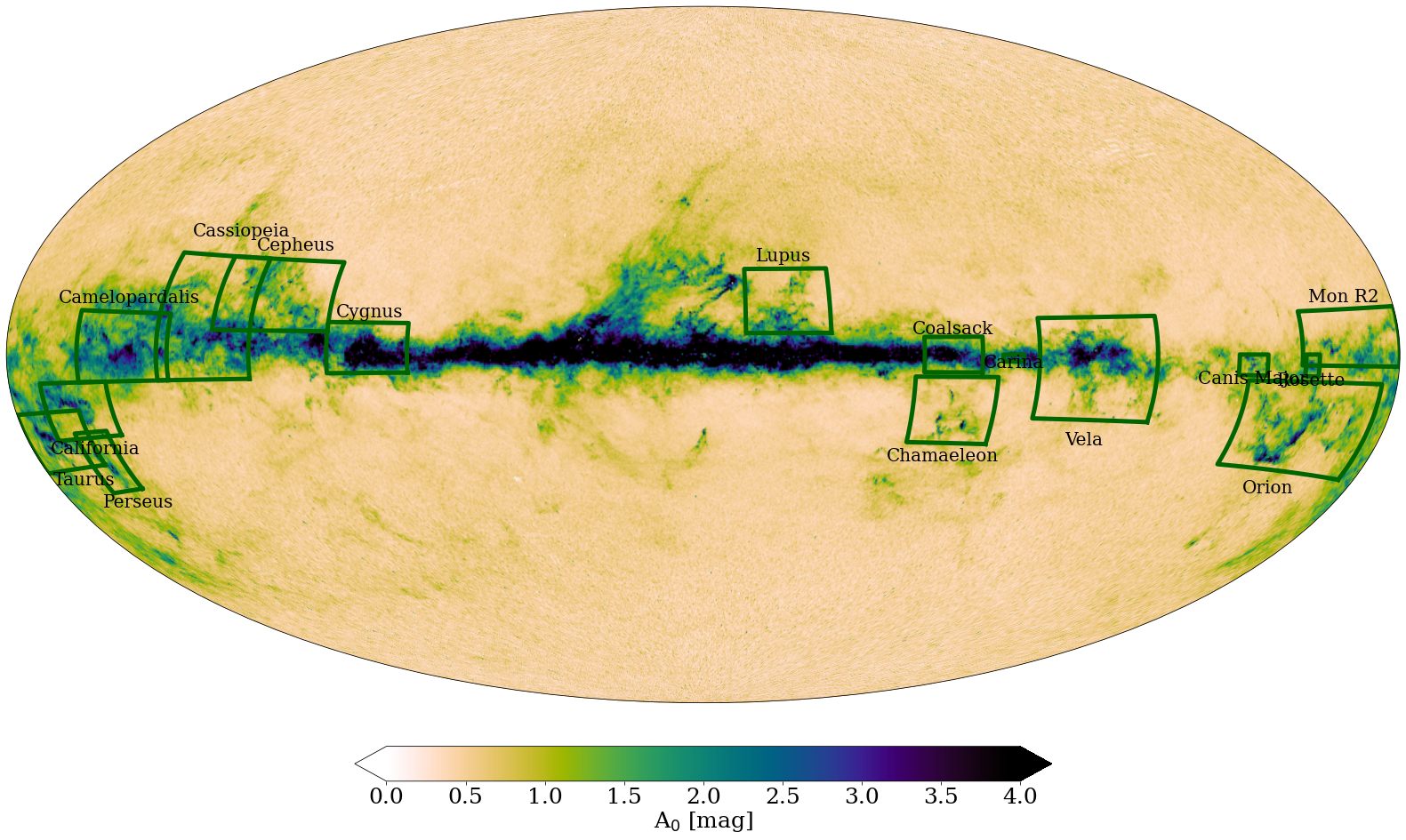}
    \caption{Extinctions as a function of Galactic coordinates from the catalogue of \citet{Fouesneau2022_LBol} with the molecular clouds and cloud complexes analysed in this work highlighted.}
    \label{fig:lbol_inputdata}
\end{sidewaysfigure*}

\begin{figure*}
\centering
\begin{subfigure}{0.8\textwidth}
  \centering
 \includegraphics[width=\textwidth]{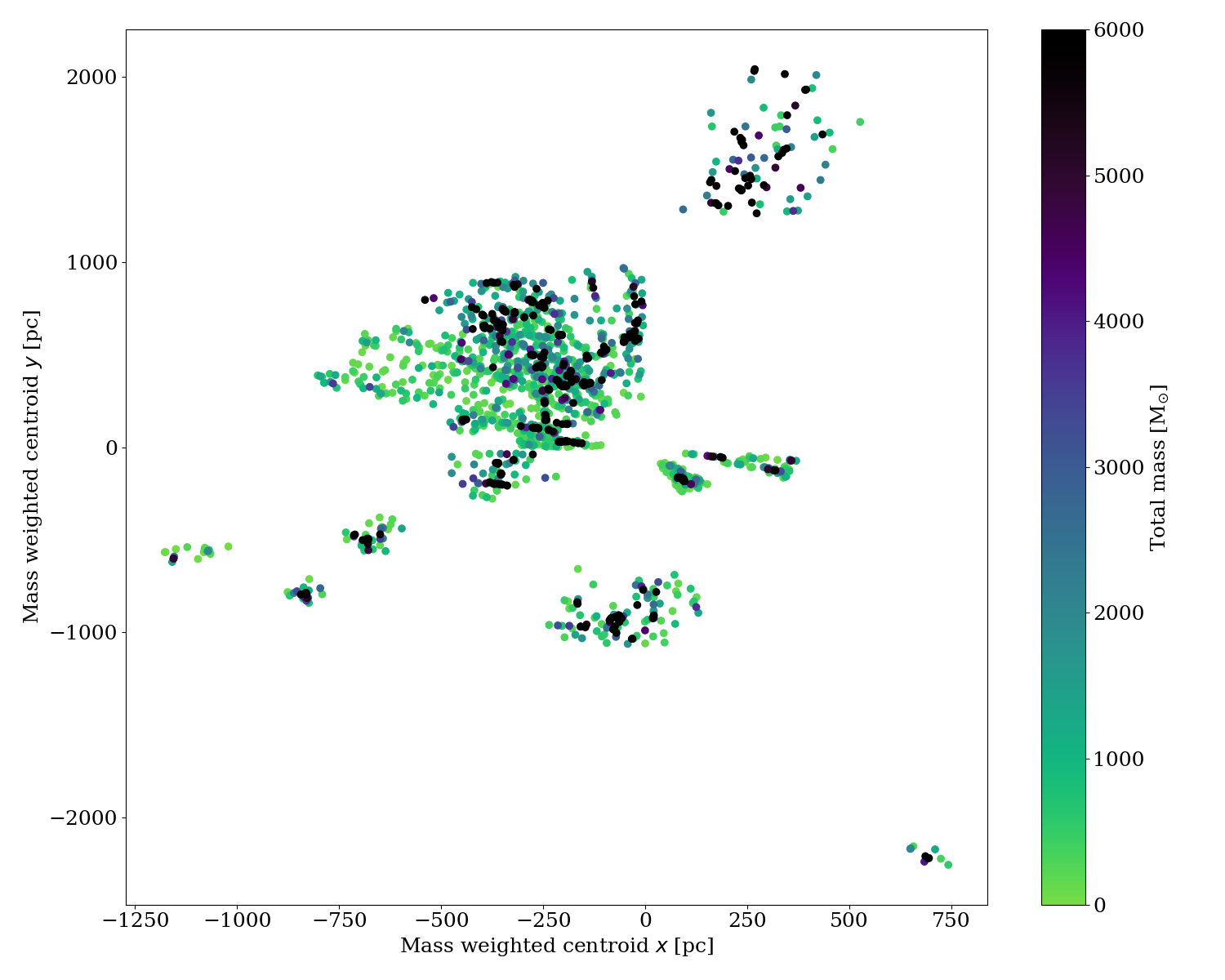}
  \end{subfigure}
  
\begin{subfigure}{0.8\textwidth}
  \centering
   \includegraphics[width=\textwidth]{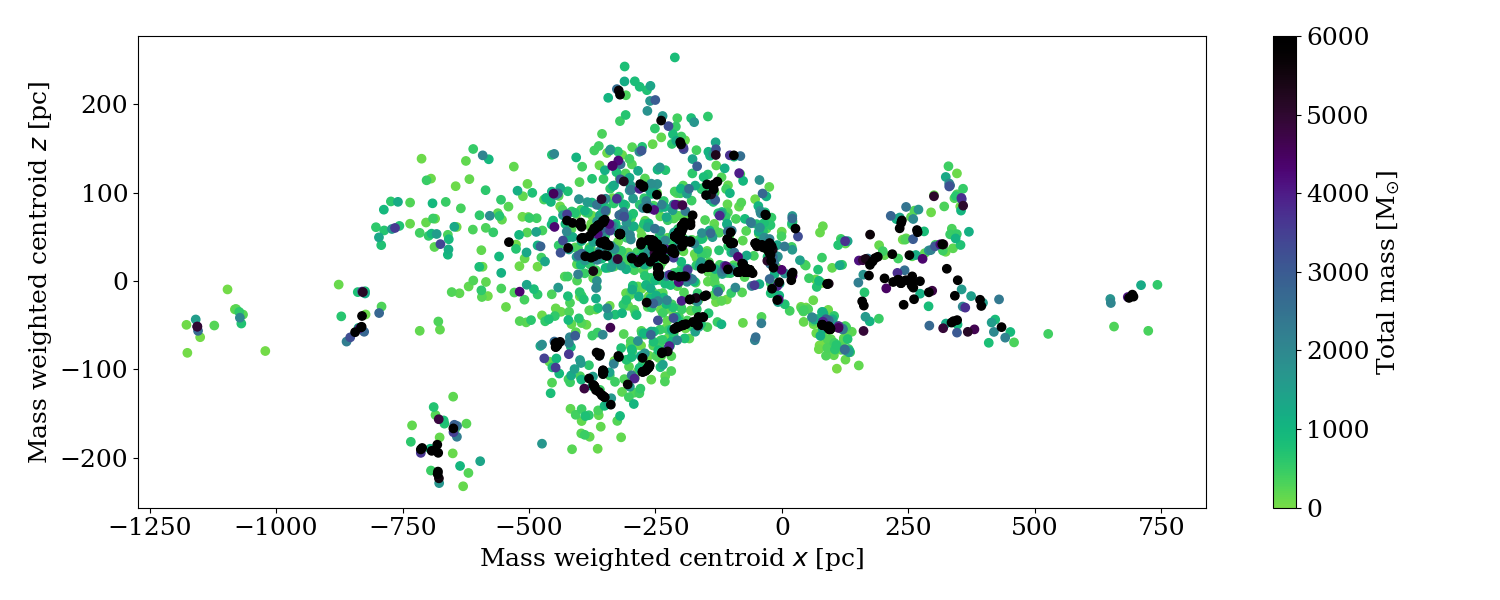}
  \end{subfigure}
  
\begin{subfigure}{0.8\textwidth}
  \centering
 \includegraphics[width=\textwidth]{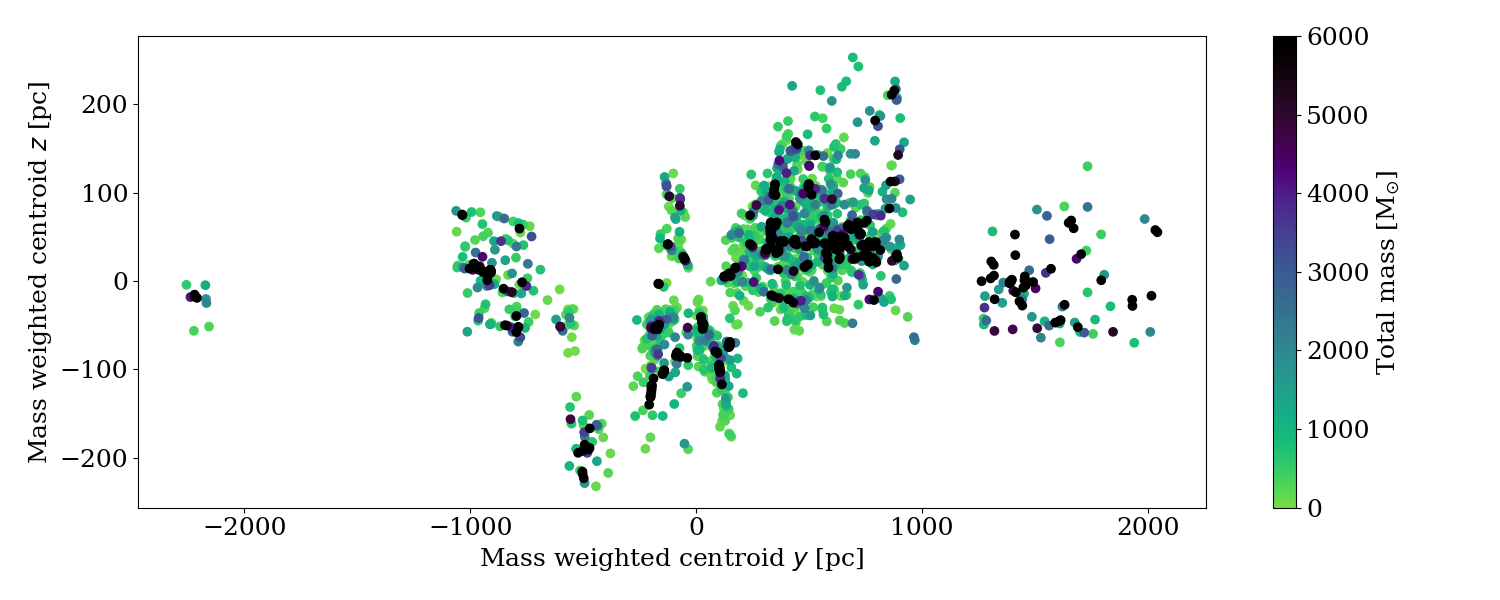}
  \subcaption{}
  \end{subfigure}
 \caption{Maps of the Mass weighted centroids of all structures identified by \texttt{astrodendro} in Heliocentric Cartesian coordinates.}
\label{fig:AllStructures_Centroid_SkyMaps}
\end{figure*}

Figure \ref{fig:lbol_inputdata} shows the full \citep{Fouesneau2022_LBol} catalogue sky map with the mapped regions outlined. In addition, figure \ref{fig:AllStructures_Centroid_SkyMaps} shows the mass weighted centroids of all structures, indicating the spread of the clouds and the individual structure on the Galactic plane and relative to it. 

In table \ref{tab:region_params} we present the results for the extended environment of the molecular clouds. The table includes the number of trunks (including isolated trunks) and leaves identified by \texttt{astrodendro} and their extent in Galactic coordinates, along with the total volume and total mass of the regions which are calculated by combining all trunks. 

In table \ref{tab:maintrunk_params} we present the results for the primary trunk -- the most massive trunk recovered from \texttt{astrodendro} -- which typically includes the main star formation regions of the molecular cloud. For the primary trunk we once again include the $l,b,d$ extents, ellipsoidal volume, volume enclosed by the trunk's contour and the filling factor ($\frac{V_{C}}{V_{E}}$). We also include the total mass and mass weighted centroid of the primary trunk. 

Finally in table \ref{tab:leaf_params} we identify structures in the dendrogram which correspond to known or any interesting features such as cloud formations, star-formation regions, cavities, filaments, etc., in the molecular clouds and their extended environments. For each identified we give their \texttt{astrodendro} assigned index; l, b, d extents; ellipsoidal volume and the volume enclosed by the structure's contour; filling factor; total mass; and relevant notes or identification. 

The tables in the text contains the most relevant parameters for this work. The electronic tables contain the full set of derived parameters as described in appendix \ref{sec:Ap:ElecTabs} in the supplementary material.

From these extended tables it is possible to compute the aspect ratios of the structures from the magnitudes of the three principle vectors $v_{max1}, v_{max2}, v_{max3}$. 
The ratios of these magnitudes give aspect ratios along each axis of the structure; if all ratios are close to one this indicates the cloud is close to spherical, while ratios close to $>>1$ indicate high elongation along the direction of the denominator or numerator respectively.

We find aspect ratios ranging from 1 to 11, indicating a diversity of shapes. Both near-spherical and highly elongated shapes are well represented, demonstrating \texttt{astrodendro}'s ability to capture arbitrary or filamentary shapes as well as spherical structures.

\setlength{\extrarowheight}{4pt}
\begin{table*}
    \centering
    \caption{Extended environment parameters derived from dendrograms}
    \label{tab:region_params}
    \begin{threeparttable}
    \begin{tabular}{lrrrrrrrrrr}
        \hline
         Region & n$_{\rm Trunk}$ & n$_{\rm Leaf}$ & $l_{min}$  & $l_{max}$  & 
         $b_{min}$  & $b_{max}$ &
         $d_{min}$  & $d_{max}$ &
         V$_{\rm C tot}$  & 
         M$_{\rm T}$   \\
         
        &&& 
        ($^{\circ}$) & ($^{\circ}$) & 
        ($^{\circ}$) & ($^{\circ}$) & 
        (pc) & (pc) &
        ($10^{3}$ pc$^{3}$) & 
        ($10^{3}$ M$_{\odot}$) \\
         
        (1) & (2) & (3) & (4) & (5) & (6) & (7) & (8) & (9) & (10) & (11) \\

        \hline \hline
        \\
        
California & 7 & 10 & 155.0  & 169.0  & -15.00  & -5.00  & 415  & 529  & 253 & 47.5 $^{ +0.1 } _{ -0.6 }$ \\

Camelopardalis & 101 & 58 & 132.0 & 158.0 & -5.00 & 10.00  & 140 & 930  & 1207 & 102.0 $^{ +0.1 } _{ -0.1 }$ \\

Canis Major & 3 & 9 & 220.6  & 226.2  & -4.00  & -0.01  & 1073  & 1200  & 128 & 16.92 $^{ +0.01 } _{ -0.02 }$ \\

Carina & 2 & 5 & 286.5  & 288.5  & -1.50  & -0.00  & 2250  & 2399  & 130 & 11.08 $^{+ 0.05 } _{- 0.35 }$ \\

Cassiopeia & 50 & 139 & 110.0  & 135.0 & -5.00  & 20.00  & 190  & 814  & 3473 & 287.5 $^{+ 0.2 } _{- 0.4 }$ \\

Cepheus & 41 & 198 & 90.1 & 125.0  & -5.00  & 25.00  & 270  & 1000  & 10903 & 716.1 $^{+ 0.8 } _{- 2.7 }$ \\

Chamaeleon & 6 & 52 & 290.0  & 308.0  & -22.00  & -10.00  & 100  & 277  & 115 & 29.8 $^{+ 3.0 } _{- 3.0}$ \\

CoalSack & 1 & 3 & 295.0  & 305.0  & -4.00  & 4.00  & 141  & 212  & 22 & 6.5 $^{+ 0.7 } _{- 0.7 }$ \\

Cygnus & 30 & 35 & 73.0  & 86.9  & -3.81  & 4.57  & 1280  & 2091  & 4458 & 502.2 $^{+ 0.3 } _{- 0.8 }$ \\

Lupus & 7 & 29 & 333.0 & 350.0  & 5.00  & 20.00  & 100 & 395 & 237 & 26.4 $^{+ 2.6 } _{- 2.6 }$ \\

MonR2 & 7 & 14 & 210.0  & 222.0  & -17.00  & -8.11  & 760 & 922  & 302 & 37.97 $^{+ 0.02 } _{- 0.06 }$ \\

Orion & 16 & 26 & 182.5  & 217.0  & -25.50 & -3.80  & 270  & 525  & 1213 & 181.5 $^{+ 18.1 } _{- 18.1 }$ \\

Perseus & 6 & 30 & 154.0 & 164.0 & -25.00  & -14.00  & 230 & 474  & 271 & 41.2 $^{+ 4.1 } _{- 4.1 }$ \\

Rosette & 7 & 7 & 205.3  & 209.0  & -4.00  & -0.27  & 1150  & 1329  & 51 & 7.38 $^{+ 0.02 } _{- 0.07 }$ \\

Taurus & 5 & 40 & 165.0  & 180.0  & -20.00  & -10.00  & 93  & 342  & 188 & 37.2 $^{+ 0.1 } _{- 1.9 }$ \\

Vela & 9 & 64 & 255.2  & 279.8  & -4.00  & 5.00  & 670  & 1090  & 4437 & 444.2 $^{+ 0.2 } _{- 0.5 }$ \\

         \hline
    \end{tabular}
  \begin{tablenotes}
  \footnotesize
    \item (1) Region name; 
    (2) Total number of trunks; 
    (3) Total number of leaves; 
    (4),(5) Extent in $l^{\circ}$; 
    (6),(7) Extent in $b^{\circ}$; 
    (8),(9) Extent in distance along line-of-sight (pc); 
    (10) Total volume of the region held within the contours laid down by astrodendro to determine locations of the structures. This is calculated by adding up the volume within all trunks which are the largest possible structures produced by astrodendro and hence encompass the total volume of the region; 
    (11) total gas$+$dust mass of the region calculated by adding up all the densities in the same way as the total volume and converted to a total mass in solar masses using the method outlines in Sec 7 of Dharmawardena+2021b.   
  \end{tablenotes}
    
    \end{threeparttable}
\end{table*}


\setlength{\extrarowheight}{4pt}
\begin{table*}
    \caption{Parameters of the primary trunk derived from dendrograms}
    \label{tab:maintrunk_params}
    \begin{threeparttable}
    \begin{tabular}{lrrrrrrrrrrr}
        \hline
         Region & 
         Centroid &
         $l_{min}$  & $l_{max}$  & 
         $b_{min}$  & $b_{max}$ &
         $d_{min}$  & $d_{max}$ &
         V$_{\rm Etrunk}$  & V$_{\rm Ctrunk}$  & Filling & 
         M$_{\rm Ttrunk}$  \\
         
         & ($l^{\circ}$,$b^{\circ}$,d\_pc) &
        ($^{\circ}$) & ($^{\circ}$) & 
        ($^{\circ}$) & ($^{\circ}$) & 
        (pc) & (pc) & 
        ($10^{3}$ pc$^{3}$) & 
        ($10^{3}$ pc$^{3}$)  & Factor & 
        ($10^{3}$ M$_{\odot}$) \\
         
        (1) & (2) & (3) & (4) & (5) & (6) & (7) & (8) & (9) & (10) & (11) &(12) \\
        \hline \hline
        \\
        
California & 162.3 , -8.90 , 474 & 155.0  & 169.0  & -13.54  & -5.00  & 415  & 526  & 756 & 231 & 0.31 & 44.0 $^{ +0.1 } _{ -0.5 }$ \\
        
Camelopardalis & 144.5 , 2.92 , 301 & 132.0 & 158.0  & -4.97  & 9.98  & 141  & 481  & 3052 & 521 & 0.17 & 45.3 $^{ +0.04 } _{ -0.06 }$ \\ 

Canis Major & 223.8 , -1.96 , 1150 & 221.0  & 226.2  & -3.99  & -0.01  & 1073  & 1200  & 645 & 126 & 0.20 & 16.71 $^{ +0.01 } _{ -0.02 }$ \\

Carina & 287.3 , -0.47 , 2313 & 286.5  & 288.5  & -1.50  & -0.00  & 2250 & 2365  & 864 & 121 & 0.14 & 10.54 $^{+ 0.05 } _{- 0.35 }$  \\
        
Cassiopeia & 121.1 , 5.23 , 512 & 110.0  & 135.0  & -5.00  & 19.97  & 190  & 811  & 27190 & 2883 & 0.11 & 241.8 $^{+ 0.2 } _{- 0.3 }$ \\

Cepheus & 109.3 , 4.52 , 643 & 90.1  & 125.0  & -4.98  & 24.96  & 270  & 999  & 58066 & 9524 & 0.16 & 628.6 $^{+ 0.7 } _{- 2.4 }$ \\

Chamaeleon & 298.8 , -15.69 , 202 & 290.0  & 308.0  & -22.00  & -10.00  & 100  & 275  & 586 & 106 & 0.18 & 28.6 $^{+ 2.9 } _{- 2.9 }$ \\

Coalsack & 298.6 , -0.97 , 187 & 295.0  & 305.0  & -4.00  & 4.00  & 141  & 212  & 66 & 22 & 0.33 & 6.5 $^{+ 0.7 } _{- 0.7 }$ \\

Cygnus & 80.4 , -0.03 , 1410 & 73.0  & 85.3  & -3.81  & 3.80  & 1280  & 1611 & 13338 & 2825 & 0.21 & 319.5 $^{+ 0.3 } _{- 0.6 }$ \\

Lupus & 343.2 , 7.93 , 200 & 334.6  & 350.0  & 5.00  & 16.56  & 100  & 312  & 329 & 95 & 0.29 & 9.3 $^{+ 0.9 } _{- 0.9 }$ \\

MonR2 & 215.9 , -12.41 , 861 & 210.7  & 220.4  & -17.00  & -8.28  & 760  & 922  & 1705 & 278 & 0.16 & 34.92 $^{+ 0.02 } _{- 0.05 }$ \\

Orion & 201.7 , -14.85 , 394 & 182.5 & 217.0  & -25.50  & -3.80  & 270  & 524  & 4737 & 1111 & 0.23 & 170.0 $^{+ 17.0 } _{- 17.0 }$ \\

Perseus & 159.0 , -19.19 , 313 & 154.0  & 164.0  & -25.00 & -14.00  & 230  & 456  & 907 & 251 & 0.28 & 38.8 $^{+ 3.9 } _{- 3.9 }$  \\

Rosette & 207.5 , -2.26 , 1304 & 205.4  & 209.0  & -3.11  & -1.41  & 1274  & 1329  & 95 & 29 & 0.31 & 4.97 $^{+ 0.02 } _{- 0.07 }$ \\

Taurus & 172.2 , -14.17 , 221 & 165.0  & 180.0  & -20.00  & -10.00  & 93 & 342  & 544 & 175 & 0.32 & 35.2 $^{+ 0.1 } _{- 1.9 }$ \\

Vela & 266.0 , 0.80 , 912 & 255.3  & 277.3  & -4.00  & 5.00  & 670  & 1090  & 14737 & 4293 & 0.29 & 433.4 $^{+ 0.2 } _{- 0.5 }$ \\

         \hline
    \end{tabular}
  \begin{tablenotes}
  \footnotesize
    \item (1) Region name; 
    (2) Mass weighted centroid of primary trunk;
    (3),(4) Extent of primary trunk in $l^{\circ}$; 
    (5),(6) Extent of primary trunk in $b^{\circ}$; 
    (7),(8) Extent of primary trunk along distance line-of-sight (pc); 
    (9) Total volume of the ellipsoid fitted to the maximum possible 3D extents of the primary trunk orthogonal to one-another;
    (10) Total volume of the region held within the contour laid down by astrodendro to determine locations of the primary trunk; 
    (11) $\frac{ V_{Ctrunk}}{V_{Etrunk}}$ indicating how fragmented the primary trunk is; 
    (12) total gas$+$dust mass of the primary trunk calculated by adding up all the densities in the same way as the total volume and converted to a total mass in solar masses using the method outlines in Sec 7 of Dharmawardena+2021b.

  \end{tablenotes}
    
    \end{threeparttable}
\end{table*}


\setlength{\extrarowheight}{4pt}
\begin{table*}
\begin{adjustwidth}{-1cm}{}
    \centering
    \caption{Densest regions forming the main regions of the molecular clouds determined by x-matching to literature data}
    \label{tab:leaf_params}
    \begin{threeparttable}
    \begin{tabular}{lrrrrrrrrrl}
        \hline
         Region & Index &  
         Centroid &
         $l_{min}$  & $l_{max}$  & 
         $b_{min}$  & $b_{max}$ &
         $d_{min}$  & $d_{max}$ &
         M$_{\rm T}$ & Notes \\
         
         && ($l^{\circ}$,$b^{\circ}$,d\_pc) &
        ($^{\circ}$) & ($^{\circ}$) & 
        ($^{\circ}$) & ($^{\circ}$) & 
        (pc) & (pc)
        & (M$_{\odot}$) &  \\
         
         (1) & (2) & (3) & (4) & (5) & (6) & (7) & (8) & (9) & (10)  \\
        \hline \hline
        \\

\textbf{California} & & & & & & & & & & \\

Spine & L11 & 163.3 , -8.28 , 475 & 159.9  & 166.1 & -9.27  & -6.92  & 450  & 509  & 2655 $^{ 3 } _{ 5 }$ & sheet elongated along los \\
& L9 & 164.3 , -8.41 , 501 & 162.7  & 165.4 & -8.96 & -7.82  & 490 & 513 & 1343 $^{ +4 } _{ -8 }$ & extends furthest along los \\

Bubble top & L18 & 157.6 , -8.73 , 461 & 155.4 & 159.9  & -9.70  & -8.01  & 440  & 483  & 2195 $^{ +4 } _{ -6 }$ & connects to spine with C10 \\
& L12 & 161.1 , -11.50 , 490 & 159.3  & 162.3  & -12.31 & -10.14 & 464  & 513  & 780 $^{ +2 } _{ -5 }$ & hosts LDN 1462, 1463, 1464, 1469 \\

Bubble bottom & L15 & 157.8 , -12.41 , 487 & 156.7  & 159.0  & -13.28  & -11.59  & 475  & 505  & 895 $^{ +3 } _{ -9 }$ & hosts LDN 1442, 1449, 1456 \\
& L23 & 159.5 , -11.83 , 435 & 158.4  & 160.6  & -12.86  & -9.30  & 420 & 451  & 1532 $^{ +6 } _{ -14 }$ & connect bubble top and bottom  \\

Bubble back & L16 & 157.6, -10.30 , 490 & 156.9 & 158.4  & -11.06  & -9.46  & 482  & 499  & 955 $^{ +4 } _{ -8 }$ & central back edge of bubble  \\

Outer clouds & T13 & 155.6 , -14.19 , 518 & 155.0  & 156.7  & -15.00  & -12.96  & 507  & 529  & 827 $^{ +11 } _{ -185 }$ & host to LDN1434  \\
& T14 & 168.7 , -13.93 , 477 & 168.1  & 169.0  & -15.00  & -12.75  & 465  & 489 & 259 $^{ +3 } _{ -19 }$ & host to LDN1497, LDN1500 \\
& L7 & 166.9 , -10.27 , 490 & 165.3  & 168.7  & -11.72  & -8.60  & 469  & 510  & 3464 $^{ +5 } _{ -14 }$ & L20 on sky overlap; not visible in 2D ext; massive \\
& L20 & 167.3 , -9.88 , 434 & 165.8  & 169.0 & -11.07 & -8.62  & 422 & 448  & 960 $^{ +4 } _{ -10 }$ & L20 on sky overlap; not visible in 2D ext \\

\textbf{Camelopardalis} & & & & & & & & & \\

Central & L166 & 141.7 , -3.79 , 340 & 140.5  & 143.0  & -4.67  & -2.60  & 328  & 351  & 151 $^{ +1 } _{ -1 }$ & TGU 942 local arm comp \\
& L176 & 142.4 , -2.59 , 298 & 139.9  & 145.3  & -3.58  & -1.62  & 285  & 326  & 250 $^{ +1 } _{ -1 }$ & TGU 942 local arm comp; oblong cloud \\
& L179 & 144.3 , 0.61 , 219 & 139.6  & 150.6  & -1.46  & 2.31  & 162  & 270  & 2335 $^{ +5 } _{ -6 }$ & largest TGU 942 local arm comp \\
& T28 & 147.2 , -3.91 , 806 & 146.6  & 147.9  & -4.23  & -3.62  & 795  & 818  & 161 $^{ +1 } _{ -2 }$ & TGU 942 Cam OB1 layer comp \\
& T36 & 146.8 , -1.03 , 762 & 146.3  & 147.3  & -1.42  & -0.64  & 748  & 780 & 164 $^{ +1 } _{ -1 }$ & TGU 942 Cam OB1 layer comp \\
& T51 & 143.4 , -1.39 , 739 & 142.8  & 144.1  & -1.79  & -0.96  & 725  & 752  & 233 $^{ +2 } _{ -2 }$ & TGU 942 Cam OB1 layer comp \\
& T55 & 141.0 , -0.07 , 752 & 140.0  & 142.1  & -0.35  & 0.21  & 740  & 763  & 182 $^{ +1 } _{ -1 }$ & TGU 942 Cam OB1 layer comp \\

South East & L148 & 148.7 , 2.14 , 361 & 146.5  & 151.4  & 1.00  & 3.59  & 324  & 395  & 1134 $^{ +4 } _{ -6 }$ & TGU 994 \\
& L184 & 156.3 , -0.20 , 160 & 153.3  & 157.7  & -1.75  & 1.14  & 143  & 182  & 185 $^{ +1 } _{ -1 }$ & TGU 1041, 1045, 1056 \\
& L154 & 156.1 , 0.17 , 274 & 153.9  & 157.9  & -1.51  & 1.95  & 222  & 341  & 914 $^{ +3 } _{ -3 }$ & -- \\
& T15 & 154.3 , 3.20 , 751 & 152.8  & 155.5  & 2.02  & 4.43  & 695  & 799  & 3386 $^{ +13 } _{ -24 }$ & large clouds not visible in 2D ext  \\
& T2 & 155.8 , 4.11 , 841 & 152.6  & 158.0  & 3.31  & 5.11  & 804  & 884  & 3575 $^{ +10 } _{ -15 }$ & large clouds not visible in 2D ext \\

North West & L138 & 134.3 , 7.46 , 481 & 132.2  & 137.1  & 6.72  & 8.38  & 440  & 520  & 1544 $^{ +4 } _{ -6 }$ & TGU 878 local arm comp \\
& T128 & 134.9 , 7.62 , 488 & 132.0  & 138.6  & 6.33  & 10.00  & 433  & 534  & 4167 $^{ +8 } _{ -12 }$ & TGU 878 local arm comp \\
& T42 & 133.3 , 9.37 , 874 & 132.0  & 134.7  & 8.62  & 10.00  & 841  & 913 & 2088 $^{ +9 } _{ -17 }$ & TGU 878 Cam OB1 comp \\
& T25 & 141.4 , 4.60 , 881 & 140.9  & 141.9  & 4.30  & 4.92  & 872  & 889  & 115 $^{ +1 } _{ -1 }$ & TGU 937 \\
& T31 & 138.8 , 1.95 , 878 & 137.8  & 139.9  & 1.43  & 2.46  & 857  & 896  & 968 $^{+ 6 } _{ -10 }$ & TGU 912, Sh 2-202 \\
& L32 & 140.4 , 3.50 , 859 & 139.6  & 141.1 & 3.12  & 3.91  & 844  & 877  & 346 $^{ +2 } _{ -2 }$ & TGU 929 comp \\
& T19 & 140.4 , 3.45 , 888 & 139.3  & 141.3  & 2.87  & 3.99  & 842  & 919  & 1533 $^{+ 6 } _{ -11 }$ & TGU 929 comp \\

\textbf{Canis Major} & & & & & & & & & \\

East & C9 & 224.5 , -0.61 , 1162 & 222.3  & 226.2  & -1.28  & -0.01  & 1113  & 1199  & 5177 $^{+ 5 } _{- 9 }$ & composed of L13, L14, L15; host LDN 1658 \\

Central & L5 & 222.6 , -1.93 , 1184 & 222.1  & 223.4  & -2.41  & -1.51  & 1170  & 1197  & 648 $^{+ 1 } _{- 2 }$ & host LDN 1657, 1570 \\
& L8 & 222.4 , -3.36 , 1168 & 221.3  & 223.0  & -4.00  & -2.62  & 1148  & 1189  & 2235 $^{+ 3 } _{- 7 }$ & -- \\
& L11 & 224.5 , -2.83 , 1160 & 223.4 & 225.3  & -3.54  & -1.91  & 1129  & 1187  & 2745 $^{+ 2 } _{- 4 }$ & host LDN 1657, 1570 \\
& L17 & 223.7 , -1.88 , 1102 & 222.7  & 224.5  & -2.62  & -1.01  & 1075  & 1122  & 2744 $^{+ 7 } _{- 16 }$ & -- \\

North & L16 & 220.9 , -1.98 , 1089 & 220.6  & 221.1  & -2.38  & -1.59  & 1081  & 1097  & 93.5 $^{+ 0.3 } _{- 0.3 }$ & host LDN 1653-1656; unclear in 2D extinction \\

\hline
\end{tabular}

  \begin{tablenotes}
  \footnotesize
    \item (1) Region name; 
    (2) Assigned index in dendrogram with type of structure (L:Leaf; T:Trunk; C:Child)
    (3) Mass-weighted centroid coordinates of the structure
    (4) Extent of structure in $l^{\circ}$; 
    (5) Extent of structure in $b^{\circ}$; 
    (6) Extent of structure along distance/line-of-sight (pc); 
    (7) Total volume of the ellipsoid fitted to the maximum possible 3D extents of the primary trunk orthogonal to one-another;
    (8) Total volume of the region held within the contour laid down by astrodendro to determine locations of the primary trunk; 
    (9) $\frac{ V_{Ctrunk}}{V_{Etrunk}}$ indicating how fragmented the primary trunk is; 
    (10) Total mass of structure
  \end{tablenotes}
    
\end{threeparttable}
\end{adjustwidth}
\end{table*}

\addtocounter{table}{-1}
\begin{table*}
\begin{adjustwidth}{-1cm}{}
    \centering
    \caption{\emph{cont.}}
    \begin{tabular}{lrrrrrrrrrl}
        \hline
        Region & Index &  
         Centroid &
         $l_{min}$  & $l_{max}$  & 
         $b_{min}$  & $b_{max}$ &
         $d_{min}$  & $d_{max}$ &
         M$_{\rm T}$ & Notes \\
         
         && ($l^{\circ}$,$b^{\circ}$,d\_pc) &
        ($^{\circ}$) & ($^{\circ}$) & 
        ($^{\circ}$) & ($^{\circ}$) & 
        (pc) & (pc)
        & (M$_{\odot}$) &  \\
         
         (1) & (2) & (3) & (4) & (5) & (6) & (7) & (8) & (9) & (10)  \\
        \hline \hline
        \\

\textbf{Carina} & & & & & & & & & \\

North & L6 & 287.0 , -0.44 , 2340 & 286.5  & 288.1  & -0.78  & -0.00  & 2308  & 2359  & 4082 $^{+ 5 } _{- 14 }$ & extends to part of South cloud \\
& L0 & 286.9 , -1.31 , 2254 & 286.5  & 287.4  & -1.50  & -0.85  & 2250  & 2263 & 300 $^{+ 1 } _{- 2 }$ & lower filamentary section of North cloud \\
& L2 & 286.7, -0.51, 2264 & 286.5  & 287.2  & -0.92    &  -0.03  & 2250  & 2312   & 1752 $^{+ 44 } _{- 336 }$ & nearby comp. of North cloud \\

South & L8 & 288.0 , -1.38 , 2338 & 287.8  & 288.3  & -1.50  & -1.02 & 2315  & 2357 & 335 $^{+ 1 } _{- 1 }$ & southern pillar \\

Outer cloud & L7 & 288.1 , -0.11 , 2285 & 287.6  & 288.5  & -0.40  & -0.01  & 2259  & 2312 & 1261 $^{+ 2 } _{- 3 }$ & large leaf invisible in 2D ext \\

\textbf{Cassiopeia} & & & & & & & & & \\

Cas OB5 region & L243 & 115.7 , -2.81 , 528 & 114.8  & 117.3  & -4.40  & -1.40  & 507  & 549  & 917 $^{+ 5 } _{- 6 }$ & overlaps on sky with L282 \\
& L282 & 117.4 , -3.08 , 373 & 115.4  & 119.3  & -3.74  & -2.27  & 347  & 405 & 488 $^{+ 2 } _{- 3 }$ & host LDN 1265 \\

Cas OB4 region & L84 & 123.8 , -0.61 , 658 & 121.7  & 125.9  & -1.12  & 0.20  & 618  & 681  & 724 $^{+ 3 } _{- 4 }$ & host L1293 \\
& L236 & 124.4 , -0.41 , 408 & 122.6 & 127.5  & -1.48  & 0.86  & 365  & 446  & 874 $^{+ 4 } _{- 5 }$ & host LDN 1297 \\

Cas OB6 region & L189 & 131.4 , 4.34 , 434 & 128.2  & 133.7  & 3.96  & 4.82  & 425  & 444  & 424 $^{+ 3 } _{- 3 }$ & host LDN 1335 \\

Cep OB4 shell  & L68 & 121.5 , 3.42 , 760 & 120.4  & 123.0 & 2.87  & 4.25  & 747  & 773  & 1116 $^{+ 6 } _{- 10 }$ & host M120.1+0.3 cloud \\
region &&&&&&&&&&\\

Cep flare shell & L265 & 126.8 , 13.97 , 328 & 123.2 & 129.7  & 11.90  & 15.71  & 282  & 379  & 2318 $^{+ 9 } _{- 13 }$ & wraps around Cep flare shell \\
region & T38 & 130.2 , 11.48 , 719 & 129.0  & 131.5  & 10.88  & 12.05  & 702  & 736  & 687 $^{+ 6 } _{- 11 }$ & host LDN 1306 \\
& T1 & 133.5 , 8.59 , 663 & 132.0  & 135.0  & 6.69 & 12.24  & 568  & 763  & 4534 $^{+ 9 } _{- 14 }$ & host LDN 1355, 1358 \\
& T141 & 118.9 , 10.76 , 577 & 112.5  & 125.5  & 7.70  & 16.70  & 496  & 669  & 9553 $^{+ 21 } _{- 38 }$ & behind Cep OB4 shell; \\
&&&&&&&&&&large arc-like structure connecting Cas to Cep\\

\textbf{Cepheus} & & & & & & & & & \\

Cep flare region & L344 & 101.9 , 16.65 , 427 & 99.0 & 105.1  & 12.79 & 19.94  & 381 & 465  & 3980 $^{+ 18 } _{- 29 }$ & host NGC 7023, LDN 1157 \\
& C292 & 111.3 , 15.66 , 385 & 105.0  & 116.4 & 11.03 & 20.17  & 335  & 435  & 7639 $^{+ 31 } _{- 48 }$ & Largest region in Cepheus Flare \\
& T100 & 110.3 , 12.88 , 947 & 108.1  & 112.5  & 11.11  & 14.75  & 868  & 997  & 9910 $^{+ 56 } _{- 143 }$ & dense clouds behind Cep flare clouds \\

Cep OB4 shell & L256 & 118.2 , 12.76 , 451 & 116.9 & 119.2  & 12.06  & 13.46  & 444  & 460  & 183 $^{+ 2 } _{- 3 }$ & host LDN 1262 \\
region &&&&&&&&&&\\

Cep bubble region & T307 & 98.4 , 7.38 , 877 & 96.8  & 100.2  & 3.26  & 10.58  & 775  & 968  & 10334 $^{+ 43 } _{- 91 }$ & host IC 1396; wraps around Cep bubble \\
& T182 & 106.8 , 12.39 , 847 & 103.4 & 110.7  & 9.87  & 14.43  & 809  & 892  & 5729 $^{+ 25 } _{- 50 }$ & host NGC 7129 \\
& L199 & 107.3 , 5.35 , 881 & 106.7 & 107.9  & 4.66 & 5.94  & 849  & 906  & 1501 $^{+ 13 } _{- 19 }$ & host PDR S140 \\

Cep OB3 region & L198 & 110.4 , 2.25 , 680 & 108.2  & 112.6 & 1.06  & 3.39  & 628  & 725  & 7413 $^{+ 59 } _{- 129 }$ & host Cep OB3 and PDR S155 \\
& L96 & 115.2 , 5.05 , 848 & 114.4  & 115.9  & 4.31  & 5.73  & 835  & 860  & 999 $^{+ 8 } _{- 13 }$ & host M115.5+40 \\

Cep OB4 region & C50 & 118.4 , 4.84 , 895 & 117.4  & 119.5  & 3.99  & 5.78  & 848  & 932  & 3531 $^{+ 19 } _{- 30 }$ & host Cep OB4 \\

Cep OB2 region & L348 & 98.3 , 13.97 , 587 & 96.8  & 99.9  & 12.38  & 15.61  & 553  & 630 & 2564 $^{+ 40 } _{- 85 }$ & host Cep OB2 \\

\textbf{Chamaeleon} & & & & & & & & & \\

Cha I & L36 & 295.5 , -14.64 , 170\ & 292.8 & 298.4 & -16.59  & -13.18 & 161 & 175  & 780 $^{+ 8 } _{- 17 }$ & nearby Northern comp. \\
& L54 & 297.1 , -16.21 , 187 & 295.4  & 298.2 & -18.90  & -15.07  & 181  & 191  & 645 $^{+ 14 } _{- 27 }$ & distant Southern comp. \\

Cha II, III & L69 & 303.2 , -15.78 , 172 & 302.1  & 304.4  & -18.04  & -13.64  & 165  & 181  & 747 $^{+ 7 } _{- 17 }$ & Cha II only \\
& C64 & 303.0 , -16.25 , 175 & 300.3  & 305.7  & -19.06  & -13.08  & 160  & 194 & 1549 $^{+ 155 } _{- 155 }$ & outlines Cha II, III; edge comp. of Cha III L74 \\
& L74 & 302.0 , -17.20 , 187 & 301.0  & 302.9 & -17.78  & -16.60  & 182  & 191  & 134 $^{+ 2 } _{- 3 }$ & edge comp. of Cha III \\

Outer clouds & L53 & 296.6 , -11.42 , 188 & 293.9  & 299.1  & -13.27  & -10.00  & 174  & 200  & 969 $^{+ 11 } _{- 25 }$ & above Cha I; nearby part of L53 + L58 cloud \\
& L58 & 297.3 , -13.51 , 199 & 294.9  & 299.2  & -14.97 & -12.41  & 191  & 207  & 730 $^{+ 4 } _{- 6 }$ & distant part of L53 + L58 cloud \\

\hline
\end{tabular}
\end{adjustwidth}
\end{table*}

\addtocounter{table}{-1}
\begin{table*}
\begin{adjustwidth}{-1cm}{}
    \centering
    \caption{\emph{cont.}}
    \begin{tabular}{lrrrrrrrrrl}
        \hline
        Region & Index &  
         Centroid &
         $l_{min}$  & $l_{max}$  & 
         $b_{min}$  & $b_{max}$ &
         $d_{min}$  & $d_{max}$ &
         M$_{\rm T}$ & Notes \\
         
         && ($l^{\circ}$,$b^{\circ}$,d\_pc) &
        ($^{\circ}$) & ($^{\circ}$) & 
        ($^{\circ}$) & ($^{\circ}$) & 
        (pc) & (pc)
        & (M$_{\odot}$) &  \\
         
         (1) & (2) & (3) & (4) & (5) & (6) & (7) & (8) & (9) & (10)  \\
        \hline \hline
        \\

\textbf{Coalsack} & & & & & & & & & \\

Coalsack & L4 & 298.9 , -0.84 , 192 & 295.0  & 304.5  & -4.00  & 4.00  & 168 & 209 & 5392 $^{+ 539 } _{- 539 }$ & largest leaf \\

\textbf{Cygnus X} & & & & & & & & & \\

CygX North & L19 & 82.2 , 0.97 , 1319 & 81.0  & 83.4  & -0.86  & 2.80  & 1286 & 1357 & 16611 $^{+ 48 } _{- 81 }$ & most massive comp of CygX North \\
& L20 & 83.0 , 2.12 , 1423 & 81.6 & 83.9 & 0.85  & 3.64  & 1403 & 1444  & 5615 $^{+ 26 } _{- 46 }$ & overlaps on sky with L31 \\
& L31 & 82.1 , 2.70 , 1569 & 81.3  & 82.9  & 2.21 & 3.42  & 1542  & 1604 & 2801 $^{+ 10 } _{- 11 }$ & furthest along los \\

CygX South & L36 & 77.8 , -0.00 , 1292 & 75.8 & 80.7 & -1.78 & 1.80  & 1280  & 1327& 15161 $^{+ 63 } _{- 136 }$ & largest area on sky \\
& L39 & 78.8 , -0.87 , 1347 & 77.3  & 80.3  & -2.13  & 0.30  & 1312  & 1384  & 22247 $^{+ 108 } _{- 312 }$ & most massive leaf; main comp of CygX South \\
& L50 & 77.9 , 2.40 , 1343 & 77.3  & 78.7  & 1.92  & 2.86  & 1332  & 1354  & 842 $^{+ 5 } _{- 6 }$ & northern most edge \\

\textbf{Lupus} & & & & & & & & & \\

Lup II -- IX & C5 & 341.9 , 8.54 , 121 & 336.4  & 346.8  & 5.00  & 13.16 & 100 & 155  & 1285 $^{+ 7 } _{- 57 }$ & densest comp of Lupus II - IX \\

Outer clouds & L8 & 343.3 , 8.23 , 179 & 338.9  & 349.7  & 5.00  & 11.90  & 147  & 235 & 2396 $^{+ 19 } _{- 345 }$ & massive cloud behind Lupus clouds \\
& T23 & 338.4 , 6.99 , 343 & 333.0  & 343.1  & 5.00  & 10.82  & 268  & 395  & 6557 $^{+ 656 } _{- 656 }$ & largest sec. trunk \\

\textbf{Mon R2} & & & & & & & & & \\

North & L5 & 214.0 , -12.72 , 883 & 212.8 & 215.3  & -13.44  & -11.76& 842 & 915  & 3638 $^{+ 5 } _{- 8 }$ & host NGC 2170 \\

West & L18 & 216.1 , -15.26 , 869 & 215.6 & 216.5  & -15.95  & -14.42 & 843 & 886 & 1695 $^{+ 4 } _{- 6 }$ & host NGC 2149 \\

East & L24 & 217.5 , -12.28 , 829 & 216.6 & 218.3  & -12.89  & -11.45 & 777  & 863 & 2238 $^{+ 3 } _{- 4 }$ & host TGU 1523 \\
& C13 & 219.3 , -10.11 , 891 & 217.9  & 220.4  & -11.80  & -8.32  & 861 & 920 & 4811 $^{+ 7 } _{- 13 }$ & host LBN 1015, 1017 and LDN 1652 \\

Outer clouds & L11 & 217.4 , -12.24 , 895 & 216.4 & 218.2  & -13.07  & -10.77  & 881  & 909  & 953 $^{+ 2 } _{- 3 }$ & not seen in 2D ext \\
& L21 & 216.9 , -10.68 , 851 & 216.1  & 217.7  & -11.41 & -10.05  & 838 & 866  & 793 $^{+ 2 } _{- 2 }$ & not seen in 2D ext \\

 \textbf{Orion} & & & & & & & & & \\
 
Ori A & L46 & 211.5 , -19.42 , 421 & 204.5 & 215.6  & -21.05  & -16.71  & 382  & 444  & 21797 $^{+ 172 } _{- 1888 }$ & L33, L41, L50 background clumps not included \\

Ori B & L30 & 206.6 , -15.55 , 454 & 204.9  & 208.5  & -16.84  & -13.68 & 422  & 484  & 4825 $^{+ 37 } _{- 127 }$ & largest leaf of Ori B; part of comp1 in D22 \\
& L54 & 211.1 , -15.04 , 411 & 209.6  & 212.3  & -16.40  & -14.00  & 403  & 422  & 444 $^{+ 4 } _{- 6 }$ & nearby part of comp 1 in D22 \\
& L48 & 205.4 , -11.92 , 408 & 204.1  & 206.6 & -14.04  & -9.78  & 391  & 422  & 1972 $^{+ 15 } _{- 22 }$ & nearby part of comp 2 in D22 \\
& L26 & 205.2 , -12.39 , 464 & 204.3  & 206.3  & -14.66  & -10.61  & 453  & 478  & 1476 $^{+ 11 } _{- 20 }$ & center of Ori B \\
& L23 & 205.5 , -7.77 , 457 & 204.0  & 206.9 & -9.50  & -5.83  & 443  & 482  & 2967 $^{+ 25 } _{- 153 }$ & host LDN 1629, 1628, 1633, 1623 \\
& L43 & 202.1 , -8.66 , 409 & 200.5  & 203.6 & -10.07  & -7.52  & 399  & 419  & 477 $^{+ 6 } _{- 11 }$ & closest along los; edge of Ori B \\
& L15 & 203.9 , -11.31 , 499 & 202.4  & 205.2  & -13.90  & -9.51  & 483  & 512  & 3560 $^{+ 33 } _{- 169 }$ & furthest along los; distant part of comp 2 in D22 \\
& L24 & 209.1 , -13.44 , 493 & 208.3  & 209.8  & -14.44  & -12.45  & 481 & 504  & 467 $^{+ 5 } _{- 7 }$ & distant part of comp 1 in D22 \\
& L16 & 201.7 , -10.33 , 461 & 199.9  & 204.0  & -11.96  & -8.76  & 442  & 480  & 3614 $^{+ 24 } _{- 75 }$ & edge of Ori B \\

$\lambda$ Ori & L53 & 195.2 , -16.31 , 368 & 193.7  & 196.5  & -17.76  & -14.27  & 349  & 386  & 1383 $^{+ 10 } _{- 18 }$ & bottom back edge of bubble; joined to L58 by C52 \\
& L58 & 193.0 , -13.09 , 335 & 191.3  & 194.9  & -16.80  & -11.30  & 325  & 347  & 732 $^{+ 5 } _{- 8 }$ & bottom front edge of bubble; joined to L53 by C52 \\
& L39 & 193.0 , -12.13 , 385 & 188.3  & 197.3  & -15.60  & -9.36 & 356  & 418  & 10878 $^{+ 62 } _{- 356 }$ & right edge of bubble \\
& L45 & 198.3 , -8.62 , 389 & 196.6  & 199.8  & -9.68  & -7.67  & 377  & 402  & 487 $^{+ 4 } _{- 5 }$ & left top bubble edge; holds LDN 1598 \\
& L42 & 197.9 , -15.39 , 407 & 193.9  & 201.8  & -16.44  & -14.14  & 395  & 421  & 1479 $^{+ 9 } _{- 16 }$ & left bottom edge of bubble \\
& L19 & 192.5 , -8.40 , 434 & 188.8  & 195.2  & -11.94  & -6.41  & 416  & 451  & 2028 $^{+ 15 } _{- 86 }$ & furthest along los \\
& L31 & 199.6 , -12.39 , 432 & 198.1  & 201.4  & -13.96  & -10.84  & 417  & 448 & 1594 $^{+ 12 } _{- 40 }$ & furthest along los \\

$\lambda$ fil & L60 & 188.1 , -17.35 , 291 & 183.4  & 194.7  & -25.50 & -5.42 & 274  & 330  & 12851 $^{+ 39 } _{- 99 }$ & main fil comp. \\
& L59 & 187.5 , -21.52 , 326 & 185.3  & 189.7  & -25.35  & -18.13  & 307  & 342  & 1636 $^{+ 9 } _{- 17 }$ & bottom edge; furthest along los \\

\textbf{Perseus} & & & & & & & & & \\

Perseus & L39 & 160.3 , -18.67 , 333 & 159.3 & 161.3  & -20.87 & -16.29 & 302  & 355  & 2428 $^{+ 33 } _{- 88 }$ & includes IC348 \\
& L46 & 158.4 , -22.02 , 325 & 156.8  & 159.9  & -24.16  & -20.89  & 306  & 339  & 539 $^{+ 3 } _{- 4 }$ & includes NGC 1333 \\

\hline
\end{tabular}
\end{adjustwidth}
\end{table*}

\addtocounter{table}{-1}
\begin{table*}
\begin{adjustwidth}{-1cm}{}
    \centering
    \caption{\emph{cont.}}
    \begin{tabular}{lrrrrrrrrrl}
        \hline
        Region & Index &  
         Centroid &
         $l_{min}$  & $l_{max}$  & 
         $b_{min}$  & $b_{max}$ &
         $d_{min}$  & $d_{max}$ &
         M$_{\rm T}$ & Notes \\
         
         && ($l^{\circ}$,$b^{\circ}$,d\_pc) &
        ($^{\circ}$) & ($^{\circ}$) & 
        ($^{\circ}$) & ($^{\circ}$) & 
        (pc) & (pc)
        & (M$_{\odot}$) &  \\
         
         (1) & (2) & (3) & (4) & (5) & (6) & (7) & (8) & (9) & (10)  \\
        \hline \hline
        \\

\textbf{Rosette} & & & & & & & & & \\

outer cloud & T9 & 207.6 , -1.96 , 1210 & 206.2  & 208.9  & -2.63  & -1.18  & 1181  & 1232  & 2032 $^{+ 2 } _{- 2 }$ & foreground cloud; largest sec. trunk \\

\textbf{Taurus} & & & & & & & & & \\

TMC 1 & L72 & 168.3 , -15.88 , 176 & 165.0  & 170.4  & -16.62  & -15.18  & 162  & 186  & 642 $^{+ 2 } _{- 4 }$ & connected to L73 by C71 \\
& L73 & 172.8 , -13.94 , 171 & 169.5  & 174.9  & -15.59  & -12.83  & 163  & 180  & 598 $^{+ 2 } _{- 2 }$ & connected to L72 by C71 \\

TMC 2 & L66 & 170.8 , -18.41 , 205 & 168.8  & 175.2  & -20.00  & -15.88  & 192  & 226  & 761 $^{+ 2 } _{- 3 }$ & furthest away along los; contains L1498 \\
& L75 & 174.1 , -13.98 , 150 & 171.7  & 175.8  & -16.64  & -12.50  & 134  & 162  & 702 $^{+ 3 } _{- 4 }$ & main comp. of TMC 2; contains L1498 \\

 \textbf{Vela} & & & & & & & & & \\
 
Vela A & L94 & 270.0 , 0.74 , 989 & 269.2  & 270.9  & -0.48  & 1.77  & 959  & 1027  & 4372 $^{+ 9 } _{- 14 }$ & most massive leaf in Vela A; distant comp; perp. to Gal.plane \\
& L101 & 271.3 , 0.38 , 923 & 269.8  & 273.0  & -0.78  & 1.10  & 892  & 951  & 4414 $^{+ 7 } _{- 9 }$ & nearby comp; parallel to Gal.plane; forms L shape with L94 on sky \\

Vela C & L76 & 266.0 , 0.80 , 1027 & 265.4 & 266.5  & 0.29  & 1.27  & 1000 & 1042  & 2858 $^{+ 9 } _{- 14 }$ & center of cloud \\
& L72 & 265.1 , 1.31 , 907 & 264.2  & 265.9  & 0.63  & 2.12  & 889  & 941  & 1779 $^{+ 4 } _{- 5 }$ & Northern right edge comp \\
& L79 & 266.3 , -0.34 , 927 & 265.5  & 267.0  & -1.00  & 0.29  & 906  & 944  & 2172 $^{+ 5 } _{- 7 }$ & Northern right edge comp \\
& L86 & 268.4 , 1.52 , 1034 & 267.7  & 269.1  & 0.87  & 2.15  & 1019  & 1051  & 776 $^{+ 3 } _{- 4 }$ & Northern right edge comp \\
& L85 & 268.7 , -0.57 , 853 & 266.7 & 270.7  & -1.72  & 0.75  & 836  & 876  & 6385 $^{+ 9 } _{- 11 }$ & Southern left edge; nearby and most massive leaf in Vela mol.ridge \\

Vela D & L53 & 261.4 , 0.99 , 968 & 260.4  & 262.4  & -0.17  & 1.84  & 904  & 1025  & 5314 $^{+ 6 } _{- 7 }$ & most massive leaf in Vela D \\
& L38 & 259.1 , 0.75 , 982 & 257.9  & 260.1  & 0.25  & 1.34  & 943  & 1019  & 3572 $^{+ 5 } _{- 6 }$ & central comp \\
& L49 & 261.5 , 0.86 , 1043 & 260.2  & 262.4  & 0.13  & 1.91  & 1026  & 1056  & 1745 $^{+ 4 } _{- 5 }$ & distant comp \\
& L70 & 264.6 , -0.43 , 972 & 263.9  & 265.3 & -1.06  & 0.18  & 956  & 988  & 1996 $^{+ 5 } _{- 9 }$ & nearby comp \\

Outer cloud & L55 & 262.5 , -3.04 , 922 & 261.6 & 263.8  & -4.00  & -1.58  & 904  & 935  & 1074 $^{+ 2 } _{- 3 }$ & cloud south of Vela D on sky \\

\hline
\end{tabular}
\end{adjustwidth}
\end{table*}


\section{Analysis of Individual molecular clouds and their extended environments}
\label{sec:IndGMCs}

In the following sub-sections detailing individual molecular clouds and their extended environments we refer to leaf substructures of interest as L\#\# where \#\# represent the index number assigned by \texttt{astrodendro}. Similarly trunks are denoted with T\#\# and children with C\#\#. Figures for all mapped regions showing the predicted extinction are found in appendix~\ref{sec:Ap:2Dext} in the supplementary material, and figures showing the predicted 3D density structures are in appendix~\ref{sec:Ap:3Ddense} in the supplementary material.

\subsection{California}
\begin{figure}
    \centering
    \includegraphics[width=0.55\textwidth]{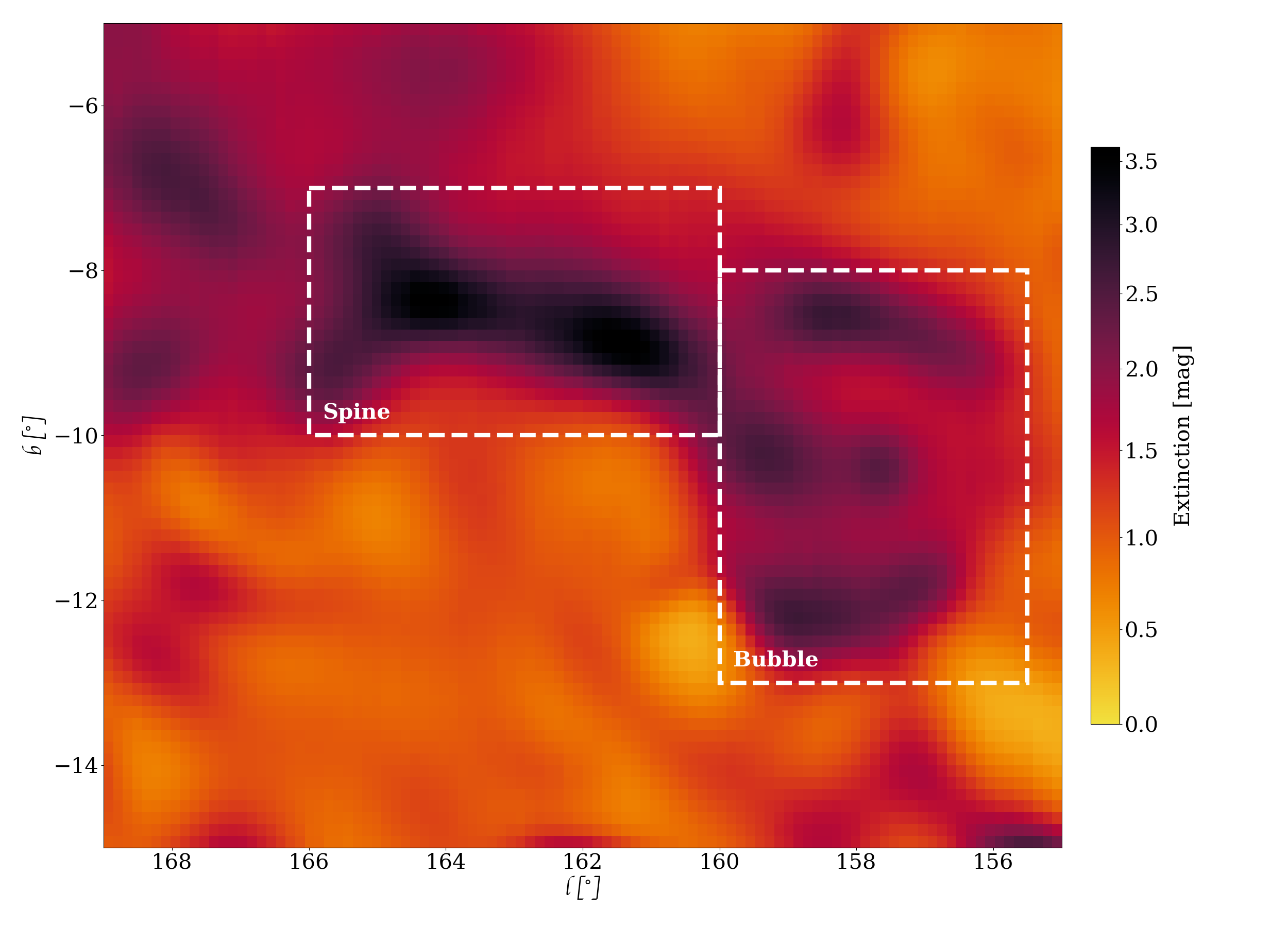}
    \caption{The predicted integrated 2D extinction of the California molecular cloud complex as seen from the Sun (i.e: on the plane of the sky). The California spine and bubble are highlighted}
    \label{fig:Calif_2D}
\end{figure}

\begin{landscape}
\begin{figure*}
\begin{adjustwidth}{-7.5cm}{0cm}
    \centering
    \begin{multicols}{2}
    \includegraphics[width=0.75\textwidth]{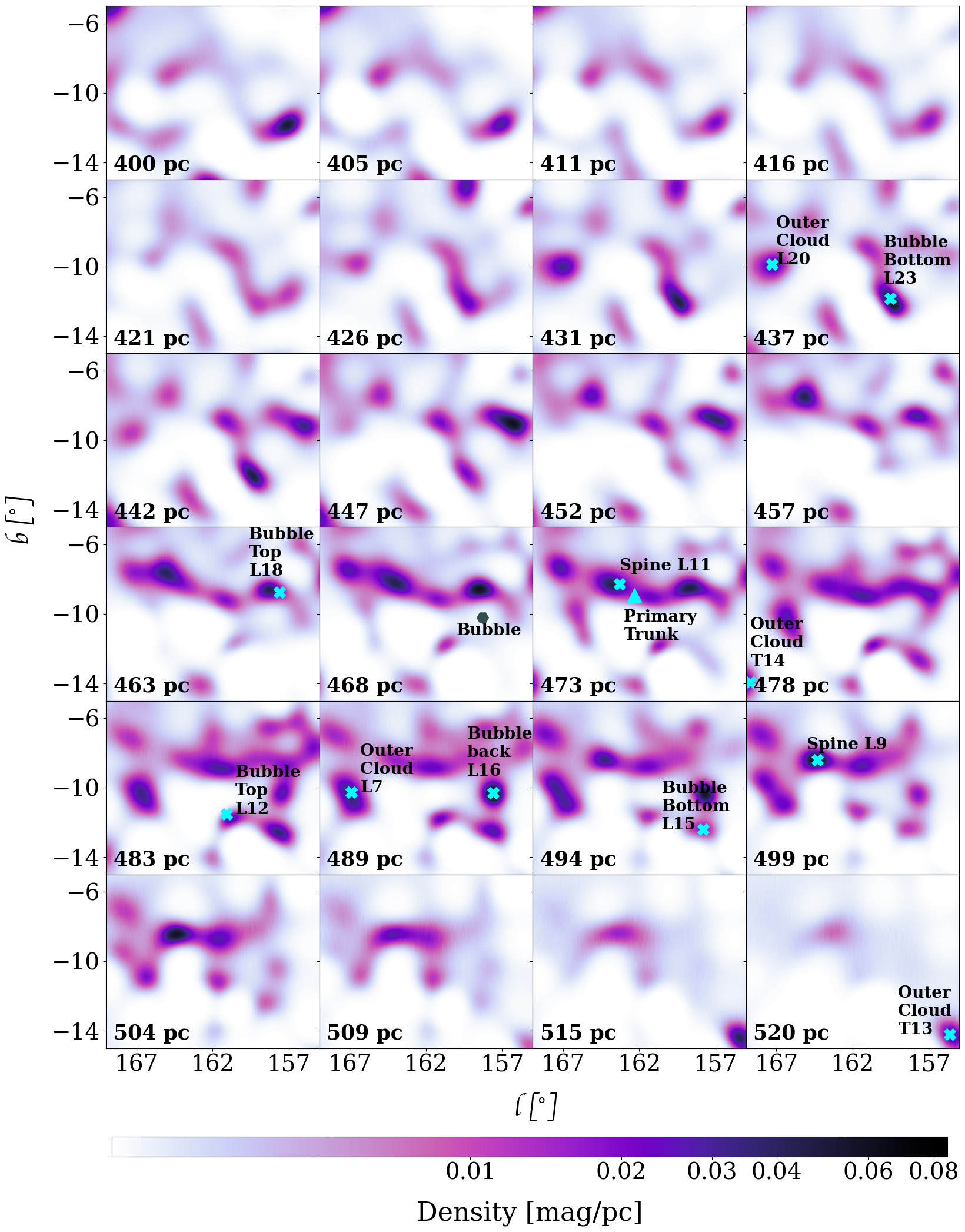}\vfill
    
   \hfill
    \includegraphics[width=0.65\textwidth]{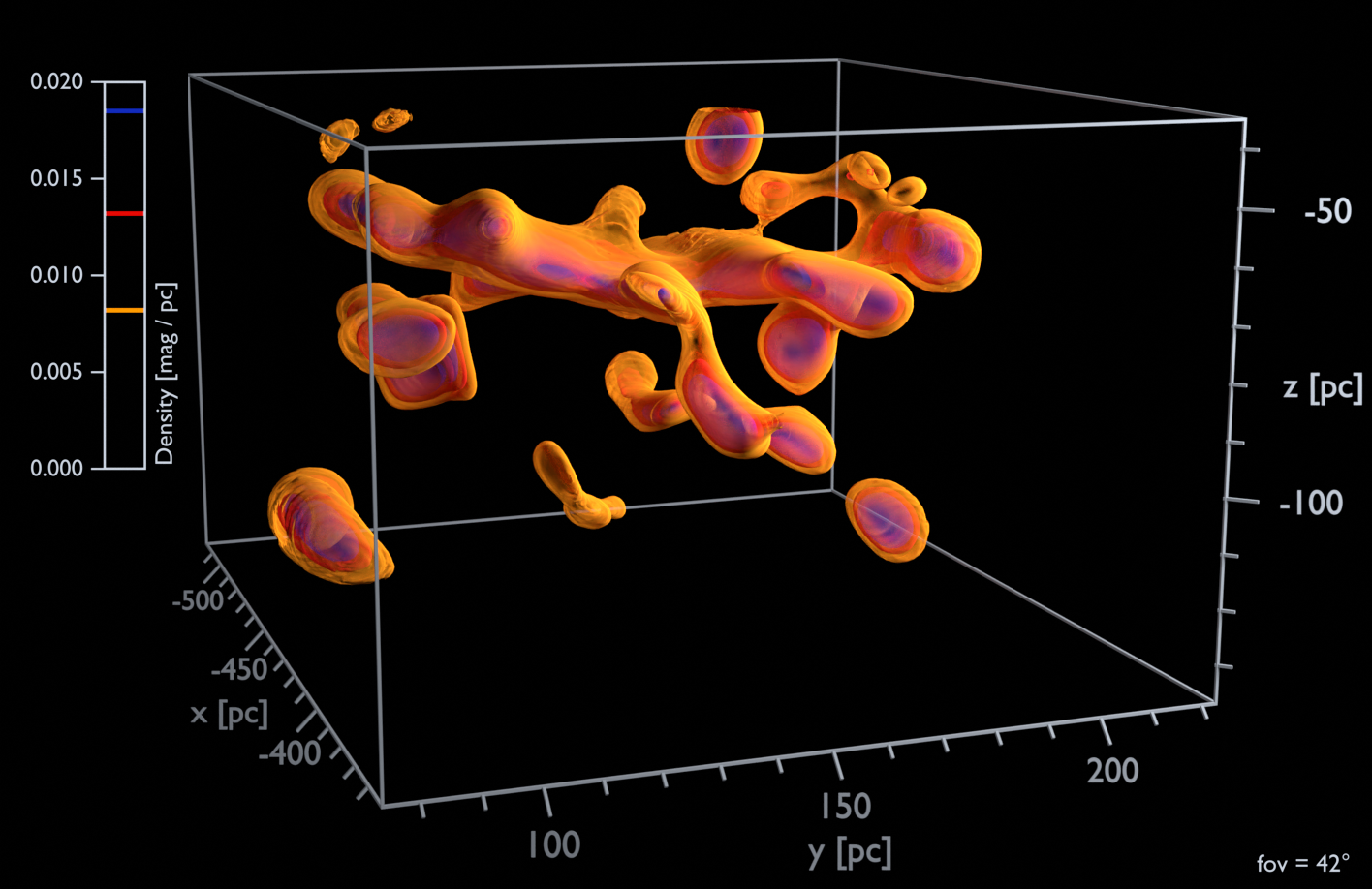}
    
    \hfill
    \includegraphics[width=0.65\textwidth]{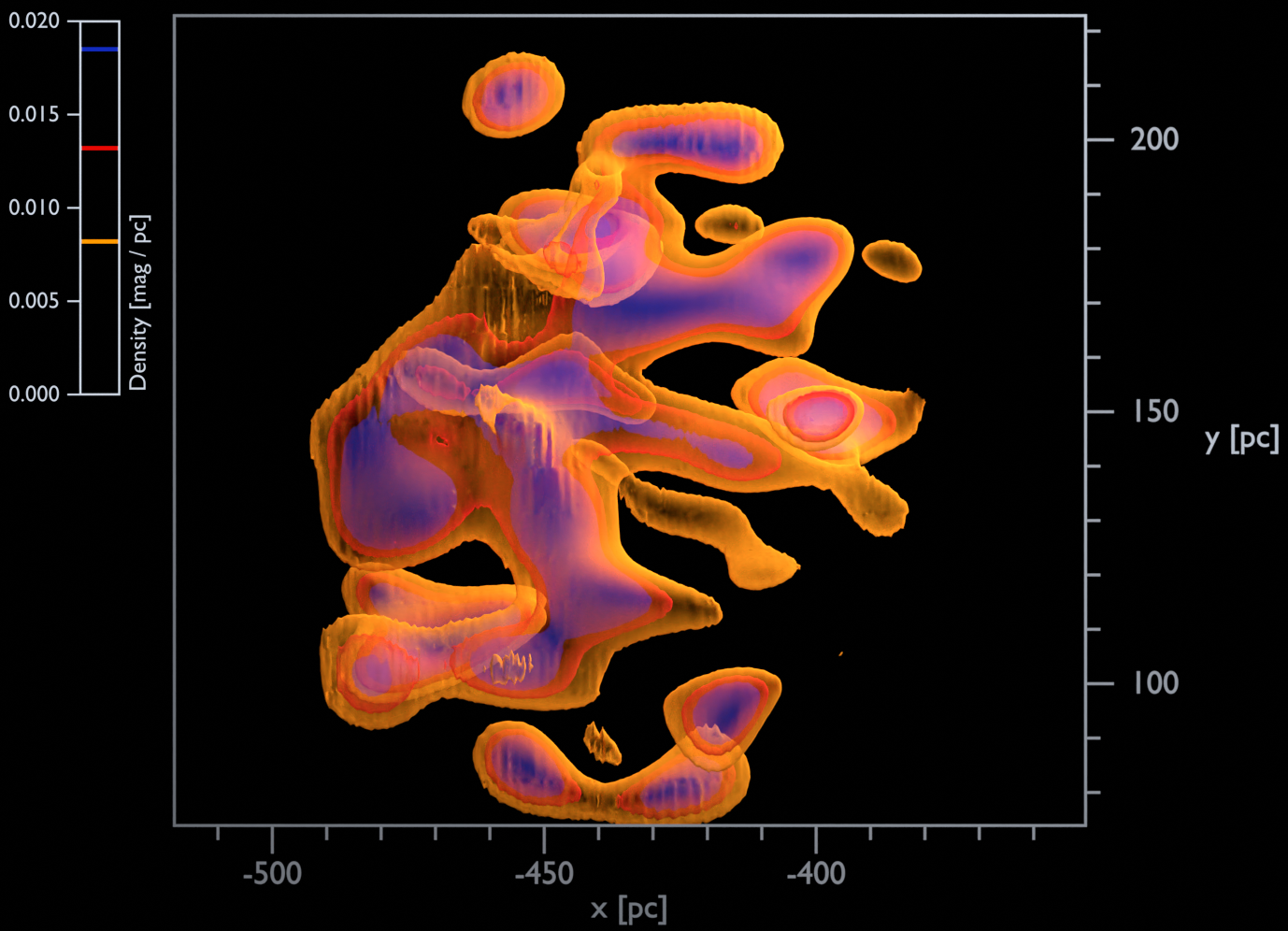}
    \end{multicols}
    \vspace{-0.8cm}
    \caption{ Predicted 3D density structure of the California molecular cloud complex. Left: Slices along the line-of-sight of the predicted 3D density structure. With the Cyan triangle we have marked the mass weighted centroid of the primary trunk (as given in Table~\ref{tab:maintrunk_params} placed on the closest distance slice included in the plot. The Cyan $\times$s mark the mass weighted centroids of the interesting features discussed in Sec.\ref{sec:IndGMCs} and highlighted in Table~\ref{tab:leaf_params} placed on the closest distance slice included in the plot. The grey hexagon marks the centre of the cavity known as the California bubble; Right top: Video of a volume rendering of the predicted 3D density structure which begins from the view as seen from the Sun. It then rotates anti-clockwise about an axis perpendicular to the initial viewing angle. The semi-transparent iso-surfaces mark three different density levels with orange being the least dense to blue being the most dense as shown by the colour bar.; Right bottom: Still image showing the top down view of the predicted 3D density structure of the molecular cloud region using identical rendering to the preceding video.}
    
\label{fig:calif_3Ddens}
\end{adjustwidth}   
\end{figure*}
\end{landscape}

California is one of the largest GMCs within 500 pc \citep{Lada2017}. Looking at California on the plane of the sky, the molecular cloud includes a \emph{bubble} and a \emph{spine}. We see this clearly in our predicted integrated 2D extinction in Fig.~\ref{fig:Calif_2D}. Exploring the three-dimensional density, shown in figure \ref{fig:calif_3Ddens}, we see that the spine is a flattened, sheet-like structure extended in the Galactic plane while the bubble is a hollow cavity. The spine has a width of $\sim 100$ pc along the LOS and has a width of 24 pc along the z axis. L09 forms the densest part of the spine with a peak density of $0.07$ mag pc$^{-1}$ at 447 pc from the sun. This matches well with literature distances of $450$ pc and $436 - 466$ pc for California put forward by \citet{Lombardi2010_Perseus_Taurus} and \citet{Zucker2020} based on the density of stars along the line-of-sight. As suggested by \citet{Rezaei2022} the shape of California along the line-of-sight is peculiar given its thin filamentary appearance on sky.

The centroid of the cavity lies at $x,y,z \sim -428, 170, -83$ pc and it has a radius of $\sim 30$ pc. The shape of the cavity is also peculiar given there is no evidence for a source that could have created the bubble shape. Typically a supernova or the UV radiation from a massive star is thought to be the cause of bubbles in star formation regions and molecular clouds. The bubble shape may simply be a chance alignment of the clouds outlining the cavity or we have not detected the possible SN-remnant which gave rise to the bubble.

\cite{Lada2009_Calif} suggested a similarity between the appearance on sky, kinematics and mass derived from 2D extinction measurements of California and Orion A. They therefore raised the question of why  Orion A has a much higher star formation rate. We find that the mass of Orion A (see Sect.~\ref{Sec:Ori}) is half that of our revised mass for California (which is itself 50\% lower than previous estimates, e.g. \citealt{Lombardi2010_Perseus_Taurus}), while the volume of California is seven times larger than Orion A. Hence, the mean density is 3.5 times lower in California and California is a much more diffuse region than Orion A. If we believe that relations such as the Kennicutt-Schmidt law applies to individual star-forming regions and in 3D, not merely 2D, this presents a simple solution to the conundrum; the two regions are simply not that similar after all, when considering their mass, extents and shape. This showcases the importance of including 3D information, particularly on the extent of clouds, when interpreting the relationships between environment and star formation. 

The discrepancy between our mass for California and that of \citet{Lombardi2010_Perseus_Taurus} is primarily related to our ability to consider the mass only within the dendrogram structures, rather than having to integrate over the entire line of sight. 
This excludes foreground and background dust unrelated to the molecular cloud, hence reducing the mass.
In \citet{Dharmawardena2022}, we performed a detailed comparison of mass estimates derived from extinction, to determine what part of the discrepancy was caused exclusively by the choice of e.g. extinction-to-mass conversion or the extinction measurements themselves. 
Using the same extinction-to-mass conversion and extinction estimates as this work, we found masses that were 1.3 -- 1.5 times higher than other estimates. 
On the other hand, our mass estimate here, accounting for the three-dimensional structure of the cloud, is now 50\% lower than previous estimates, suggesting that a comparison with the same conversion factors could result in still lower mass estimates.




\subsection{Camelopardalis}
The less well studied Camelopardalis molecular cloud region is formed of several layers of molecular clouds extending from nearby clouds associated with the local arm to clouds much further away located in the Perseus arm ($\sim 2$\,kpc). Many clouds are also located in between the two extremes in the Cam OB1 association layer at $800-1000$ pc. \citep{Straizys2007_Camel_ppr, Straizys2008_Camel_SFHB}. Our predicted dust density cube extends only up to the Cam OB1 layer so we do not attempt to recover any clouds in the Perseus arm.

Camelopardalis stands out in our sample as the region with the most trunks (101) and so is one of our most fragmented regions as seen in Fig.~\ref{fig:Camel_3Ddens}. This is reflected by the measured filling factor which is only 0.17. The primary trunk extends from 140--480 pc and covers only the nearby local arm clouds and has a total mass of $4.5 \times 10^{4}$ M$_{\odot}$. This is only $44\%$ of the total mass of the mapped region. The diversity of trunks shows that Camelopardalis comprises distinct groupings of clouds and is not a singular inter-connected complex: not only are the nearby clouds distinct from the Cam OB1 layer, even the clouds in the Cam OB1 layer are not necessarily related. These findings are consistent with layering behaviour described in \citet{Straizys2007_Camel_ppr, Straizys2008_Camel_SFHB} based on kinematic distances.

We recover several massive clouds in the form of leaves within the main trunk. One of the most prominent is the TGU 942 cloud \citep{Dobashi2005_TGUclouds} which is primarily formed of leaf L179, which has a peak density of $0.010^{+0.001}_{0.001}$ mag pc$^{-1}$ at $l,b,d= 144.2^{\circ}, 0.31^{\circ}, 211$ pc. Further behind we find clouds extending out to 920 pc within the Cam OB1 layer. One of the more massive such clouds is trunk T02 which is not visible in the 2D extinction map ( fig.~\ref{fig:Camel_2D}). It has a peak density of $0.008^{+0.001}_{0.001}$ mag pc$^{-1}$ at $lbd= 156.5^{\circ}, 3.80^{\circ}, 866$ pc and a total mass of $3.6 \times 10^{3}$ M$_{\odot}$ and extending from 804 -- 884 pc.

\subsection{Canis Major}
In Canis Major shown in Fig~\ref{fig:CanMaj_3Ddens}, we see that the most dense parts of the region are not in its three-dimensional centre. Canis Major forms an incomplete shell with a radius of 30 pc, centred at $x,y,z \sim -824, -796, -30$. If we assume that the front of this shell is defined by L17 and the back by L5, we arrive at a radius of $24 \pm 1$ pc. Both these estimates are consistent with the suggestion of \citet{Gregorio-Hetem2008_CanMaj_SFHB} that the current rate of star formation was triggered by the expansion of a supernova. 

Canis Major extends between distances of 1070 -- 1200 pc. The density of the primary trunk for CMa reaches its peak of $0.027^{+0.004}_{-0.006}$ mag pc$^{-1}$ at 1100 pc while the mass weighted centroid of the primary trunk is at 1150 pc. The measured distance agrees well with those determined by \citet{Lombardi2011_Orion} and the range by \citet{Zucker2020}. The mass we recover for the primary trunk of CMa is 12 times smaller than the total mass recovered by \citet{Lombardi2011_Orion} who used the 2D column extinction to determine their masses. 

\texttt{Astrodendro} separates the main body of CMa into three components, which we refer to as North, East and Central. 
The Northern component is the most massive region in CMa with a total mass of $5.18 \times 10^{3}$ M$_{\odot}$ and a mass weighted centroid of $l,b,d = 224.5^{\circ}, -0.61,^{\circ}, 1162$. We tie several leaves to known LDN clouds. L14, the closest leaf along the LOS in the Northern part of CMa hosts LDN 1658. T16 corresponds to the easternmost clouds from the main body of CMa; its leaves L8 and L12 are associated with LDN 1561 and 1565.

\subsection{Carina}

The Carina molecular cloud complex stands out from the rest of our sample in being the furthest away from the Sun. 
It is one of the best-known nebulae, home to $\eta$ Carinae, the most luminous star in the Milky Way. The main trunk and hence the star-formation region (SFR) has an extension along the LOS of 150 pc starting from 2250 pc. The mass weighted centroid lies at 2310 pc and is consistent with the distance predicted for the Carina star-formation region and $\eta$ Carina  in literature \citep{Smith2008_Carina_SFHB, Contreras2019}. 

The location of $\eta$ Car is clearly visible in the density maps shown in Fig~\ref{fig:Carina_3Ddens}, with the northern cloud wrapping around the bubble that it has created. The southern cloud and Southern Pillar, meanwhile, are at larger distances than $\eta$ Car, suggesting a distance gradient across the cloud from north to south. There appears to be a mid-plane between the two clouds whose axis goes through the location of $\eta$ Car, indicating how dramatically the star has shaped the nebula.
We decompose the Northern cloud into two components; the upper main part of the cloud is separate from the lower, filamentary part of the cloud. Their appearance as a single cloud is actually a chance alignment. The filamentary section is a cloud located closer to us at 2250 pc and outlined well by L0. Additionally the main upper component is layered along distance with a nearby component, L2 and a more distant component, L6.

Although Carina is home to a large population of massive stars, and therefore must have had a very large gas (and hence dust) reservoir originally, we recover a relatively low dust mass for the complex. Similarly, measurements of molecular-line emission from star-forming clumps in the complex show rather weak emission, indicating small present-day molecular gas reservoirs \citep{Contreras2019}. This presumably indicates the effects of the extreme radiation field of $\eta$ Car, whose extensive HII region may trigger the sputtering of dust grains and the photodissociation of molecules.

\subsection{Cassiopeia}
\label{Sec:Cas}

The Cassiopeia and Cepheus molecular clouds regions can essentially be considered a single giant molecular cloud complex, however on sky they are separated by their neighbouring parent constellations. While massive, they are not well studied, perhaps due to their proximity to the Galactic plane. To the best of our knowledge, this is the first time both regions are studied in detail in  3D (excluding CO studies that use l,b,velocity). Both regions spread from as close as 200 pc in the local arm to the Perseus arm ($>2$ kpc). We limit our mapped region to within the local arm and find this section of Cassiopeia to extend from 190 -- 810 pc revealing a variety of clouds and filaments as seen in Fig~\ref{fig:Cas_3Ddens}. One of the three most massive regions in our sample with a total region mass of $2.9 \times 10^{5}$ M$_{\odot}$, it is also highly fragmented, with the lowest filling factor in our sample (0.11).

This molecular cloud region is thought to be composed of several expanding shells as a result of either supernovae or OB associations. One of the largest such shells on sky is the Cepheus flare shell which we will also discuss in the following section of the Cepheus GMC region. We find several clouds lying on the surface of the Cepheus flare shell in the Cassiopeia region. Most notably the LDN 1333 cloud which has literature distances from 180 pc \citep{Kun2008_Cas_SFHB} to 283 pc \citep{Zucker2020}. We associate this cloud with L265 which extends from 282 --	379  pc and has a mass of $2.3 \times 10^{3}$ M$_{\odot}$. 

\begin{figure}
    \centering
    \includegraphics[width=0.55\textwidth]{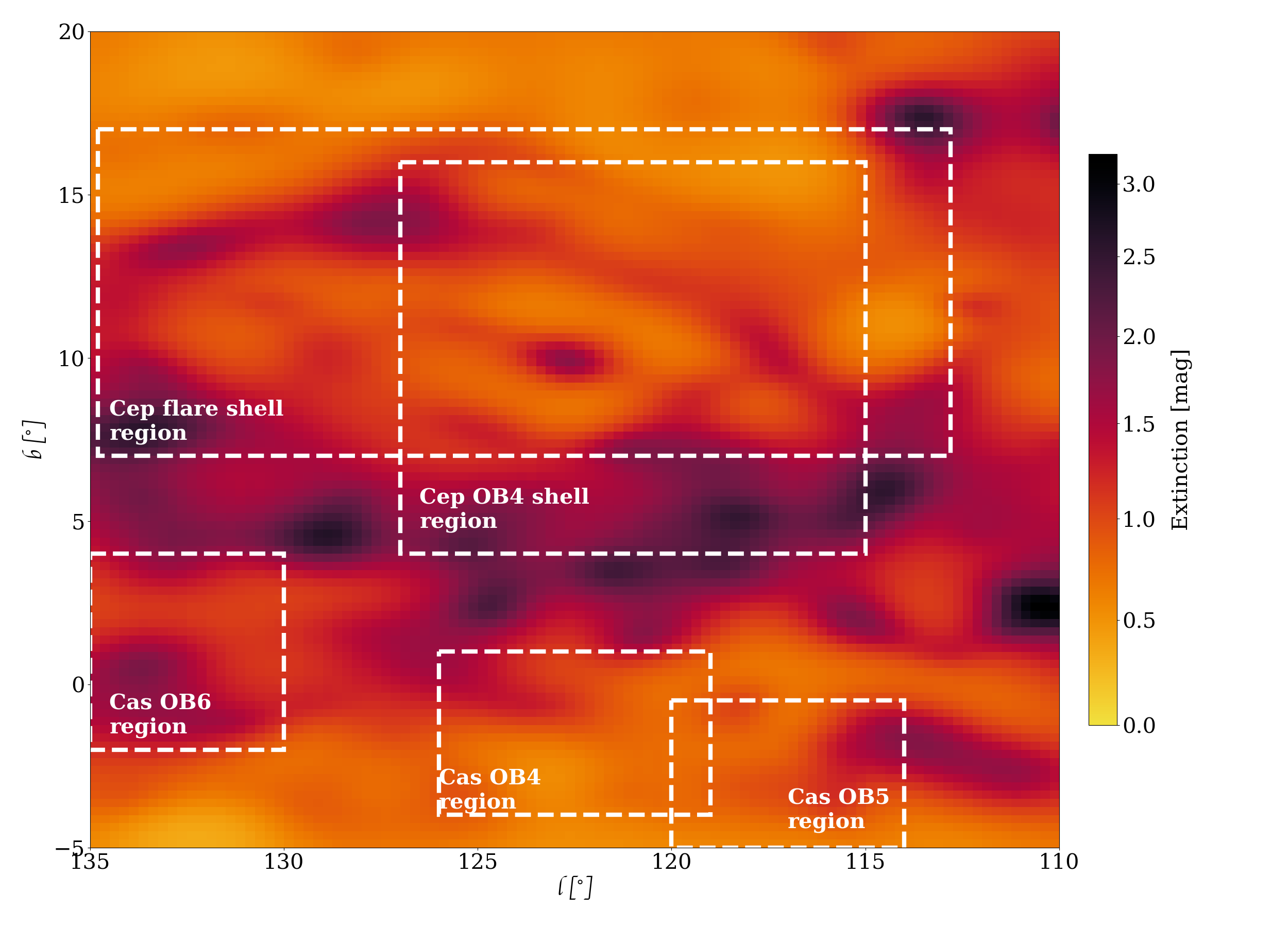}
    \caption{The predicted integrated 2D extinction of the Cassiopeia molecular cloud complex as seen from the Sun (i.e: on the plane of the sky). The regions highlighted are based on nearby features of the cloud complex as seen on sky. The arc-like structure discussed in Sec.~\ref{Sec:Cas} can be found within the region marked Cep flare shell region.}
    \label{fig:Cas_2D} 
\end{figure}

\begin{landscape}
\begin{figure*}
\begin{adjustwidth}{-7.5cm}{0cm}
    \centering
    \begin{multicols}{2}
    \includegraphics[width=0.75\textwidth]{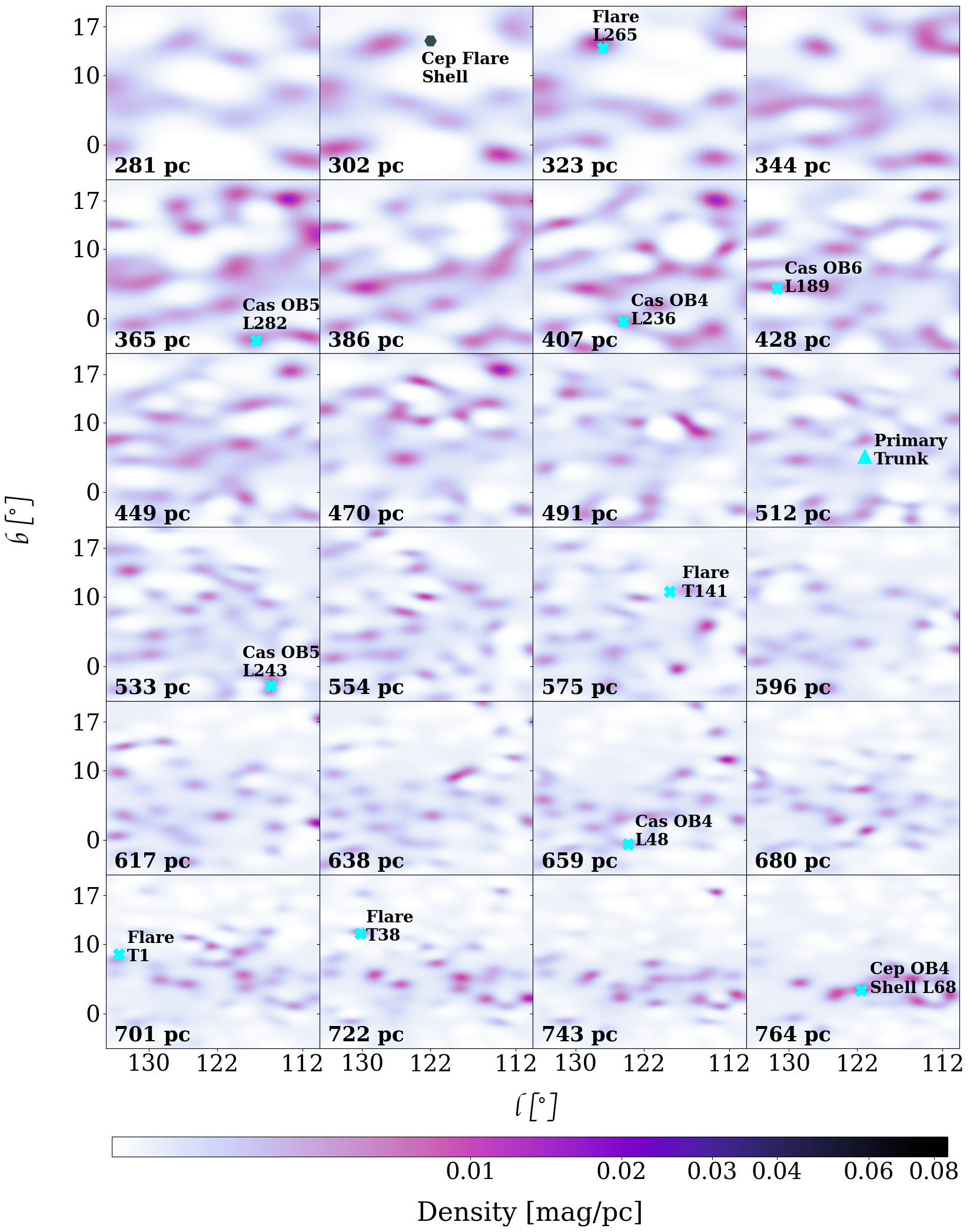}
    
    \hfill
    \includegraphics[width=0.6\textwidth]{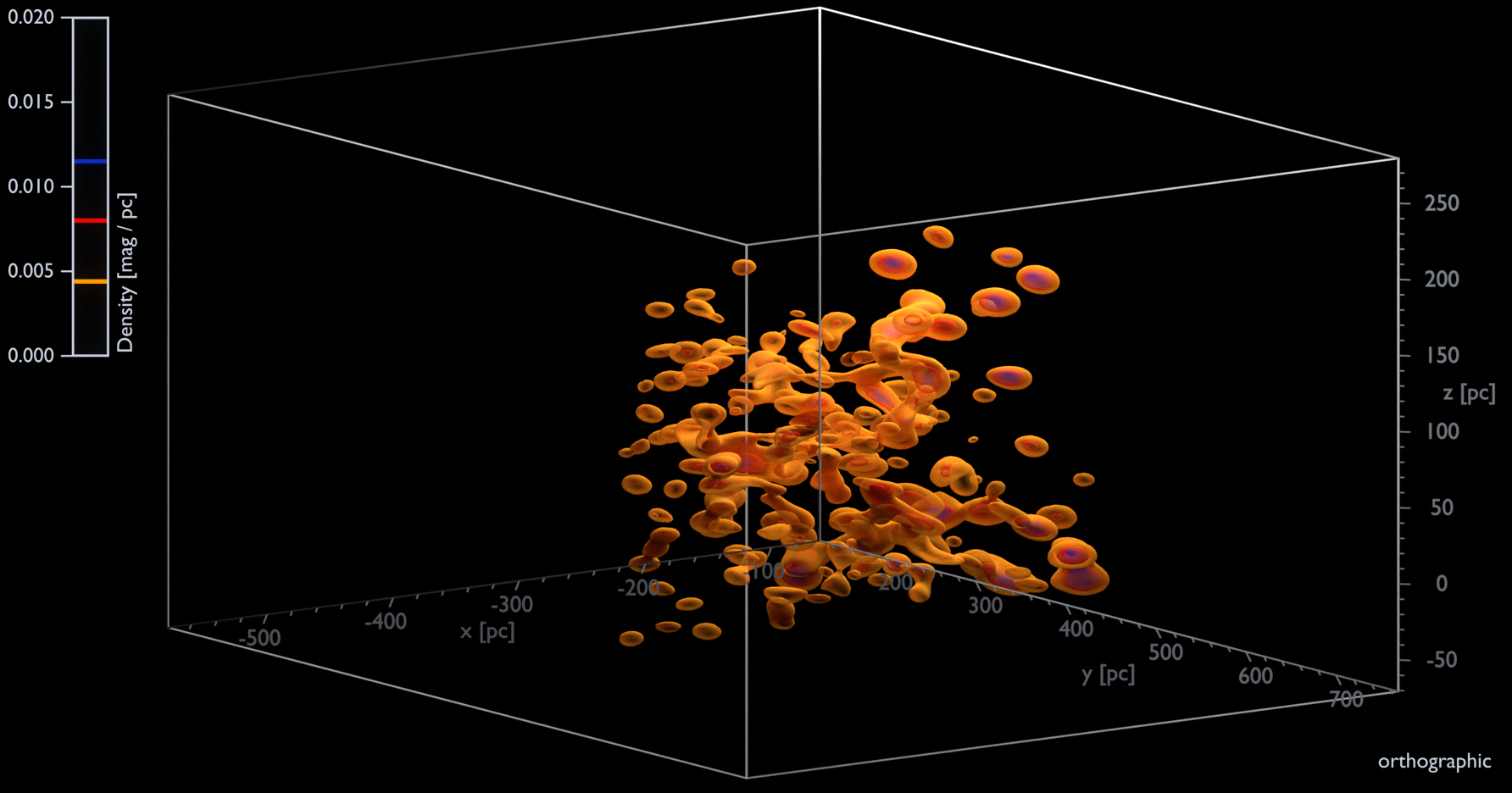}
    
    \hfill
    \includegraphics[width=0.6\textwidth]{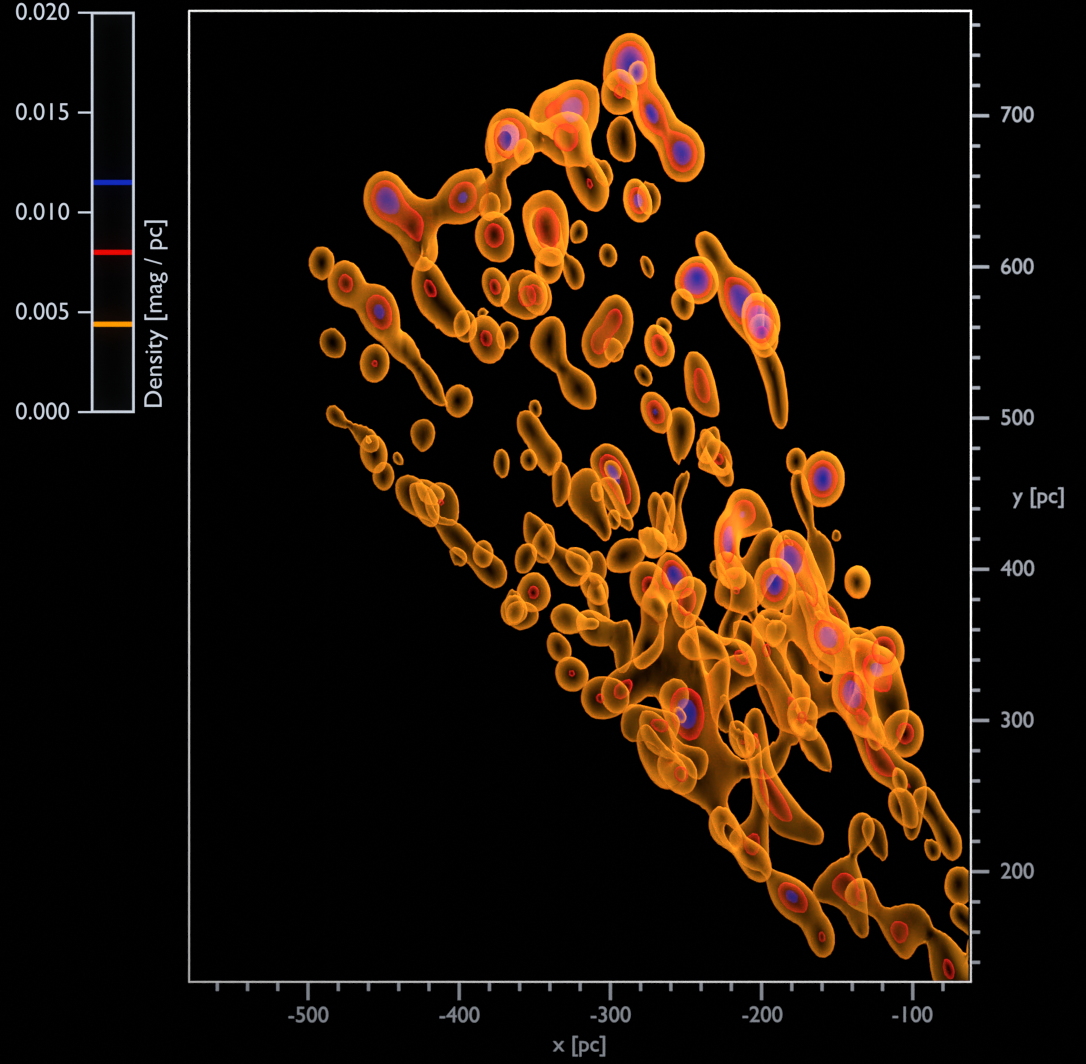}
    
    \end{multicols}
    \vspace{-0.8cm}
    \caption{ Predicted 3D density structure of the Cassiopeia molecular cloud complex. Left: Slices along the line-of-sight of the predicted 3D density structure. With the Cyan triangle we have marked the mass weighted centroid of the primary trunk (as given in Table~\ref{tab:maintrunk_params} placed on the closest distance slice included in the plot. The Cyan $\times$s mark the mass weighted centroids of the interesting features discussed in Sec.\ref{sec:IndGMCs} and highlighted in Table~\ref{tab:leaf_params} placed on the closest distance slice included in the plot. The grey hexagon marks the centre of the Cepheus flare shell; Right top: Video of a volume rendering of the predicted 3D density structure which begins from the view as seen from the Sun. It then rotates anti-clockwise about an axis perpendicular to the initial viewing angle. The semi-transparent iso-surfaces mark three different density levels with orange being the least dense to blue being the most dense as shown by the colour bar.; Right bottom: Still image showing the top down view of the predicted 3D density structure of the molecular cloud region using identical rendering to the preceding video.}
    
\label{fig:Cas_3Ddens}
\end{adjustwidth}   
\end{figure*}
\end{landscape}

We also observe a large scale structure forming a thin arc-like shape in 2D extinction (Fig.~\ref{fig:Cas_2D}) extending between the Cassiopeia and Cepheus region. A segment of this arc as is well outlined by T141 and is located behind Cep OB4 bubble at 496 -- 669 pc and it it is highly filamentary at almost 200 pc in length from the two furthest apart points. Similarly, another loop-like structure is recovered by T1 that is also $\sim$ 200 pc in length. This is similar in structure to the $\lambda$ filament in Orion discussed below. These filamentary structure can be viewed in Fig.~\ref{fig:Cas_3Ddens}.  

Both the Cepheus and Cassiopeia regions contain many shells and bubbles, several of which are well-known  \citep{Kun2008_Cas_SFHB, Kun2008_Cep_SFHB}. We recover the Cepheus flare shell centred at a distance of 320 pc ($x,y,z = -156, 249, 79$). The shell has a radius of $\sim 50$ pc. We see this consistently in the Cep 3D maps as well. The Cep flare clouds which are the edge of this bubble observed in our Cepheus map described below wraps nicely around this bubble. 

\subsection{Cepheus}

Similar to its sibling closer to the Galactic anti-centre, the Cepheus GMC also extends from the local arm all the way out to the Perseus arm and has several large scale bubbles present as a result of supernovae explosions or OB associations \citep{Kun2008_Cep_SFHB}.  It contains both star-forming and non-star-forming clouds with the upper Cepheus flare layer located closest to us at 200 -- 450 pc and the lower Galactic plane clouds associated with several OB associations located at 600 -- 800 pc, making up the local arm region \citep{Kun2008_Cep_SFHB}. 

We map the local arm region as shown in Fig.~\ref{fig:Cep_3Ddens} and find Cepheus to extend from 270 -- 1000 pc. The cloud complex has a total mass of $7.16 \times 10^{5}$ M$_{\odot}$, making it the most massive GMC in our sample. 
We find the primary trunk of Cepheus wraps around a large-scale low-density structure which we associate with the Cepheus bubble centred at $x,y,z = -160, 785, 98$ pc.  We do not recover any dust associated with NGC 7160 which has an estimated distance \citep[800 pc;][]{Dib2018} which would place it somewhere close to the center of the Cepheus bubble. This positioning, combined with the apparant lack of dust, indicates this could be the ionising source giving rise to the Cepheus bubble. 

We recover the Cepheus flare clouds with C292 which extends from 335 -- 435 pc and has a mass of $7.64 \times 10^{3}$ M$_{\odot}$. We further recover several clouds in the lower latitude regions at much further distances ($\sim 800$\,pc range) such as C50 which encloses Cep OB4. 


\subsection{Chamaeleon}

The Chamaeleon molecular cloud comprises three separate dark clouds Chamaeleon I,II,III. Although they are considered to be part of the same complex, they have quite different levels of star-formation activity. Cha I has a large population of YSOs and is continuing to form stars, while Cha II has fewer YSOs and a large median age. Cha III, on the other hand, is quiescent, with no evidence for either YSOs or current star formation \citep{Oliveira2014_Cham}. 

Chamaeleon's closest edge is 100 pc away, making it is one of the closest molecular cloud complexes to us, and the region extends to 277 pc from the Sun. 
Individually we recover Cha I with two leaves L26 and L54 which coincide with the northern and southern parts of the cloud, respectively. These two leaves are located at different line-of-sight distances, suggesting a north-south distance gradient in Cha I. We recover Cha II and III together as a single structure, C64, and find that L69 (a child of C64) corresponds to the densest part of Cha II. This places the two clouds at similar distances from us with a peak-density distance of 169 pc and extending from 160 -- 194 pc. The distances determined by \citep{Voirin2018_Cham} fall well within these ranges, albeit the upper end. 

Interestingly we find Cha I and II - III are separated by a central elongated cavity approximately 30 pc long which they wrap around ($x,y,z = 80, -140, -50$). There is no obvious explanation (e.g., young supernova remnant, massive star or HII region) for the presence of this cavity. 

Finally, while our mapped region covers the lower part of Musca molecular cloud,  we do not successfully map it in 3D. This is likely due to a combination of its very thin filamentary structure and the relatively small number of stars within that area in the \citep{Fouesneau2022_LBol} catalogue. The 3D density structure of Chamaeleon cloud complex is shown in Fig.~\ref{fig:Cham_3Ddens}.

\subsection{Southern Coalsack}

The Southern Coalsack (or simply Coalsack) is a single dark molecular cloud lying just above the Chamaeleon cloud complex, closer to the Galactic plane. It is believed to not have formed any stars \citep{Nyman2008_CoalSack}. 

While CO observations show it to be very fragmented with many small-scale clumps, we recover it as a single large structure (see Fig.~\ref{fig:Coalsack_3Ddens} with a mass weighted centroid at $l,b,d = 298.6^{\circ}, -0.97^{\circ}, 187$ pc. It is expected that we do not recover the small-scale clumps in our 3D maps as they are much smaller than the scale length (5 pc) we adopt. The Coalsack differs from the other region in our sample for several reasons: 1) It is the only cloud in our sample to have a single trunk; 2) It is the smallest in size with an $x,y,z$ extent of $=51 \times 70 \times 29$; and 3) It is also the second smallest in mass and volume with a total mass of $6.5 \times 10^{3}$ M$_{\odot}$. We place the Coalsack at a distance of 141 – 212 pc, where the density peaks at 158 pc. This imples that the literature hypothesis that the Coalsack may extend from $\sim 150 - 200$ pc is correct \citep{Nyman2008_CoalSack}.

\subsection{Cygnus X}
\label{Sec:CygX}

In \citet{Dharmawardena2022} we mapped the 3D dust density distribution of Cygnus and suggested it to be one large structure with multiple small scale structures spreading behind the primary structure. We found Cygnus X to extend $\sim 1300 - 1500$ pc in our previous work. Upon the application of \texttt{astrodendro} in this work, we confirm that Cygnus X comprises  a single structure (the primary trunk) extending from $lbd =$ $73.0^{\circ} - 85.3^{\circ}$, $-3.81^{\circ}, 3.80^{\circ}$, $1280 - 1611$ pc encompassing $63\%$ of the total volume of the region map by the dendrogram. 

The full region of Cygnus spreads out to $2091$ pc from us with many smaller dense structures as shown in Fig.~\ref{fig:Cyg_3Ddens}. These structures extracted by the remaining trunks are concentrated behind the primary trunk. The isolated trunks in Cygnus holds $14 \%$ of the total mass which is the most mass held by isolated trunks in our sample. 

We derive a total mass of $5.02 \times 10^{5}$ M$_{\odot}$ for Cygnus X, making it the second most massive region in our sample. The mass derived in this work however, is 13 times smaller than the total mass derived in \citet{Dharmawardena2022}. This difference can be understood from the fact that we now only take into account mass within the structures defined by \texttt{astrodendro}, which may represent a more accurate definition of the boundary of the cloud. See Sec.~\ref{sec:MassDiff} for further discussion on these mass differences and their implications. 

\subsection{Lupus}

Several large dark clouds, Lupus I -- VI, and a few more recently discovered small dark clouds, Lupus VII -- IX, form the Lupus molecular cloud complex. Similar to Chamaeleon, the Lupus clouds are in several different stages of star formation: Lupus I -- IV show active star-formation while the others do not \citep{Galli2020_Lupus}.  

We find the extended environment of the Lupus clouds from 100 pc out to 312 pc. However analysis of the recovered structures from dendrogram allow us to narrow down the densest regions (likely the well known star-formation regions) of the Lupus II -- IX clouds to be enveloped by a single child, C05 which extends from 100 -- 155 pc, indicating they are part of a single complex consistent with literature \citep{Galli2020_Lupus, Zucker2021}. The 3D structure of Lupus is shown in Fig.~\ref{fig:Lup_3Ddens}.

The primary trunk of Lupus holds only $34\%$ of the total mass of the complete region which is the smallest mass and volume for a primary trunk in our sample. This mass and volume fraction indicates there is significant mass spread across the extended environment. We recover several large secondary trunks including T23 which extends from 268 -- 395 pc and has a mass of $6.56 \times 10^{3}$ M$_{\odot}$.   
Our 3D map did not recover Lupus I, which similar to Musca is thin and filamentary. Once again this is likely due to the lack of sources to recover extinction parameters.

\subsection{Monoceros R2}

Monoceros R2 (Mon R2) is one of the nearest massive star formation regions to us. Based on our \texttt{astrodendro} analysis, 
we recover a total region mass of $3.80 \times 10^{4}$ M$_{\odot}$, consistent with that predicted by literature CO measurements \citep{Carpenter2008_MonR2}. Fig.~\ref{fig:MonR2_3Ddens} depicts the top down view and view from the sun of Mon R2.

We split the GMC to 4 distinct regions based on the most massive structure including L05, L18 and L24. With this we also identify distances to many small scale clumps such as TGU, LDN and LBN clouds. Further we recover two quite large distant leaves which are not recovered in 2D extinction.

\subsection{Orion}
\label{Sec:Ori}

In \citep{Dharmawardena2022} we derived the 3D dust density and extinction of this well-studied molecular cloud. Applying astrodendro to this precalculated map we find 
the largest of the trunks encompasses the four main star forming regions of the Orion molecular cloud identified in our previous work: Ori A, Ori B, lamda Ori and $\lambda$ filament.  

The extended environment of Orion has a volume of $1.21 \times 10^6$ pc$^{3}$, making it one the largest regions in our sample in both volume and mass. The wider environment extends from 270 -- 526 pc along the line-of-sight, consistent with our previous work and includes the foreground $\lambda$ filament identified in it. We recover a mass of $1.70\times10^{5}$ M$_{\odot}$ for the primary trunk of which $44\%$ lie within the dense leaves. 

Orion A clearly corresponds to a single leaf as shown in Fig~\ref{fig:Ori_3Ddens}. In the cases of Orion B, $\lambda$ Ori and the $\lambda$ filament several leaves come together to recover the densest regions of these clouds. Orion A is outlined by leaf L56 with extents: $xyz =$ -376 --	-298; -239 -- -170; -154 -- -125. In distance along LOS it extends from 382 pc to 444 pc. This is similar to extents presented by \citep{Zucker2021}. We recover a mass of $2.18 \times 10^{4}$ M$_{\odot}$ which is two-thirds that recovered using the NICEST method by \citet{Zucker2021}. Conversely our mass is 2.3 times larger than that recovered from the 3D mapping and skeletanisation by \citep{Zucker2021}. This may be due to Orion molecular clouds lying at the edge of the \citep{Leike2020} maps as well as the skeletonisation method not including the complete cloud.

Such differences in mass are observed between the two studies for Orion B and $\lambda$ Ori. We recover a masses of $1.98 \times 10^{4}$ M$_{\odot}$ and $1.86 \times 10^{4}$ M$_{\odot}$ for Ori B and $\lambda$ Ori respectively. Orion B extends from 391 to 512 pc in distance and $\lambda$ Ori extends from 325 to 451 pc. Orion B consists of multiple components whose distribution along the line-of-sight results in an X-shape, with two distance components at each of the upper and lower ends of the cloud \citep{Dharmawardena2022}. Additionally we find the $\lambda$ filament coincides with two leaves elongated along the z axis, L39 and L59, having a total z extent of 109 pc. This makes it the largest structure perpendicular to the Galactic plane in the Orion region. 

Comparing the total extended environment mass calculated in this work to that calculated in \citet{Dharmawardena2022}, we find the mass here is six times smaller than our previous work. Similar to the discussion on Cygnus, we attribute this difference to the use of \texttt{astrodendro} to define the regions. 

As described for Cygnus X in Sec. \ref{Sec:CygX} above, in \citet{Dharmawardena2022} we took into account the contribution of all points within the modelled 3D dust density cube when calculating mass, whereas here we only take into account the region outlined by the dendrograms. Therefore we now exclude any background/surrounding diffuse ISM material, which was not the case in our earlier work.

An interesting feature we recover is the dust ring identified by \citet{Schlafly2015_OrionDust} partly formed by Orion A and B. \citet{Schlafly2015_OrionDust} recovers this dust ring using a pseudo-3D method which used the extinction slices at nearby, Orion region and distant extents. However in our 3D maps we observe this "ring" to actually be a cavity. We will refer to this bubble as the Orion AB cavity. Orion A and B lie on the Orion AB cavity's surface; Orion A lies along the "front" surface of this cavity (i.e. closest to us) and Orion B traces along one side. This explains the X-shape of Orion B described above. The center of the cavity lies at distance of 480 pc with a 50 pc radius matching well with that predicted by \citet{Schlafly2015_OrionDust}. 

Several suggestions have been put forward in literature for the origin of this cavity. It could be a bubble caused by the wind of massive stars. The massive stars in the Orion OB1 associations have been ruled out as possibilities for the origin for the Orion AB cavity by \citet{Schlafly2015_OrionDust} already and they suggest NGC 2232 as a possible candidate. However, NGC 2232 is at a distance of 330 pc \citep{Dib2018} placing it on the foreground of this cavity. Therefore should this be a bubble, the origins of the Orion AB bubble remains a mystery. It has also been suggested that the Orion AB bubble is not a bubble at all and simply a section of the Radcliffe Wave and chance alignment makes Orion A and B appear as if they are wrapping around the cavity \citep{Alves2020_RadWave}. 

We see a density depression roughly consistent with the $(l, b)$ location of Barnard's loop between distances of 440 -- 460 pc, consistent with \citep{ODell2008_OriBernardLoop}. This places the loop physically behind Orion A, and cutting across the upper part of Orion B. The density is even lower on the other side of Orion A and B, suggesting that we only see half of the loop because there is not enough material present for the other half to emit strongly. 


\subsection{Perseus}

Using the 3D density map created in \citet{Dharmawardena2022} we find the extended environment of Perseus 
extends out to 474 pc along the LOS, well beyond the distance estimated for the Perseus SFR itself in literature \citep{Bally2008_Perseus, Pavlidou2021_Perseus}. 

While the bulk of the material associated with the SFR itself extends up to 355 pc the filamentary nature of the regions means our dendrogram recovers many structures beyond this. 
The primary trunk determined by \texttt{astrodendro} encompasses both the SFR as well as these background structures, meaning they are connected at at least the minimum density level of $0.003~mag~pc^{-1}$. We determine a total mass of $4.1\times10^{4}$ M$_{\odot}$ for the full region including the Perseus SFR and the background clouds. Similar to Cygnus X and Orion, this mass is 4.5 times smaller than what we previously determined in \citet{Dharmawardena2022} for the same region of space.

The two leaves encompassing the dense regions of the Perseus SFR (L39, L46) have a mass of $2.97 \times 10^{3}$ M$_{\odot}$. L39 forms the Northern edge of Perseus when looking on sky and extends 302 -- 355 pc. The extents of L39 are consistent with it holding the IC 348 cluster whose distance in literature is estimated to be $321 \pm 10$ by \citet{OrtizLeon2018_Perseus} and $324 \pm 5$ by \citep{Kuhn2019}. The Southern edge of Perseus is formed by leaf L46 which has a mass-weighted centroid at $l, b, d = 158.4^{\circ}, -22.02^{\circ}, 325$ pc and holds cluster NGC 1333.

\subsection{Rosette}

The Rosette molecular clouds stand out in our sample for being the second smallest in mass and volume with a region mass of only $7.4\times10^{3}$ M$_{\odot}$. We recover the well-known Rosette star formation region in the primary trunk at a distance range of 1274 – 1329 pc, consistent with \citet{Zucker2019, Zucker2020}. 

The Rosette star-formation region is seen wrapping around a horse shoe shaped cavity with the center at $x,y,z = -1167, -584, -44$. This is inline with \citet{Schneider2010_Rose} who suggested that the Rose-like shape and the star-formation in Rosette is driven by the influence of nearby OB associations. Specifically, NGC 2244 has been long-thought of as causing the observed shape of Rosette. However, following Gaia DR2, the revised distance for NGC 2244 of $1550 \pm 100$ pc from \cite{Kuhn2019} places NGC 2244 in the background of Rosette, meaning based on these measurements it is not the driver causing the Rose-like shape observed in the cloud complex. 

Interestingly we also observe a large foreground cloud directly in front of Rosette along the line-of-sight at 1210 pc. It is outlined by the largest secondary trunk, T09. The cloud is more diffuse than the primary trunk and encompass $27\%$ of the total region mass. 
Its presence directly in front of Rosette along the line-of-sight means caution is required when analysing dust column density and 2D extinction measurements in the region.

\subsection{Taurus}

The Taurus molecular cloud comprises two main components TMC 1 and TMC 2 \citep{Lombardi2010_Perseus_Taurus}. In \citet{Dharmawardena2022} we mapped Taurus and found it to be composed of two layers separated along the z axis and found the  TMC 2 component to be closer one to us. 
We find TMC 2 to be more extended along the LOS and formed of two components L75 and L66. The most distant component L66 extends from 192 to 226 pc. It is also the most massive component in volume and mass of the main components of the Taurus molecular cloud. In total TMC 2 has a mass of $1.46 \times 10^{3}$ M$_{\odot}$.  

The Per-Tau shell introduced by \citet{Bialy2021} is located behind Taurus and they show Taurus as an extended region wrapped around it. They also introduce the Tau ring, an elliptical ring located in Taurus. The region we mapped includes a small fraction of the region covered by the Per-Tau bubble, but this overlapping region is too small for us to draw any conclusions about it. 
On the other hand, we observe a low-density region coinciding with the part of the Tau ring covered by our density map, although only about half of the ring is within our map. 

\subsection{Vela}

The Vela molecular ridge is a well known molecular cloud complex. It is also one of the closest regions forming intermediate-mass stars to us. The molecular ridge is separated into four regions: Vela A, B, C and D. Vela A, C, D are thought to be a common complex with distances at $\sim 700 - 900$ pc while Vela B is completely separate at a distance of $\sim 2$ kpc \citep{Massi2019_Vela}.

In our 3D reconstruction we map the nearby Vela clouds A, C and D out to 1.3 kpc leaving out the more distant Vela B. Our full complex has a distance range of 670 -- 1090 pc. The Vela complex is the second most massive region in our sample with a primary trunk mass of $4.33 \times 10^{5}$ M$_{\odot}$. 

\begin{figure}
    \centering
    \includegraphics[width=0.55\textwidth]{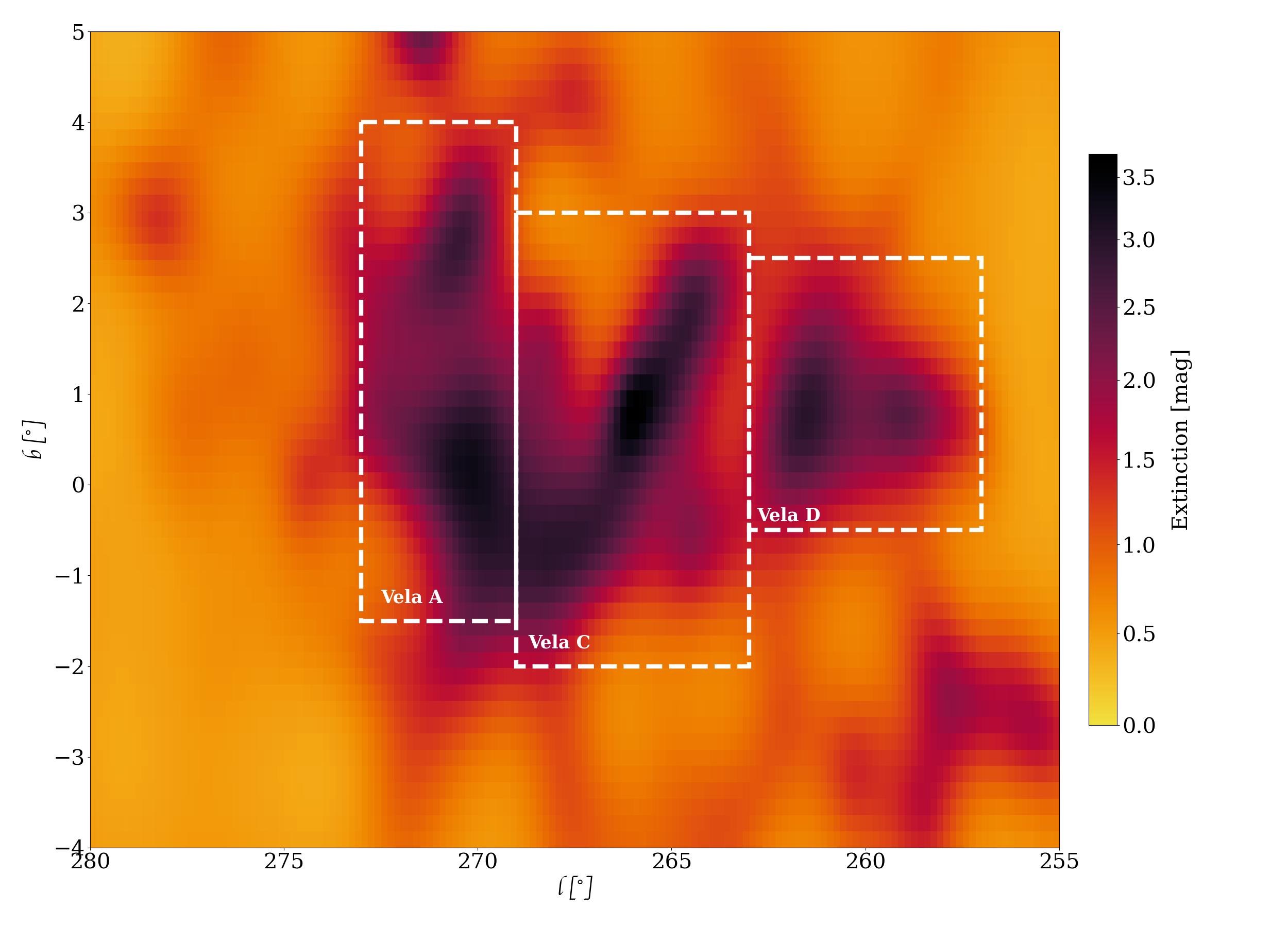}
    \caption{ The predicted integrated 2D extinction of the Vela molecular ridge as seen from the Sun (i.e: on the plane of the sky). Vela A, C and D are highlighted.}
    \label{fig:Vela_2D} 
\end{figure}

\begin{landscape}
\begin{figure*}
\begin{adjustwidth}{-7.5cm}{0cm}
    \centering
    \begin{multicols}{2}
    \includegraphics[width=0.75\textwidth]{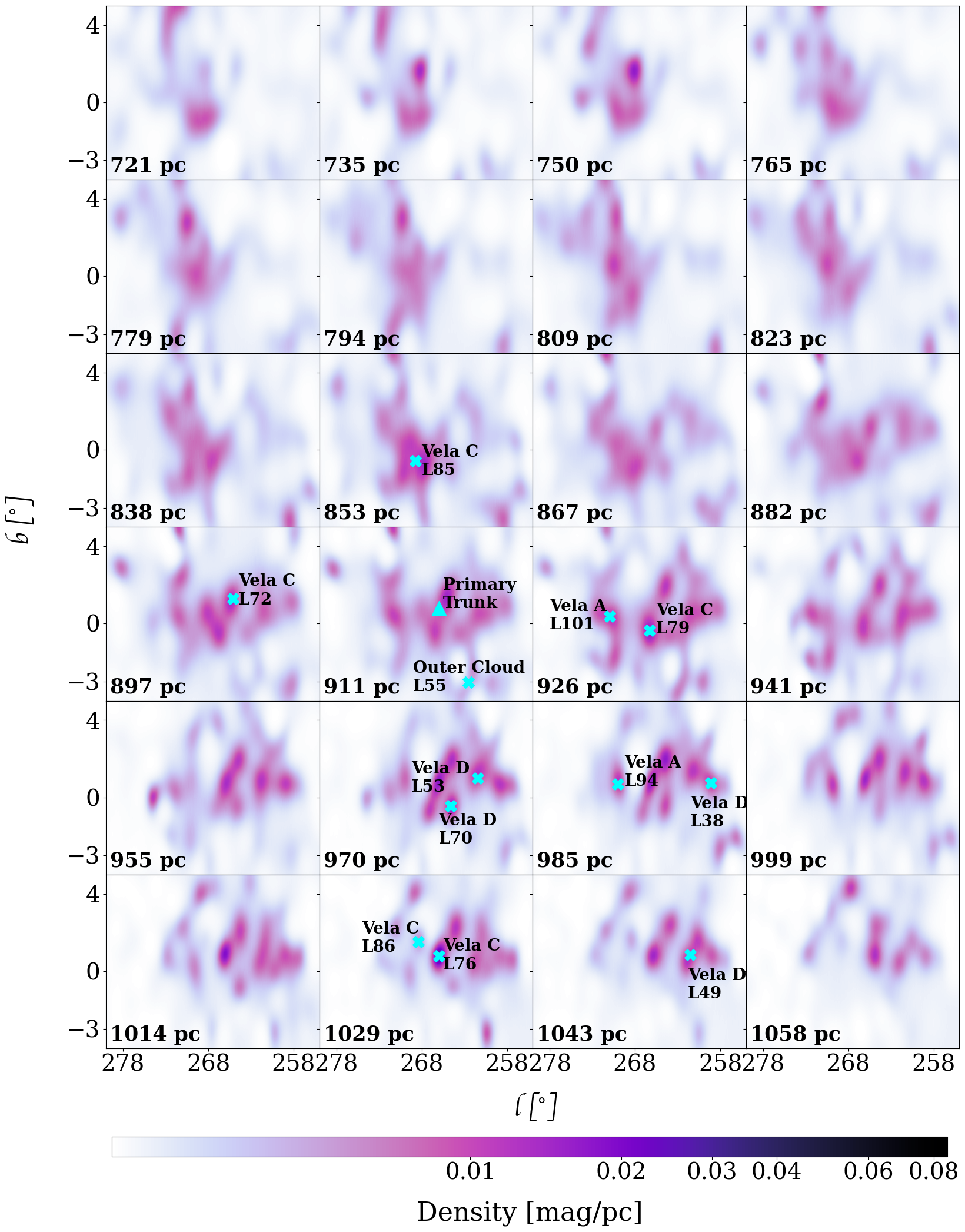}
    
    \hfill
    \includegraphics[width=0.6\textwidth]{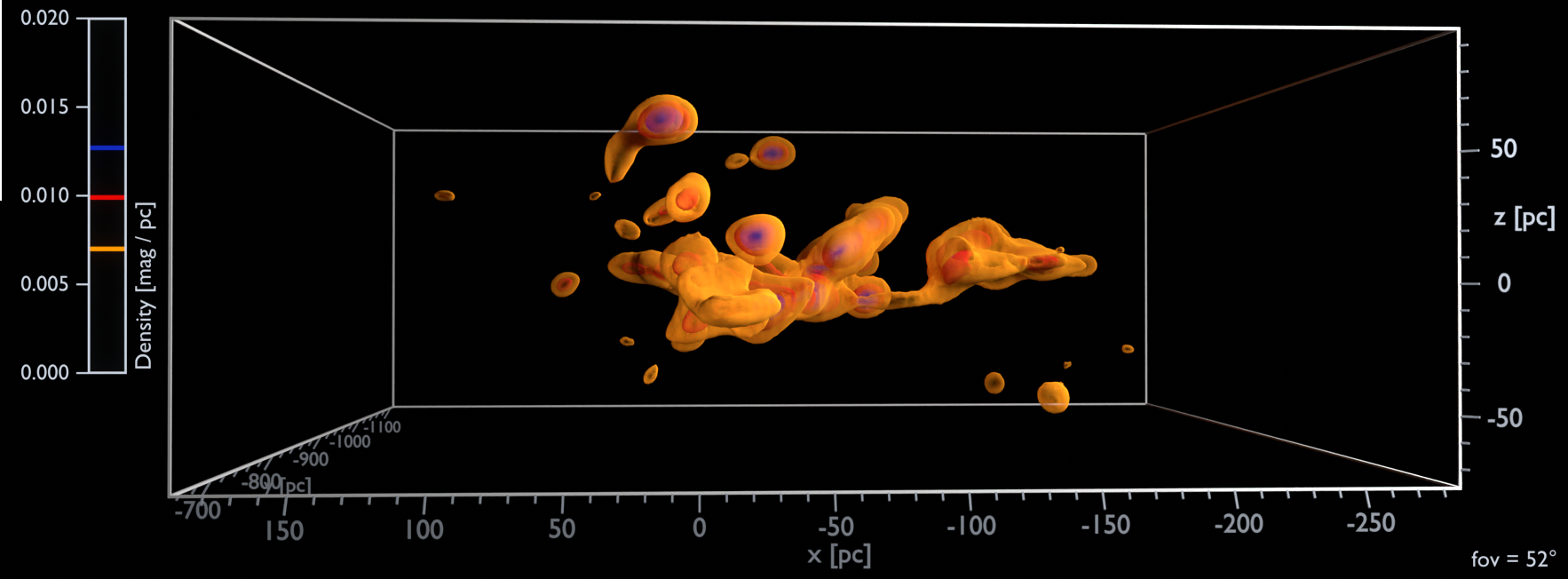}
    
    \hfill
    \includegraphics[width=0.6\textwidth]{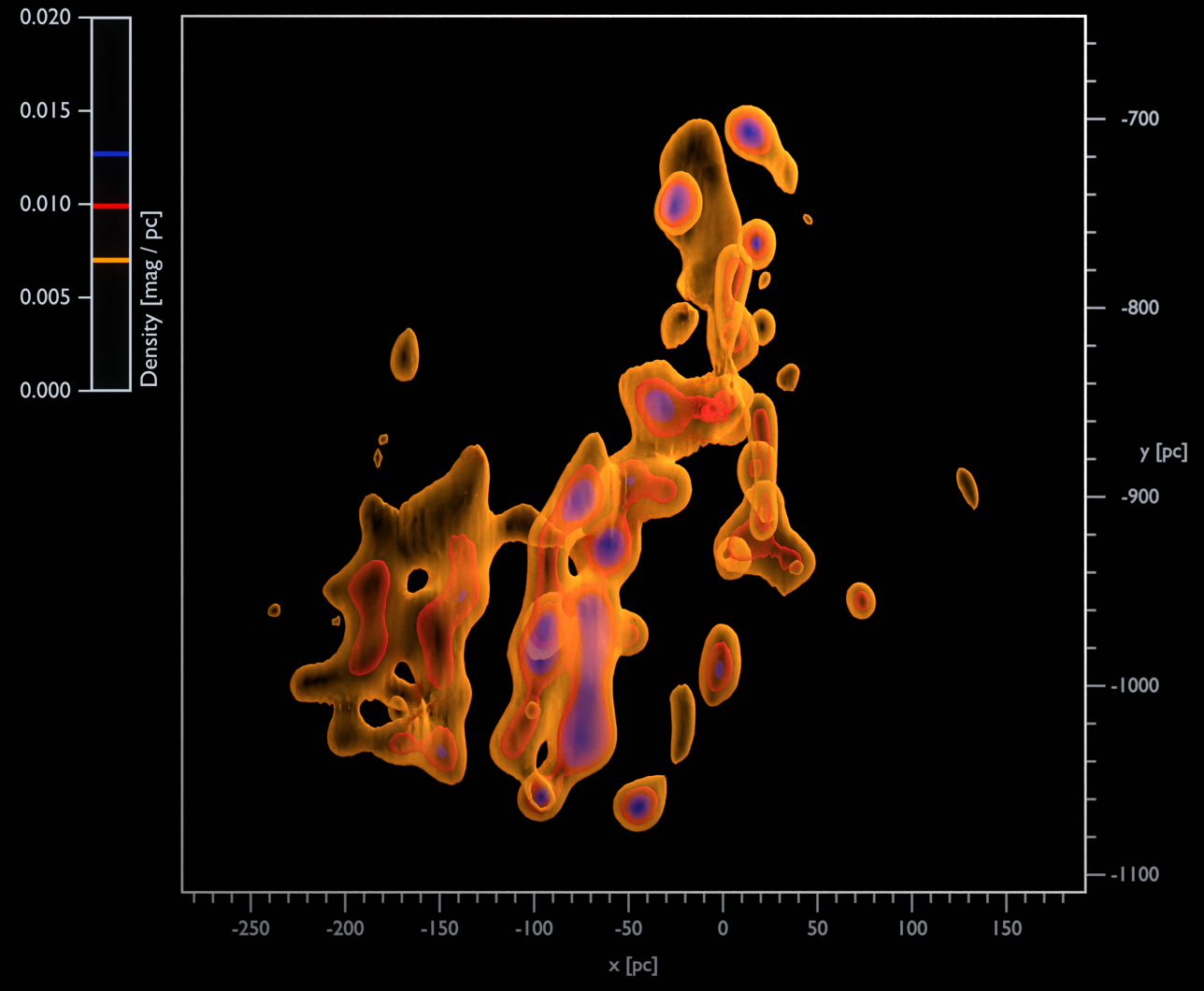}
    
    \end{multicols}
    \vspace{-0.8cm}
    \caption{Predicted 3D density structure of the Vela molecular ridge. Left: Slices along the line-of-sight of the predicted 3D density structure. With the Cyan triangle we have marked the mass weighted centroid of the primary trunk (as given in Table~\ref{tab:maintrunk_params} placed on the closest distance slice included in the plot. The Cyan $\times$s mark the mass weighted centroids of the interesting features discussed in Sec.\ref{sec:IndGMCs} and highlighted in Table~\ref{tab:leaf_params} placed on the closest distance slice included in the plot.; Right top: Video of a volume rendering of the predicted 3D density structure which begins from the view as seen from the Sun. It then rotates anti-clockwise about an axis perpendicular to the initial viewing angle. The semi-transparent iso-surfaces mark three different density levels with orange being the least dense to blue being the most dense as shown by the colour bar.; Right bottom: Still image showing the top down view of the predicted 3D density structure of the molecular cloud region using identical rendering to the preceding video.}
    
\label{fig:Vela_3Ddens}
\end{adjustwidth}   
\end{figure*}
\end{landscape}

We recover Vela A, C and D with several massive leaves as shown in Figs.~\ref{fig:Vela_2D} and \ref{fig:Vela_3Ddens}. We notice the majority of the cloud mass lies beyond 900 pc along the line-of-sight, however the single most massive leaf in the region (part of Vela C) is in fact closer. We observe Vela C to have a strong distance gradient with the Northern right edge (as viewed on sky) to be beyond 900 pc and the Southern left edge to be closer. The largest leaf of the entire region which forms the Southern left edge of Vela C is closer to us along the line-of-sight at 836 - 876 pc and has a mass of $6.38 \times 10^{3}$ M$_{\odot}$. Therefore we find Vela C to be the most massive component of the Vela molecular ridge with a distance range of $836 -- 1051$ pc and a mass weighted centroid at $l,b,d = 268.7^{\circ} , -0.57^{\circ} , 853$ pc for the most massive leaf. 

The distance range we measure for Vela C is approximately 60 pc greater than that predicted by \citet{Zucker2020}. \citet{Massi2019_Vela} predicted a distance of 950 pc using Gaia DR2 data which falls well within our range. Recent work by \citet{Hottier2021_Vela} suggest the Vela clouds are a much larger complex with some clouds with extents larger than 2 kpc along the line-of-sight. We do not find evidence for this much greater extension, which would be  unprecedented in the Solar Neighbourhood. This may be a manifestation of the fingers-of-god effect, which our forward-model approach is more robust against.

\section{Discussion}
\label{sec:Discussion}

\subsection{Systematic differences in mass estimates} 
\label{sec:MassDiff}

We find the masses calculated in this work for Orion, Taurus, Perseus and Cygnus to be 4 -- 22 times smaller when compared to the masses in \citet{Dharmawardena2022}. While both that and the present work used the same 3D dust density cubes, in \citet{Dharmawardena2022} we took into account the contribution to mass from all pixels within the 3D cube. By doing so we include all the background and surrounding pixels outside of the main molecular cloud and so overestimated the mass of the clouds. In the present work we only take into account the mass of the pixels clearly defined by the \texttt{astrodendro} contours. Although we are only considering 4 regions, we see a clear trend: the more distant and more massive a region is, the large the difference in mass observed between this work and \citet{Dharmawardena2022}. For distance, two effects may contribute to this trend: 1) Clouds that are further away have more foreground interstellar dust which is excluded by our \texttt{astrodendro} analysis; 2) Clouds that are further away require a larger volume to encompass sufficient stars as the number of stars decrease with distance in our catalogue due to obscuring. Hence the masses in \citet{Dharmawardena2022} are also contaminated by more background dust. 

When comparing to dust masses derived using the NICEST method \citep[e.g;][]{Lombardi2011_Orion, Zucker2021} we find our masses to be several times smaller. This is expected given NICEST is able to recover high extinction individual stars whereas we smooth some of the high extinction and hence lose some of the highest density. However as we are using extinctions measured using photometry from optical - infrared and therefore recover higher extinctions, we expect this effect to be reduced. Further work is required to verify this. Moreover, the background included in the NICEST maps when integrating for 2D extinction along the line-of-sight also inflates the measured mass and contributes to the mass difference.

We find a more representative mass estimate when measuring mass within the contours defined by \texttt{astrodendro} when compared to methods such as extinction masses, skeletonisation and fitting Gaussian clumps. We see that the individual clumps and filaments are mostly non Gaussian in shape and therefore only considering mass within a Gaussian inflates/deflates the true mass.  


\subsection{Mass - volume relationship} 

\begin{figure*}
\centering
\begin{subfigure}{0.45\textwidth}
  \centering
    \includegraphics[width=\textwidth]{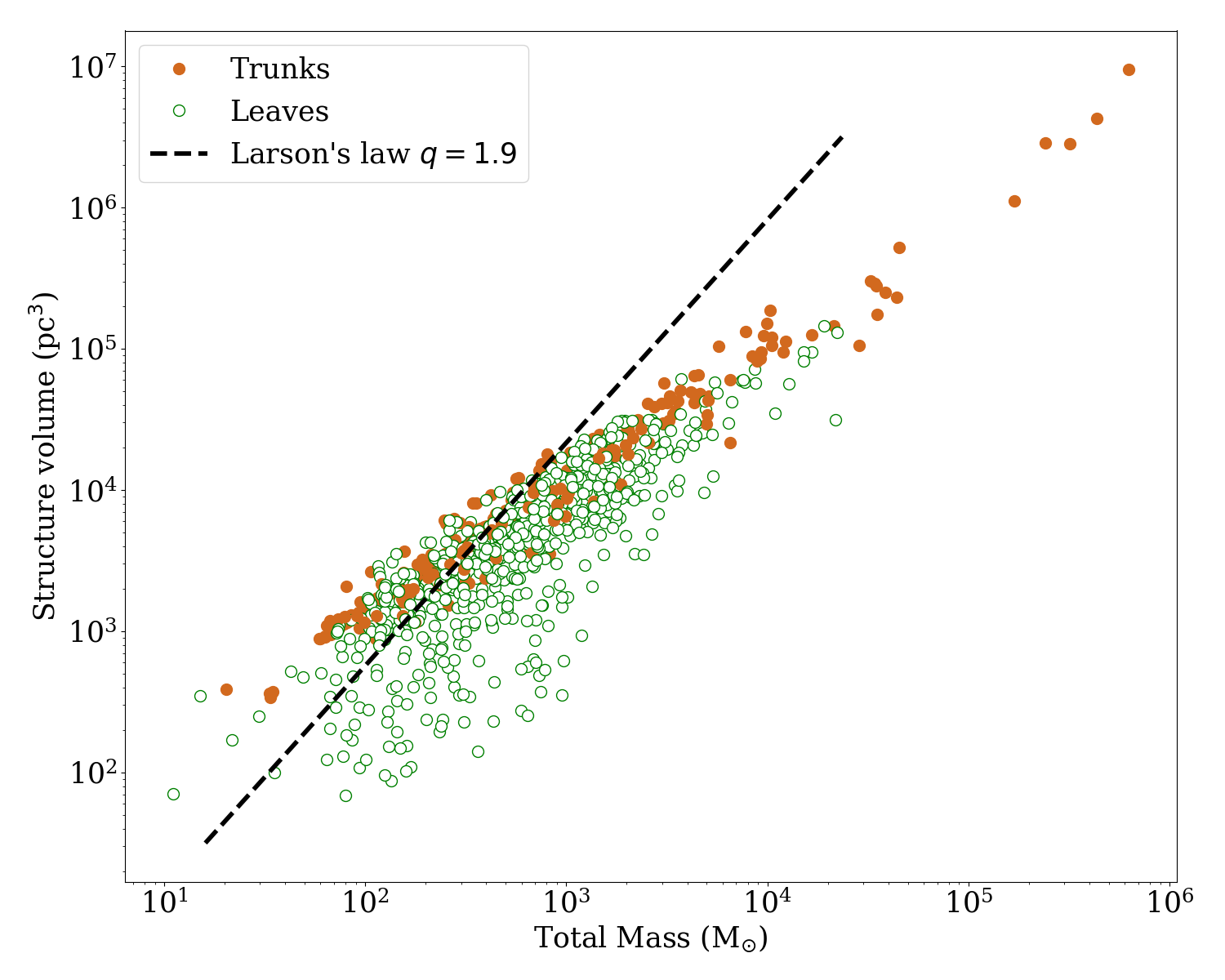}
  \label{fig:MassVol}
  \end{subfigure}
  \begin{subfigure}{0.45\textwidth}
  \centering
    \includegraphics[width=\textwidth]{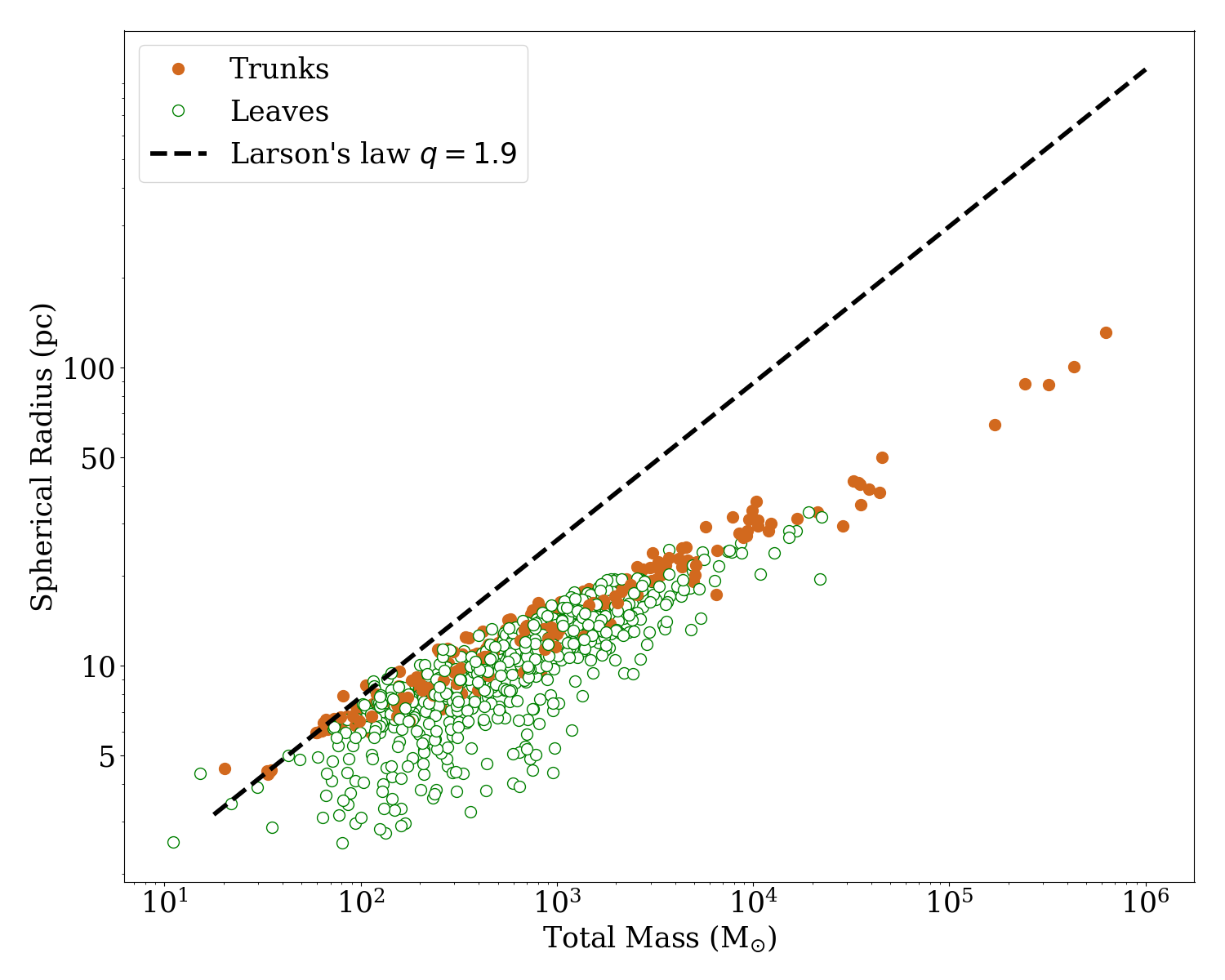}
  \label{fig:MassSphericalRad}
  \end{subfigure}
  
\begin{subfigure}{0.45\textwidth}
  \centering
    \includegraphics[width=\textwidth]{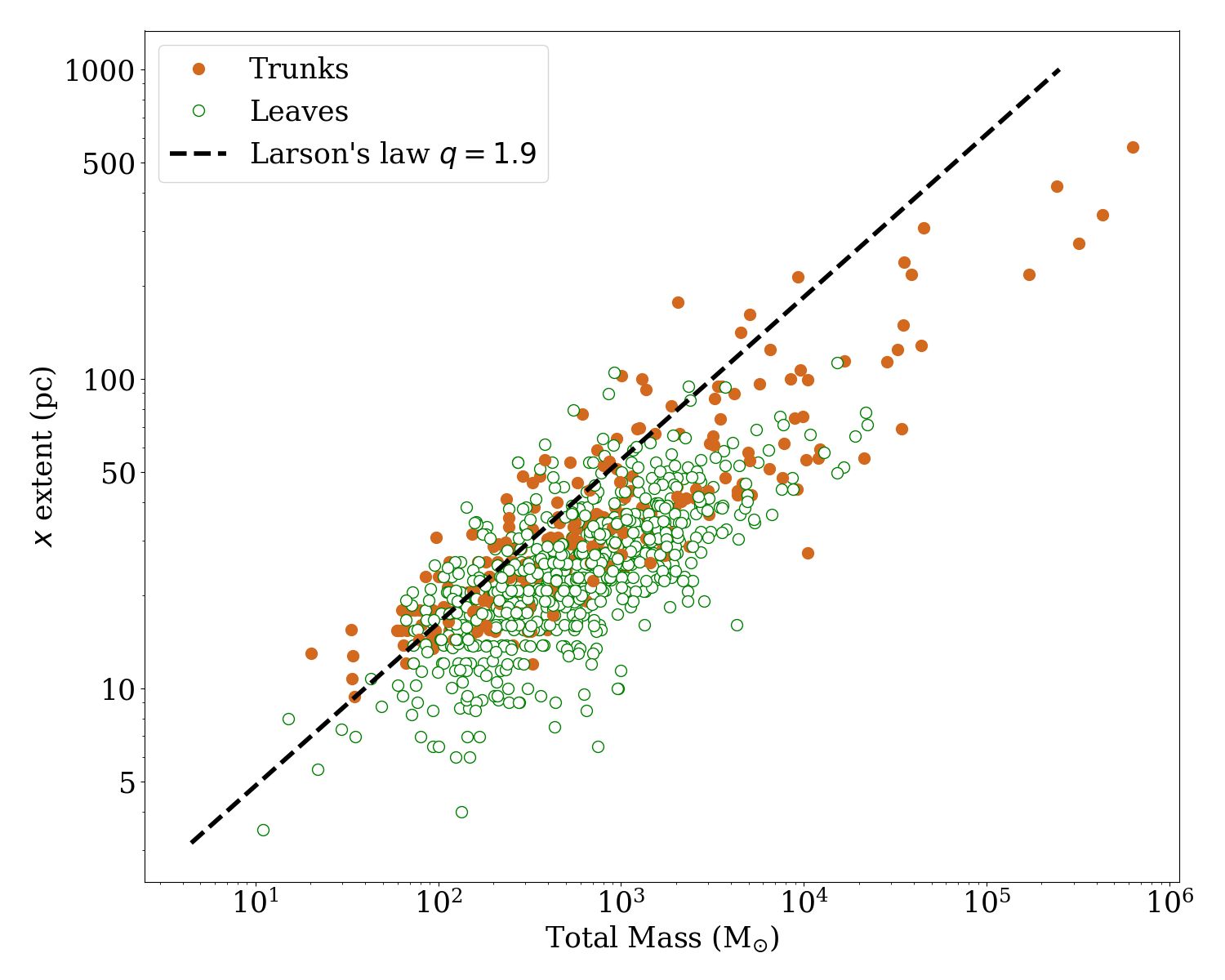}
  \label{fig:MassXextent}
  \end{subfigure}
\begin{subfigure}{0.45\textwidth}
  \centering
    \includegraphics[width=\textwidth]{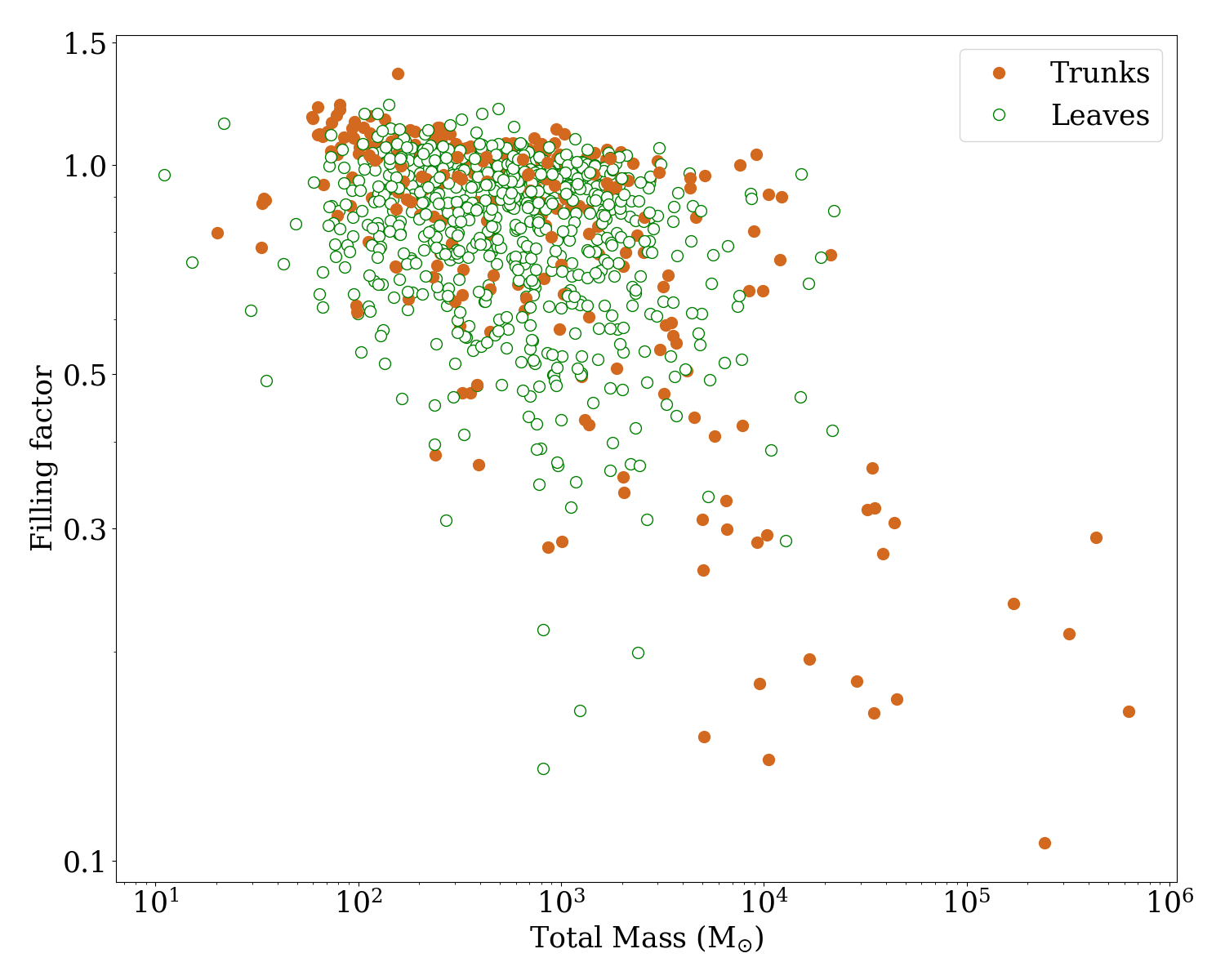}
  \label{fig:MassFillFac}
  \end{subfigure}
\caption{The total mass vs. (top left) volume of structure within contours laid down by \texttt{astrodendro}; (top right) spherical radius; (bottom left) $x$ extent; and (bottom right) filling factor of all trunks (brown filled circles) and leaves (green empty circles) for all the mapped molecular cloud complexes. In all plots the black dashed line represents the Larson's law where $M \propto R^{q}$ with $q = 1.9$.}
\label{fig:Mass_vs_ParamsComp}
\end{figure*}

Traditionally the Larson relationship predicts a linear relationship between the radius and mass measured from 2D data \citep{Larson1981}. In the top left and right panels of Fig.~\ref{fig:Mass_vs_ParamsComp} we see that the relationship between the masses of clouds in our maps and the radius of a sphere with the same volume as the cloud follows that predicted by Larson, showing that this relationship also holds when the 3D structure of clouds is accounted for. Further when we use the semi-extents of the clouds\footnote{i.e. the extent in the x, y or z direction divided by two.} instead of the radius of an equivolume sphere we find the same relationship, albeit with slightly enhanced scatter (see Fig.~\ref{fig:Mass_vs_ParamsComp} bottom left panel). This indicates that the details of how the size of a cloud is measured is relatively unimportant. Further we find that the trunks follow a tighter correlation between mass and radius compared to the leaves which are much more scattered. For the same mass, leaves tend to scatter towards smaller radii than trunks. This may indicate the difference between pressure-supported clouds and those actively undergoing collapse.

As shown in Fig.~\ref{fig:Mass_vs_ParamsComp} bottom right panel, we find an anti-correlation between the mass and filling factor with higher mass sources having a smaller filling factor. As expected, higher-mass clouds will tend to be more unstable to gravitational collapse (they are more likely to be above the Jeans' mass), and hence fragment more easily. Greater degrees of fragmentation will naturally result in lower filling factors for the largest structures.

\section{Conclusions}
\label{sec:Conclusions}

By applying astrodendro to modelled 3D dust density cubes of 16 molecular clouds we characterise their main clouds and extended environments and derive properties including mass, physical extents, volumes and filling factors. Many of the molecular clouds presented in this work  have been studied in 3D for the first time. 

\begin{enumerate}

    \item We show that, contrary to previous suggestions, the California molecular cloud is much more massive, extended and hence more diffuse than Orion A, explaining its lower star-formation rate.
    
    \item Camelopardalis is composed of a variety of distinct groupings of clouds, and is not an interconnected complex, even within a single layer.
    
    \item the Chamaeleon complex appears to be centred on a  cavity of unknown origin, which separates Cha I from Cha II and III.
    
    \item The Orion A and B molecular clouds lie along the edges of a bubble (Orion AB bubble) with a radius of roughly 50 pc. This bubble is centred behind Orion A at a distance of 480 pc. No obvious source for this bubble seems to be present; previous examinations of this bubble were unable to determine its distance precisely.
    
    \item Vela C exhibits a significant distance gradient along the cloud, and is the most massive component of the Vela Molecular Ridge.
    
    \item We demonstrate it is necessary to clearly define boundaries in 3D for these clouds to be able to accurately determine their masses. Integrating the extinction towards the cloud may overestimate the total mass by more than an order of magnitude.
    
    \item We show that the Larson relationship holds true in 3D both when assuming a spherical shape and also the true $x,y,z$ extents of the clouds. In future work we will also compare the cloud extents to the velocity dispersion to study the remaining two Larson's relationships.
    
    \item The environments of giant molecular clouds are so different, especially when you add the 3rd dimension, characterising them uniformly is not an easy task.
    
    \item We were unable to map a few small thin filamentary structures such as Lupus I and Musca. This is due to not enough sources from our stellar catalogue which required all sources have a W1 and W2 measurement. We will likely get a much better improvement with Gaia DR3 which will have extinctions using BP/RP  spectra.  
\end{enumerate}

The publicly available maps and dendrograms can be used to directly compare our dataset to others, for example by comparing the structures that are recovered, or determining the mass within the same surfaces (contours). The mass-weighted centroids, extents and cavities can be used to compare our high-resolution view of the clouds to large-scale 3D maps. The mass of structures above selected density thresholds  (read out directly from the publicly available dendrograms), for example, can be compared to masses recovered by simulations above the same density threshold.


\section*{Acknowledgements}

We thank the anonymous referee for their insightful comments which helped improve our manuscript. TED wishes to thank Andrew Gorden Wilson at New York University Courant institute of mathematical sciences and  Andreas Marek at Max Planck Computing and Data Facility, for useful discussions of the Gpytorch package and optimisations to \texttt{Dustribution}.
This project is funded by the Sonderforschungsbereich SFB 881 “The Milky Way System” of the German Research Foundation (DFG).
This research made use of \texttt{astrodendro}, a Python package to compute dendrograms of Astronomical data \url{http://www.dendrograms.org/}

\section*{Data Availability}

Our code is available at \url{https://github.com/Thavisha/Dustribution} and our results are available interactively at \url{www.mwdust.com}. All predicted 3D density and extinction data as well as the dendrogram fits files and parameter files can be downloaded from Zenodo via \url{https://doi.org/10.5281/zenodo.7061955}.



\bibliographystyle{mnras}
\bibliography{Bib_MWExt}


\newpage


\appendix

\section{Recent improvements to \texttt{Dustribution}}

Improvements made to \texttt{Dustribution} to reduce run time and memory usage. 

\label{sec:App:CodeMod}
\subsection{Memory bottleneck}
\SaveVerb{add}|add_jitter()|

In its first implementation, the memory requirements of the dust mapping program increased quadratically with the size of the training data. These translated in serious constrains on the resolution of the training grid and on the number of inducing points that could be used during the training of a GP model. Profiling the application showed a significant memory allocation in the \verb|forward()| method of the \verb|gpytorch.variational.VariationalStrategy| class, i.e. when computing the mean and the covariance of the variational distribution
$$q(f_{grid} | X_{grid}, f_{ind}, X_{ind}).$$
The implementation of this method leverages heavily GPyTorch's \verb|LazyTensors| to achieve an efficient memory utilisation and avoid storing full covariance matrices in memory. Unfortunately, one of the operations involved, namely the \UseVerb{add} method of the \verb|LazyEvaluatedKernelTensor| class, evaluates the covariance matrix on the training points. The allocation of such a big matrix explains the quadratic behaviour of the memory usage w.r.t. the size of the training dataset. As the above operation only adds a small positive value on the diagonal, there is no clear reason why the full covariance matrix should be evaluated. Moreover, the matrix is only used in the pyro integration when querying its standard deviation. Thus, only the diagonal of the covariance matrix on the training points is needed.

\begin{figure*}
\centering
\includegraphics[width=\textwidth]{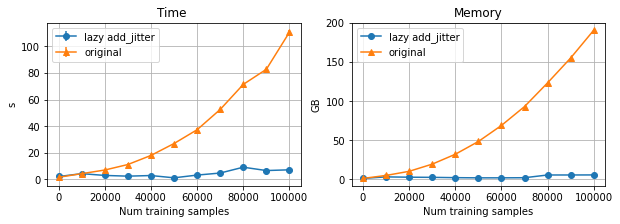}
\caption{\label{fig:pyro_int_ex}Time and peak memory required to perform 10 steps of a GP training when increasing the number of training data. The code uses the "Latent Function Inference" example from GPyTorch documentation. Both spatial and time complexity of the original GPyTorch code drop from quadratic to linear when using the "lazy" implementation of the \protect\UseVerb{add} method. Test performed on a node with 2 processors Intel Xeon IceLake Platinum 8360Y, 256 GB of main memory, PyTorch v1.9.0, GPyTorch v1.5.1, Pyro v1.7.0}
\end{figure*}

By implementing the \UseVerb{add} method in a lazy fashion, one can avoid the direct evaluation of the full covariance matrix. This, in turn, improves substantially the performances of the application. Indeed, the spatial complexity of the algorithm drops from being quadratic to linear w.r.t. the number of training data. Also the throughput of the application benefits from the lazy \UseVerb{add}, partially because of a reduced computational complexity but mainly because it allows a better memory management. This behaviour is evident from the results reported in the plots in Figure \ref{fig:pyro_int_ex}. For reproducibility purposes, the code is publicly available on GitHub \cite{gpytorch_test_gh}.

\clearpage

\section{GPytorch Input and Output parameters}
\label{sec:HPs}

Here we provide the input and output parameters of our \texttt{Dustribution} models for all 16 molecular cloud complexes as well as the parameters required to reproduce our dendrograms for them. All predicted results from \texttt{Dustribution} are predicted on grids in $x,y,z = 300 \times 300 \times 300$ cells.

\begin{table*}
    \centering
    \caption{Input and output parameters from the \texttt{Dustribution models} and input paramters for \texttt{astrodendro}}
    \label{tab:HPs}
    \begin{tabular}{llllllll}
    \hline
         Parameters 	&	California 	&	Camelopardalis	&	Canis Major	&	Carina	&	Cassiopeia	&	Cepheus	&	Chamaeleon	\\
       \hline \hline
Train $l$ bounds ($^{\circ}$)	&	155; 169	&	132; 158	&	220; 228	&	286.5; 288.5	&	110; 135	&	90; 125	&	290; 308	\\
Train $b$ bounds  ($^{\circ}$)	&	-15; -5	&	-5; 10	&	-4; 0	&	-1.5; 0.0	&	-5; 20	&	-5; 25	&	-22; -10	\\
Train $d$ bounds (pc)	&	300; 600	&	60; 1100	&	900; 1300	&	2200; 1900	&	90; 950	&	180; 1200	&	50; 300	\\
Stars incorporated from (pc)	&	370	&	110 	&	1040	&	2250	&	130	&	250	&	100	\\
Training $n_l,n_b,n_d$ (cells)	&	35; 25; 40	&	40; 40; 60	&	16; 8; 40	&	30; 30; 35	&	40; 40; 60	&	40; 40; 60	&	30; 20; 50	\\
Training $l,b, d$ resolution (pc) 	&	2.1; 2.1; 7.5	&	0.7; 0.4; 17	&	7.9; 7.8; 10 	&	2.6; 1.9; 20	&	1.0; 1.0; 14	&	2.7; 2.3; 17	&	0.5; 0.5; 5	\\
Initial $x,y,z$ scale length (pc)	&	10	&	10	&	10	&	5	&	10	&	10	&	5	\\
Final $x,y,z$  scale length (pc)	&	10.6; 8.9; 8.4	&	16.4; 13.5; 7.2	&	12.8; 12.4; 9.6	&	15.8; 17.6; 14.1	&	13.3; 16.9; 8.4	&	15.9; 15.5; 13.8	&	5.1; 6.3; 4.0	\\
Initial mean density ($log_{10}[\mathrm{mag \ pc}^{-1}]$)	&	-3.333	&	-3.333	&	-3.333	&	-2.909	&	-3.333	&	-3.333	&	-2.831	\\
Final mean density ($log_{10}[\mathrm{mag \ pc}^{-1}]$)	&	-2.954	&	 -3.003 	&	-2.887	&	-3.280	&	-3.076	&	-3.246	&	-3.459	\\
Initial scale factor (log[dex])	&	-1.215	&	-1.215	&	-1.215	&	-2.860	&	-1.215	&	-1.215	&	-1.236	\\
Final scale factor (log[dex])	&	-1.344	&	-3.665	&	-2.535	&	-3.280	&	-3.399	&	-3.235 	&	-1.348	\\
Learning rate	&	0.01	&	0.01	&	0.01	&	0.01	&	0.01	&	0.01	&	0.005	\\
Learning epsilon	&	1e-8	&	1e-8	&	1e-8	&	1e-8	&	1e-8	&	1e-8	&	1e-4	\\
Number of iterations	&	1700	&	2000	&	1000	&	2000	&	2000	&	2000	&	3500	\\
Number of inducing points	&	1000	&	1000	&	1000	&	1000	&	1000	&	1000	&	1000	\\
Predicting $d$ bounds (pc)	&	400; 530	&	140; 930	&	1040; 1200	&	2250; 2450	&	190; 820	&	270; 1200	&	100; 300	\\
Predicting $x, y, z$ resolution (pc)	&	0.6; 0.5; 0.3	&	2.6; 2.1; 0.8	&	0.7; 0.7; 0.3	&	0.5; 0.7; 0.2	&	1.7; 2.1; 1.2	&	2.3; 3.3; 2.0	&	0.5; 0.7; 0.3	\\
\texttt{astrodendro} $n_{pix}$	&	7558	&	192	&	5134	&	29269	&	232	&	116	&	636	\\
\hline
\end{tabular}

\end{table*}

\addtocounter{table}{-1}
\begin{table*}
    \centering
    \caption{\emph{cont.}}
    \begin{tabular}{llllllll}
        \hline
         Parameters 	&	Coalsack	&	Cygnus	&	Lupus	&	Mon R2	&	Orion	&	Perseus	&	Rosette	\\
       \hline \hline

Train $l$ bounds ($^{\circ}$)	&	295; 315	&	73; 87	&	333; 350	&	201; 222	&	180; 217	&	154; 164	&	205; 209	\\
Train $b$ bounds  ($^{\circ}$)	&	-4; 4	&	-4; 6	&	5; 20	&	-17; -7	&	-25.5; -3.8	&	-25; -14	&	-4; 0	\\
Train $d$ bounds (pc)	&	50; 260	&	800; 2200	&	40; 400	&	600; 1000	&	250; 550	&	180; 500	&	900; 1600	\\
Stars incorporated from (pc)	&	100	&	1200	&	80	&	730	&	270	&	200	&	1150	\\
Training $n_l,n_b,n_d$ (cells)	&	30; 15; 50	&	40; 40; 65	&	30; 30; 55	&	30; 25; 40	&	40; 40; 50	&	30; 30; 35	&	20; 20; 60	\\
Training $l,b, d$ resolution (pc) 	&	0.6; 0.5; 4	&	4.9; 3.5; 22	&	0.4; 0.3; 7	&	4.2; 4.1; 10	&	4.0; 2.3; 6	&	1.0; 1.1; 9	&	3.1; 3.1; 12	\\
Initial $x,y,z$ scale length (pc)	&	10	&	15	&	5	&	10	&	10	&	10	&	5	\\
Final $x,y,z$  scale length (pc)	&	8.9; 8.6; 8.8	&	19.1; 19.1; 19.2	&	7.0; 6.5; 6.2	&	12.2; 13.0; 11.2	&	9.5; 9.3; 9.5	&	12.8; 5.8; 6.3	&	8.3; 8.9; 5.9	\\
Initial mean density ($log_{10}[\mathrm{mag \ pc}^{-1}]$)	&	-3.333	&	-2.525	&	-2.831	&	-3.333	&	-3.333	&	-3.333	&	-2.909	\\
Final mean density ($log_{10}[\mathrm{mag \ pc}^{-1}]$)	&	-3.259	&	-2.909	&	-3.167	&	-2.897	&	-2.971	&	-3.224	&	 -3.149 	\\
Initial scale factor (log[dex])	&	-1.215	&	-2.811	&	-1.236	&	-1.215	&	-1.215	&	-1.215	&	-2.860	\\
Final scale factor (log[dex])	&	-1.378	&	-2.860 	&	-1.553	&	-2.403	&	-1.467	&	-1.227	&	 -3.149 	\\
Learning rate	&	0.01	&	0.01	&	0.001	&	0.01	&	0.01	&	0.01	&	0.01	\\
Learning epsilon	&	1e-8	&	1e-5	&	1e-4	&	1e-8	&	1e-8	&	1e-8	&	1e-8	\\
Number of iterations	&	1500	&	500	&	3500	&	1000	&	500	&	1000	&	2000	\\
Number of inducing points	&	1000	&	1000	&	1000	&	1000	&	1000	&	1000	&	1000	\\
Predicting $d$ bounds (pc)	&	130; 225	&	1280; 2200	&	100; 400	&	760; 925	&	270; 550	&	230; 500	&	1150; 1330	\\
Predicting $x, y, z$ resolution (pc)	&	0.3; 0.3; 0.1	&	1.9; 3.3; 1.3	&	1.0; 0.5; 0.4	&	0.8; 0.8; 0.6	&	1.2; 1.1; 0.7	&	0.9; 0.5; 0.5	&	0.7; 0.5; 0.3	\\
\texttt{astrodendro} $n_{pix}$	&	38091	&	460	&	616	&	2241	&	470	&	1004	&	2111	\\

\hline
\end{tabular}

\end{table*}

\addtocounter{table}{-1}
\begin{table*}
    \centering
    \caption{\emph{cont.}}
    \begin{tabular}{lll}
        \hline
         Parameters 	&	Taurus & Vela	\\
       \hline \hline

Train $l$ bounds ($^{\circ}$)	&	165; 180	&	255; 280	\\
Train $b$ bounds  ($^{\circ}$)	&	-20; -10	&	-4; 5	\\
Train $d$ bounds (pc)	&	40; 350	&	600; 1300	\\
Stars incorporated from (pc)	&	90	&	640	\\
Training $n_l,n_b,n_d$ (cells)	&	30; 30; 35	&	30; 10; 50	\\
Training $l,b, d$ resolution (pc) 	&	0.3; 0.2; 9	&	8.7; 9.4; 14	\\
Initial $x,y,z$ scale length (pc)	&	5	&	10	\\
Final $x,y,z$  scale length (pc)	&	9.3; 6.9; 4.9	&	14.9; 15.8; 13.2	\\
Initial mean density ($log_{10}[\mathrm{mag \ pc}^{-1}]$)	&	-2.831	&	-3.333	\\
Final mean density ($log_{10}[\mathrm{mag \ pc}^{-1}]$)	&	-1.992	&	-3.090 	\\
Initial scale factor (log[dex])	&	-1.236	&	-1.215	\\
Final scale factor (log[dex])	&	-1.992	&	-3.329	\\
Learning rate	&	0.005	&	0.01	\\
Learning epsilon	&	1e-4	&	1e-8	\\
Number of iterations	&	1500	&	2000	\\
Number of inducing points	&	500	&	1000	\\
Predicting $d$ bounds (pc)	&	90; 350	&	670; 1110	\\
Predicting $x, y, z$ resolution (pc)	&	0.9; 0.3; 0.3	&	1.6; 1.5; 0.6	\\
\texttt{astrodendro} $n_{pix}$	&	1818	&	1146	\\

\hline
\end{tabular}

\end{table*}

\section{Full list of parameters derived from \texttt{astrodendro} available in the electronic tables}
\label{sec:Ap:ElecTabs}

The complete set of parameters derived from astrodendro for each molecular cloud complex can be found on zenodo at \url{https://dendrograms.readthedocs.io/en/stable}. 

\begin{table*}
    \centering
    \caption{Extended environment parameters derived from dendrograms}
    \label{tab:TabCols}
    \begin{tabular}{ll}
        \hline
         Parameter & Description \\
       \hline \hline
pandas\_index & index asigned by pandas \\

astrodendro\_index & index assigned by astrodendro - used in this work to refer to structures as L\#\#, C\#\#, T\#\# \\
& for leaf, child and trunk respectively where \#\# is the assigned index \\

parent & index of parent structure \\
children & index of child structures \\
peak\_dense\_mag-per-pc & peak density in mag pc$^{-1}$ \\  
peak\_density\_err\_lower & lower peak density error \\
peak\_density\_err\_upper & upper peak density error \\
peak\_dense\_xCoord  & peak density x coordinate \\
peak\_dense\_yCoord & peak density y coordinate \\
peak\_dense\_zCoord & peak density z coordinate \\
peak\_dense\_lCoord & peak density l coordinate \\ 
peak\_dense\_bCoord & peak density b coordinate \\ 
peak\_dense\_dCoord & peak density d coordinate \\
structure\_vol\_pc**3 & volume of structure outlined by its contour in pc$^{3}$ \\
mass\_mag-pc**2 & dust mass in mag pc$^{2}$ \\
mass\_err\_lower\_mag-pc**2 & dust mass mag pc$^{2}$ lower error \\
mass\_err\_upper\_mag-pc**2 & dust mass mag pc$^{2}$ upper error \\  
mass\_grams  & dust mass in grams  \\
mass\_err\_lower\_grams  & dust mass in grams lower error  \\ mass\_err\_upper\_grams & dust mass in grams lower error  \\ mass\_Msol & dust mass in solar masses \\
mass\_err\_lower\_Msol & dust mass in solar masses lower error \\
mass\_err\_upper\_Msol & dust mass in solar masses upper error \\
mass\_Gas\_Msol & gas mass in solar masses \\ 
mass\_Gas\_Msol\_err\_lower & gas mass in solar masses lower error \\
mass\_Gas\_Msol\_err\_upper & gas mass in solar masses upper error \\
mass\_Total\_Msol & total (gas$+$dust) mass in solar masses \\
mass\_Total\_Msol\_err\_lower & total (gas$+$dust) mass in solar masses lower error \\
mass\_Total\_Msol\_err\_upper & total (gas$+$dust) mass in solar masses upper error \\ 
elliposid\_centre\_xyzCoords & $x,y,z$ center of ellipsoid fitted to the 3 vectors based on the furthest apart pixels \\
elliposid\_centre\_lbdCoords & $l,b,d$ center of ellipsoid fitted to the furthest extents in 3D \\ elliposid\_angles\_theta\_deg & angle from z axis to the vector joining to furthest extent points in radians \\
elliposid\_angles\_phi\_deg & angle of the azimuth around z to represent the ellipsoid with in radians \\
max\_dist\_structure\_orth1\_pc & vector between the two furthest away pixels within the dendrogram contours in pc \\

max\_dist\_structure\_orth2\_pc & vector orthogonal to the vector between the two furthest away pixels within the dendrogram  \\
& contours in pc\\

max\_dist\_structure\_orth3\_pc & 3rd vector orthogonal to the two perviosuly derived vectors in pc \\ max\_dist\_structure\_xyzcoords\_pcunits & coordinates of the 3 orthogonal max extent vectors \\ 
max\_vecs\_orth & vectors of the 3 orthogonal max extent vectors \\  
mass\_weighted\_centroid\_x & mass weighted centroid x \\
mass\_weighted\_centroid\_y & mass weighted centroid y \\
mass\_weighted\_centroid\_z & mass weighted centroid z \\
mass\_weighted\_centroid\_l & mass weighted centroid l \\    
mass\_weighted\_centroid\_b & mass weighted centroid b \\
mass\_weighted\_centroid\_d & mass weighted centroid d \\   
\hline
\end{tabular}

\end{table*}

\addtocounter{table}{-1}
\begin{table*}
    \centering
    \caption{\emph{cont.}}
    \label{tab:TabCols}
    \begin{tabular}{ll}
        \hline
         Parameter & Description \\
       \hline \hline

ellipsoid\_vol\_pc**3 & volume of the fitted ellipsoid \\
x\_min & minimum x coordinate \\
x\_min\_err & minimum x coordinate error \\   
x\_max & maximum x coordinate \\
x\_max\_err & maximum x coordinate error \\ 
x\_extent & extent between min and max x coordinate \\
x\_extent\_err & error of the extent between min and max x coordinate \\ 
y\_min & minimum y coordinate \\
y\_min\_err & minimum y coordinate error \\ 
y\_max & maximum y coordinate \\  
y\_max\_err & maximum y coordinate error \\   
y\_extent & extent between min and max y coordinate \\
y\_extent\_err  & error of the extent between min and max y coordinate \\ 
z\_min & minimum z coordinate \\
z\_min\_err & minimum z coordinate error \\ 
z\_max & maximum z coordinate \\  
z\_max\_err & maximum z coordinate error \\   
z\_extent & extent between min and max z coordinate \\
z\_extent\_err  & error of the extent between min and max z coordinate \\  
xy\_aspect\_ratio & aspect ratio between x and y extents \\ 
xz\_aspect\_ratio  & aspect ratio between x and z extents \\  
yz\_aspect\_ratio  & aspect ratio between y and z extents \\ 
l\_min & minimum l coordinate \\
l\_min\_err & minimum l coordinate error \\ 
l\_max & maximum l coordinate \\  
l\_max\_err & maximum l coordinate error \\   
l\_extent & extent between min and max l coordinate \\
l\_extent\_err  & error of the extent between min and max l coordinate \\  
b\_min & minimum b coordinate \\
b\_min\_err & minimum b coordinate error \\ 
b\_max & maximum b coordinate \\  
b\_max\_err & maximum b coordinate error \\   
b\_extent & extent between min and max b coordinate \\
b\_extent\_err  & error of the extent between min and max b coordinate \\
d\_min & minimum d coordinate \\
d\_min\_err & minimum d coordinate error \\ 
d\_max & maximum d coordinate \\  
d\_max\_err & maximum d coordinate error \\   
d\_extent & extent between min and max d coordinate \\
d\_extent\_err  & error of the extent between min and max d coordinate \\
fill\_fac\_pc**3 & filling factor - ratio of structure\_vol\_pc**3 / ellipsoid\_vol\_pc**3 \\

\hline
\end{tabular}

\end{table*}

\clearpage

\section{Predicted 2D extinction and 3D density plots of the molecular cloud complexes}
\label{sec:predfigs}

\subsection{Predicted 2D extinction plots of the molecular cloud complexes}
\label{sec:Ap:2Dext}

\begin{figure}
    \centering
    \includegraphics[width=0.55\textwidth]{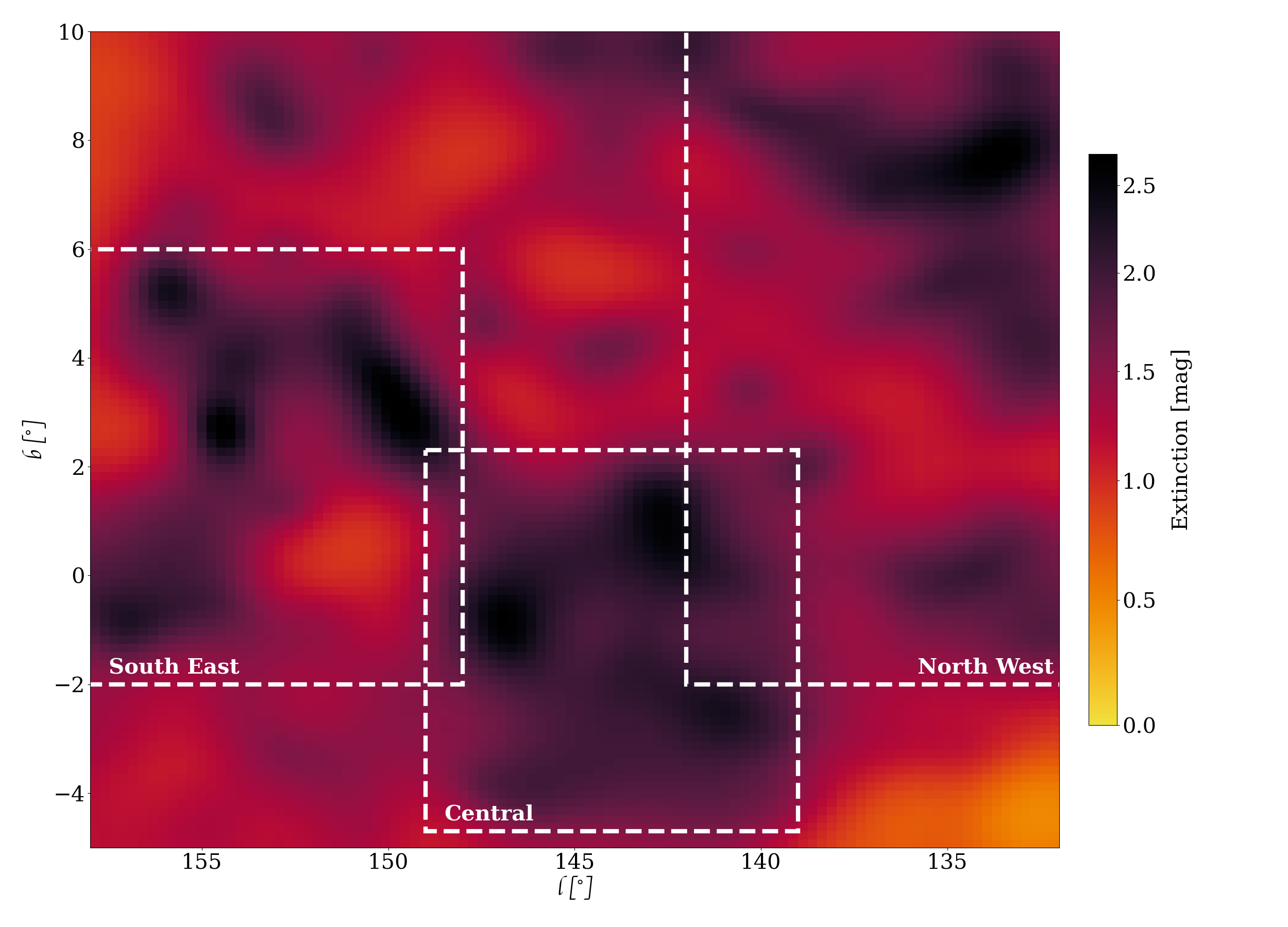}
    \caption{The predicted integrated 2D extinction of the Camelopardalis molecular cloud complex as seen from the Sun (i.e: on the plane of the sky). The regions highlighted are based on nearby features of the cloud complex as seen on sky.}
    \label{fig:Camel_2D} 
\end{figure}

\begin{figure}
    \centering
    \includegraphics[width=0.55\textwidth]{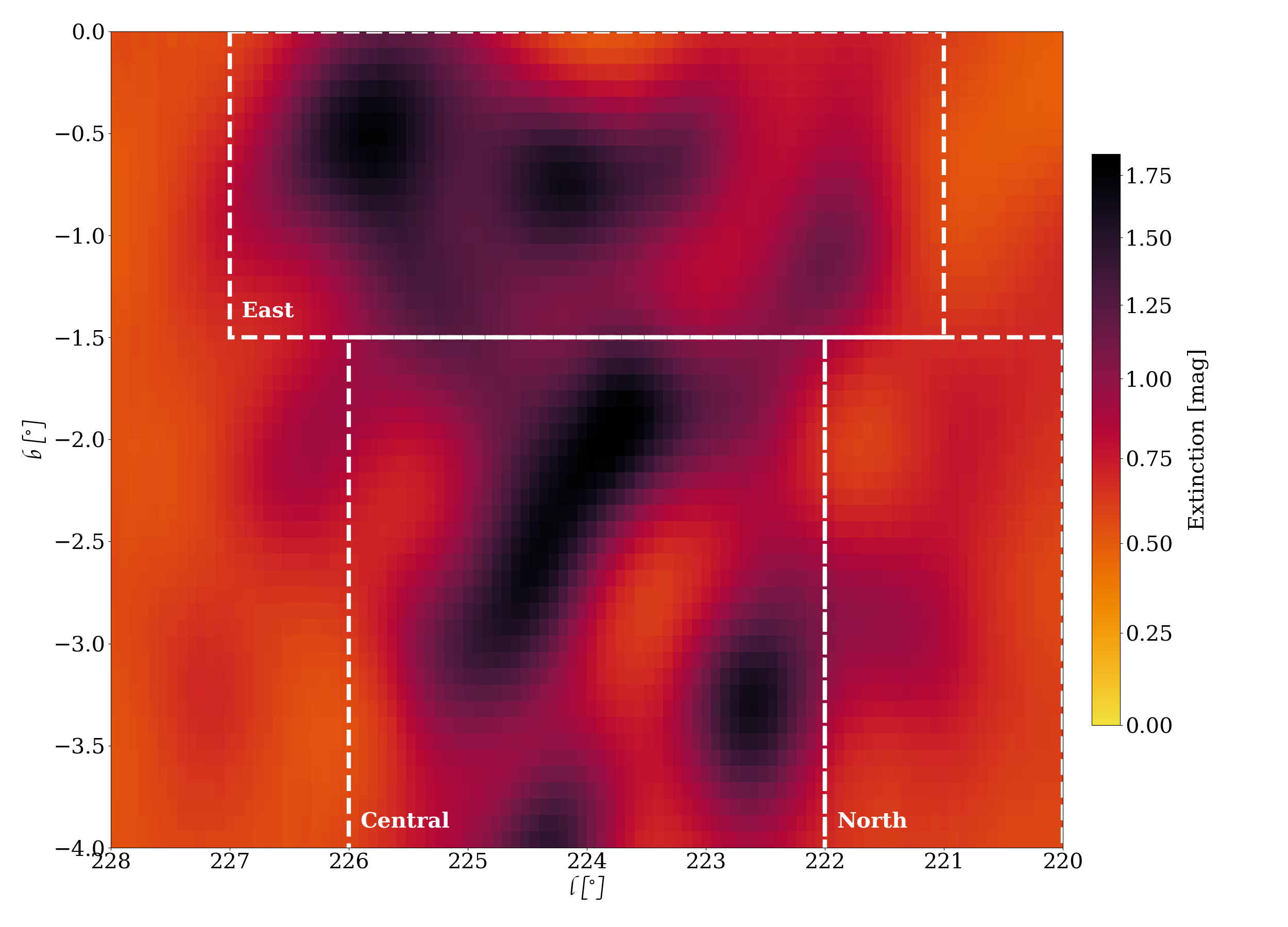}
    \caption{The predicted integrated 2D extinction of the Canis Major molecular cloud complex as seen from the Sun (i.e: on the plane of the sky). The regions highlighted are based on nearby features of the cloud complex as seen on sky.}
    \label{fig:CanMaj_2D} 
\end{figure}

\begin{figure}
    \centering
    \includegraphics[width=0.55\textwidth]{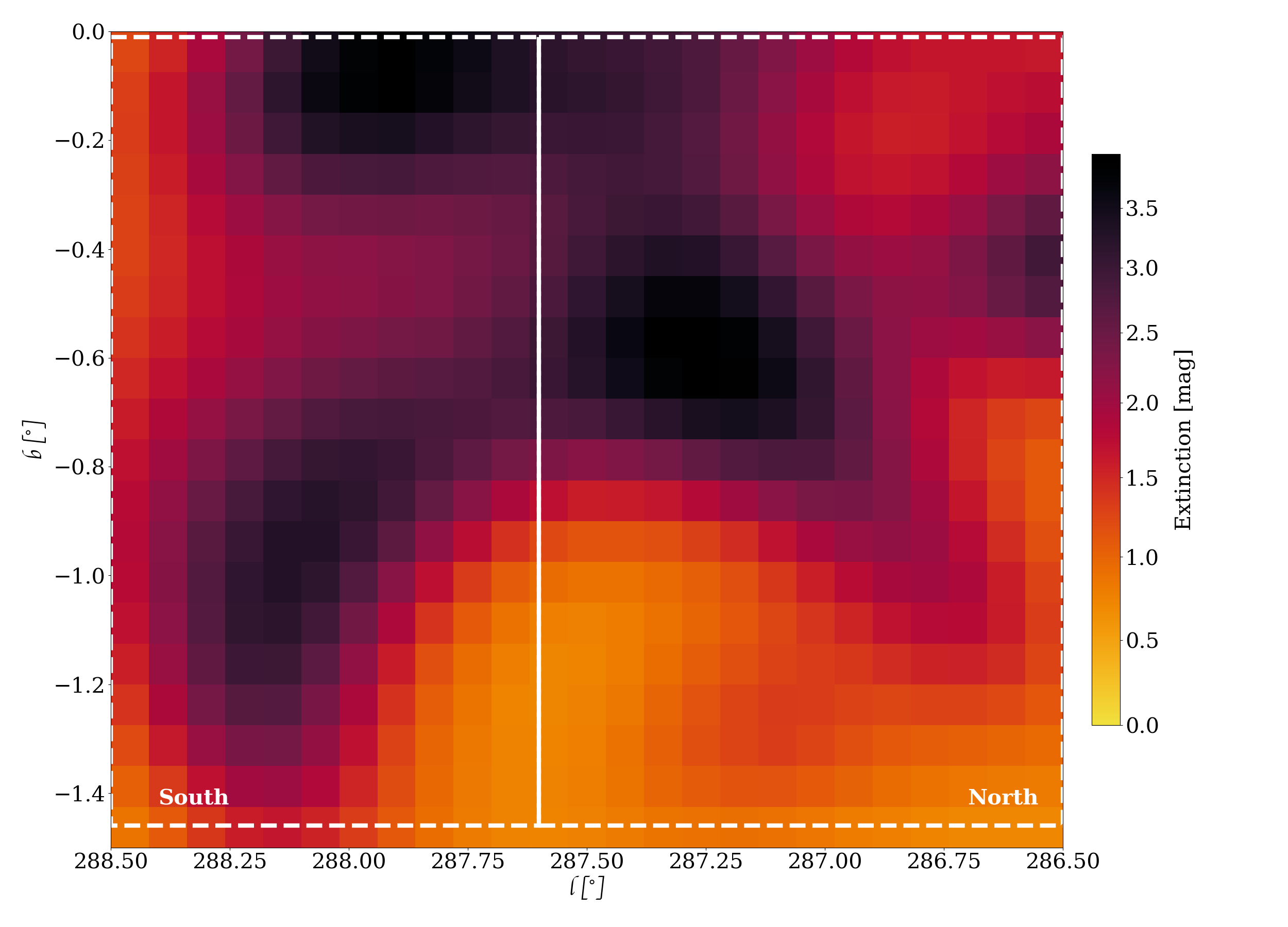}
    \caption{ The predicted integrated 2D extinction of the Carina molecular cloud complex as seen from the Sun (i.e: on the plane of the sky). The regions highlighted are based on nearby features of the cloud complex as seen on sky.}
    \label{fig:Carina_2D} 
\end{figure}

\begin{figure}
    \centering
    \includegraphics[width=0.55\textwidth]{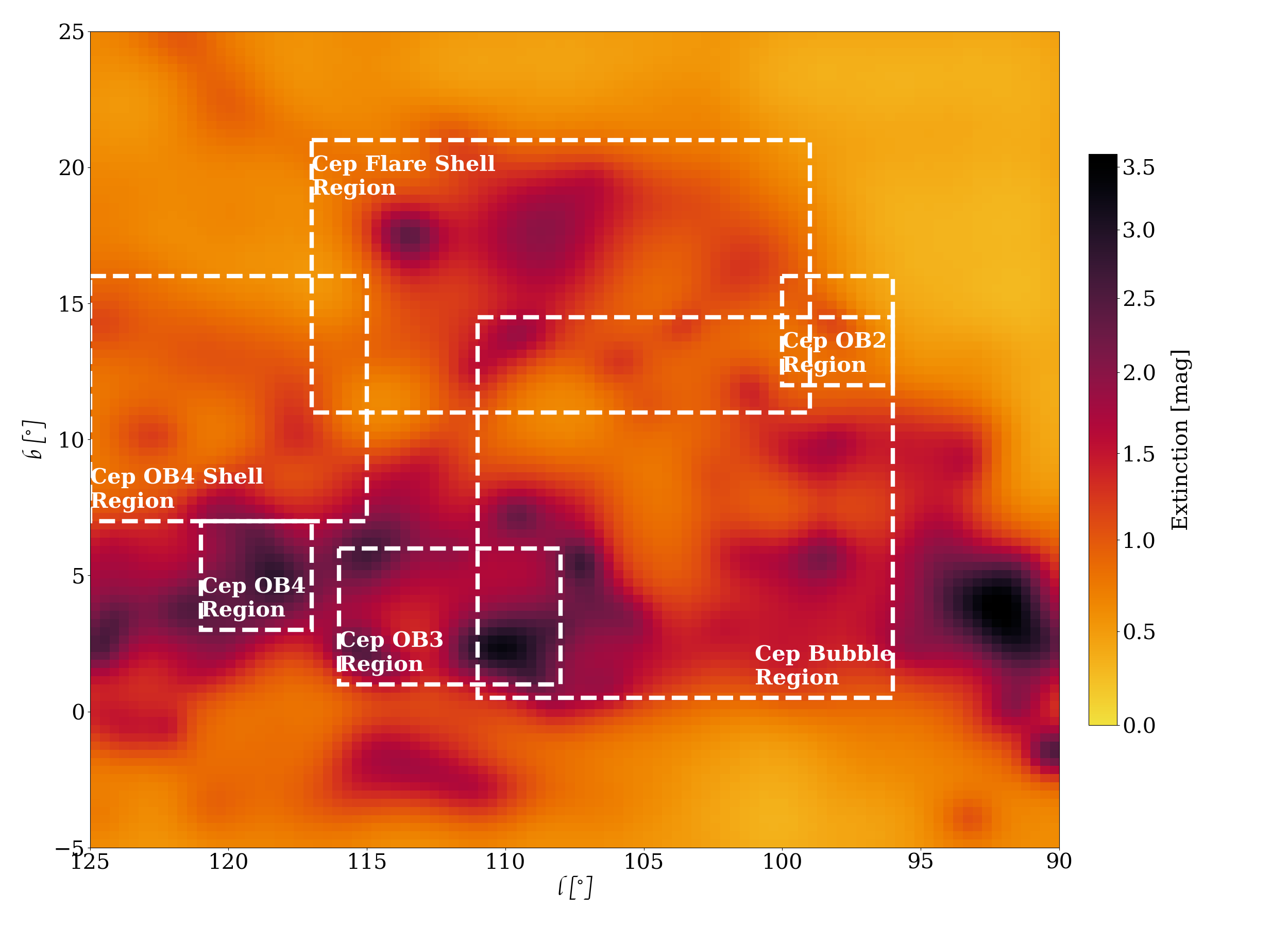}
    \caption{The predicted integrated 2D extinction of the Cepheus molecular cloud complex as seen from the Sun (i.e: on the plane of the sky). The regions highlighted are based on nearby features of the cloud complex as seen on sky.}
    \label{fig:Cep_2D} 
\end{figure}

\begin{figure}
    \centering
    \includegraphics[width=0.55\textwidth]{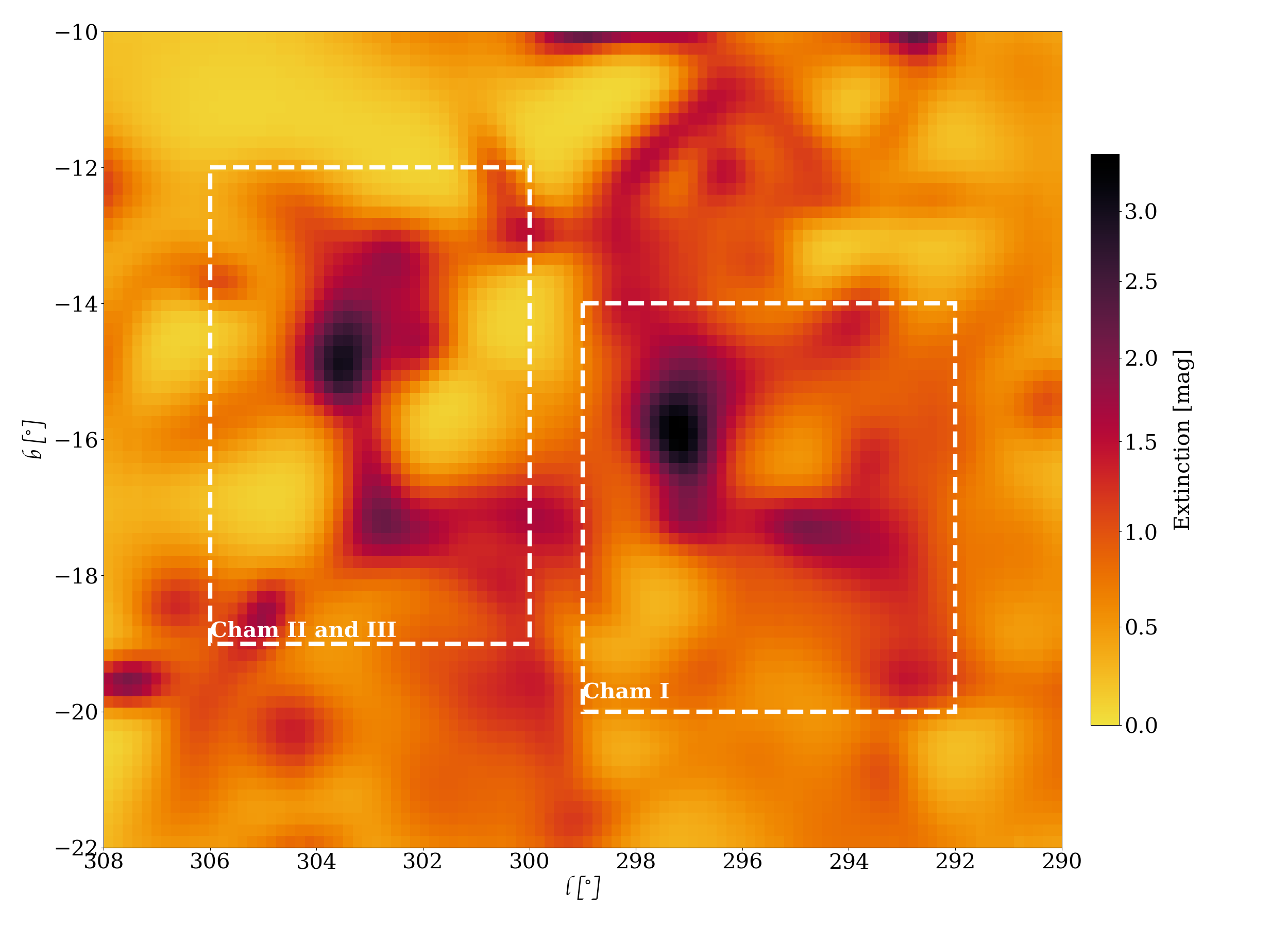}
    \caption{The predicted integrated 2D extinction of the Chamaeleon molecular cloud complex as seen from the Sun (i.e: on the plane of the sky). Cham I, II and III are highlighted.}
    \label{fig:Cham_2D} 
\end{figure}

\begin{figure}
    \centering
    \includegraphics[width=0.55\textwidth]{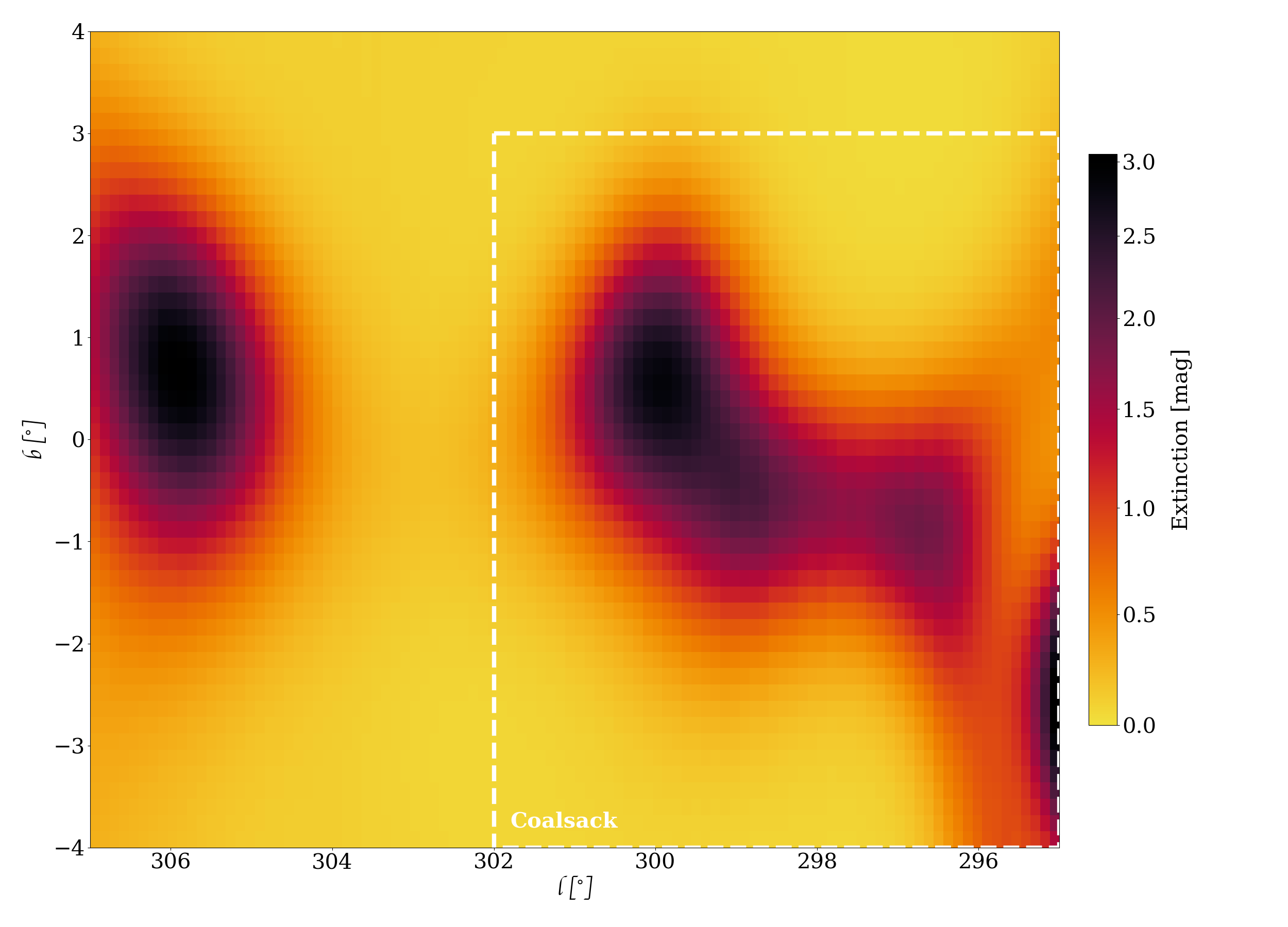}
    \caption{The predicted integrated 2D extinction of the Southern Coalsack molecular cloud as seen from the Sun (i.e: on the plane of the sky).}
    \label{fig:Coal_2D} 
\end{figure}

\begin{figure}
    \centering
    \includegraphics[width=0.55\textwidth]{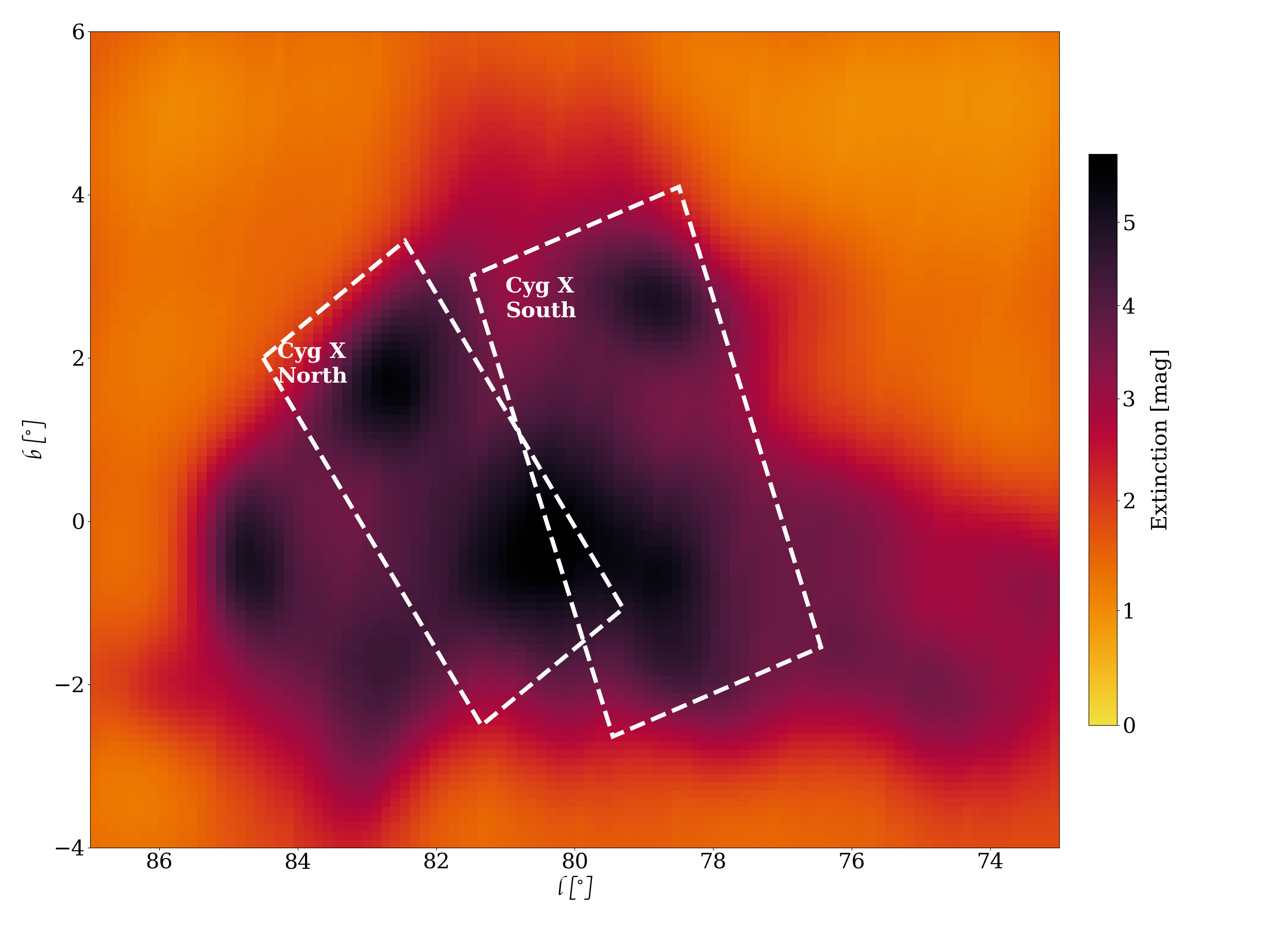}
    \caption{The predicted integrated 2D extinction of the Cygnus X molecular cloud complex as seen from the Sun (i.e: on the plane of the sky). Cyg X North and South are highlighted.}
    \label{fig:Cyg_2D} 
\end{figure}

\begin{figure}
    \centering
    \includegraphics[width=0.55\textwidth]{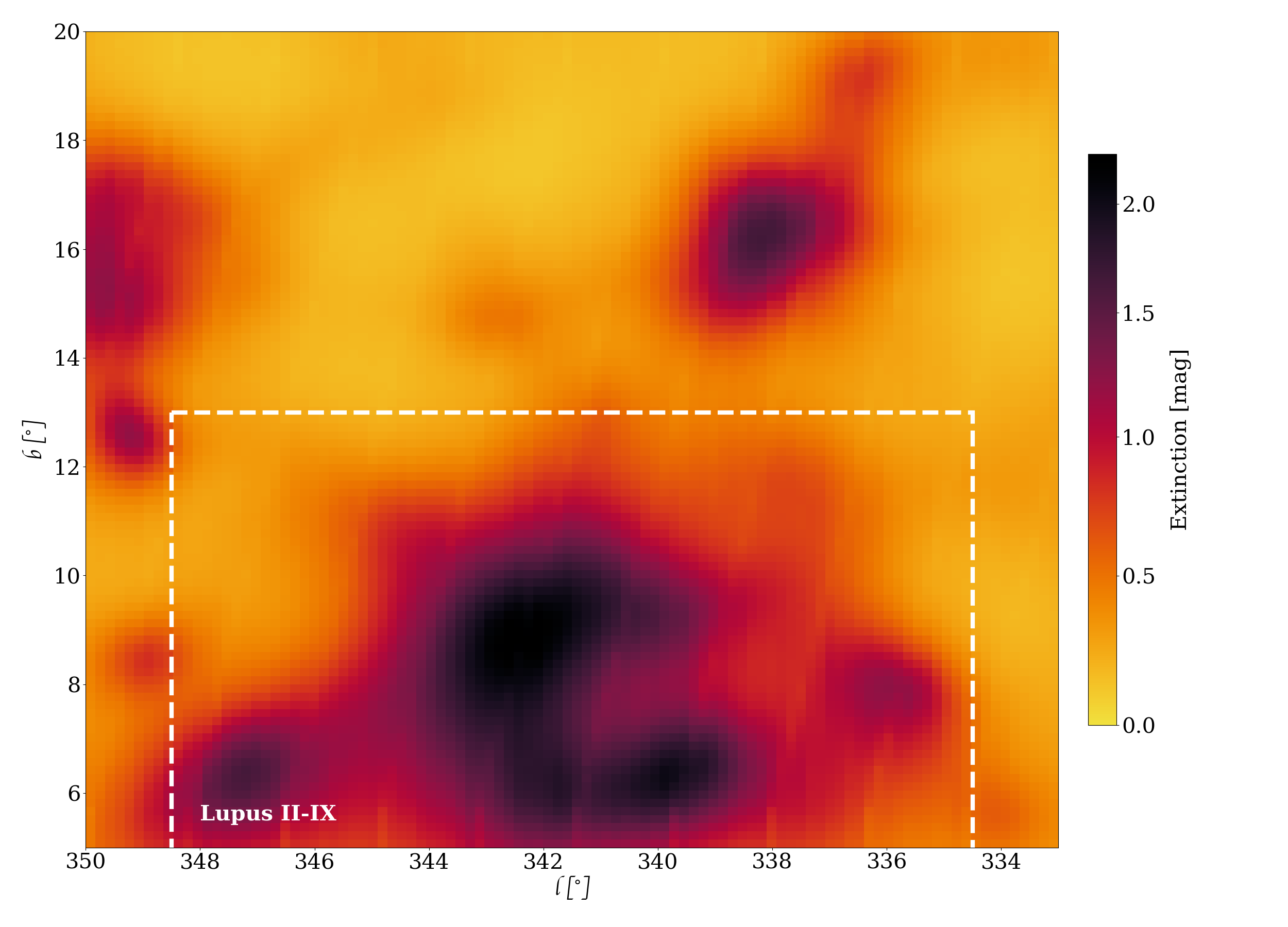}
    \caption{The predicted integrated 2D extinction of the Lupus molecular cloud complex as seen from the Sun (i.e: on the plane of the sky). Lupus II - IX are highlighted}
    \label{fig:Lup_2D} 
\end{figure}

\begin{figure}
    \centering
    \includegraphics[width=0.55\textwidth]{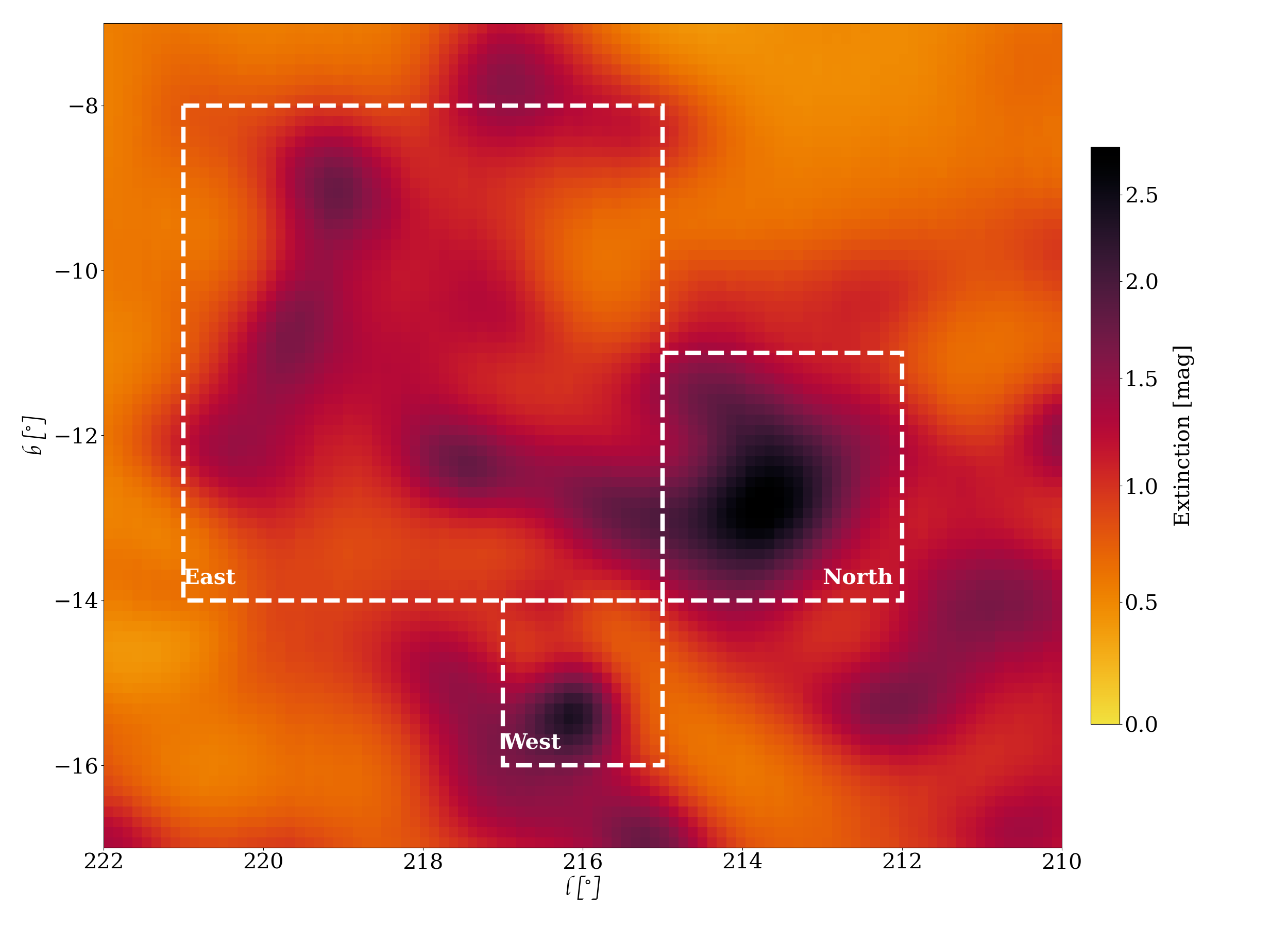}
    \caption{The predicted integrated 2D extinction of the Mon R2 molecular cloud complex as seen from the Sun (i.e: on the plane of the sky). The regions highlighted are based on nearby features of the cloud complex as seen on sky.}
    \label{fig:MonR2_2D} 
\end{figure}

\begin{figure}
    \centering
    \includegraphics[width=0.55\textwidth]{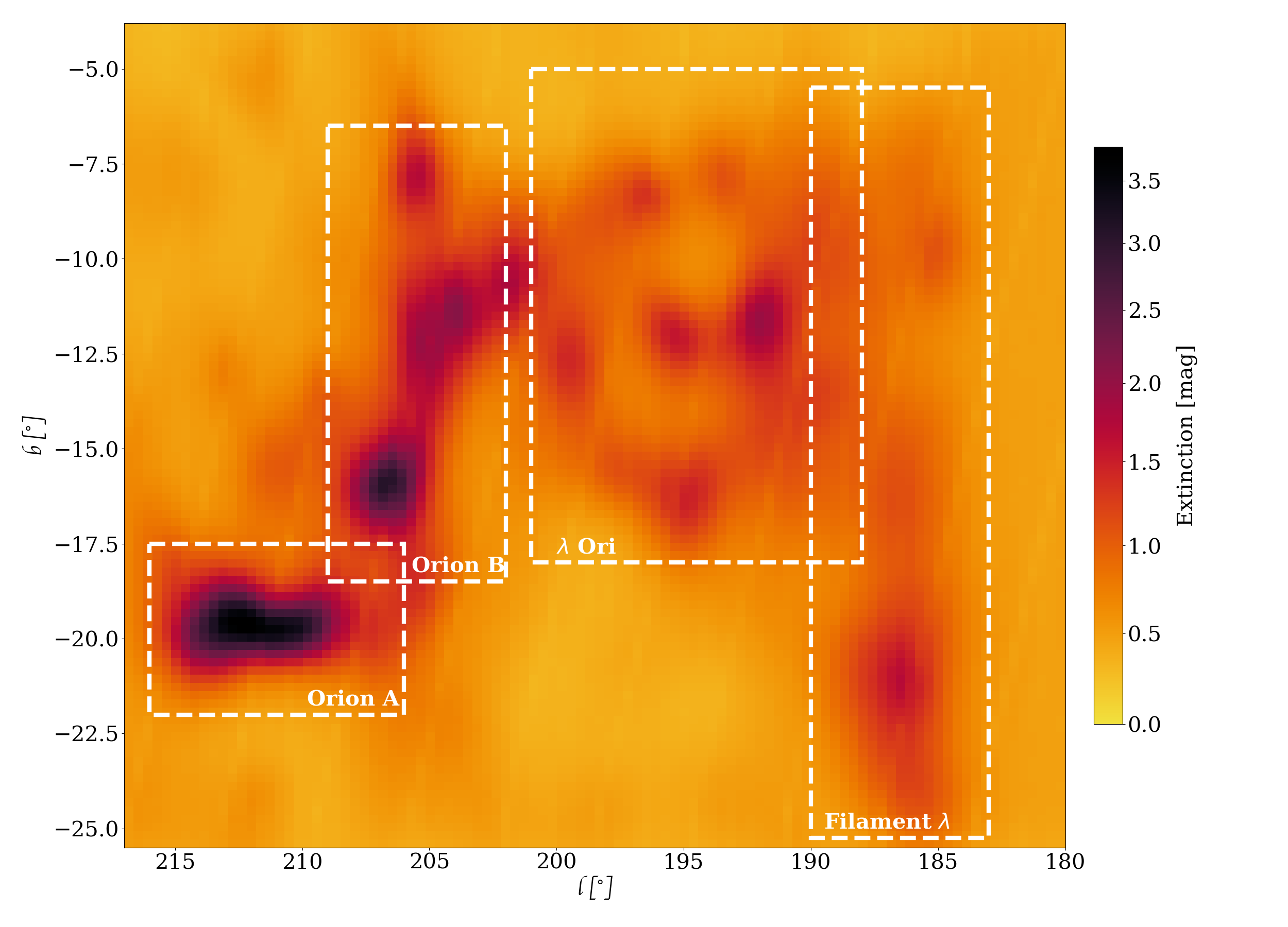}
    \caption{The predicted integrated 2D extinction of the Orion molecular cloud complex as seen from the Sun (i.e: on the plane of the sky). Orion A, B, $\lambda$ Ori bubble and the $\lambda$ filament are highlighted.}
    \label{fig:Orion_2D} 
\end{figure}

\begin{figure}
    \centering
    \includegraphics[width=0.55\textwidth]{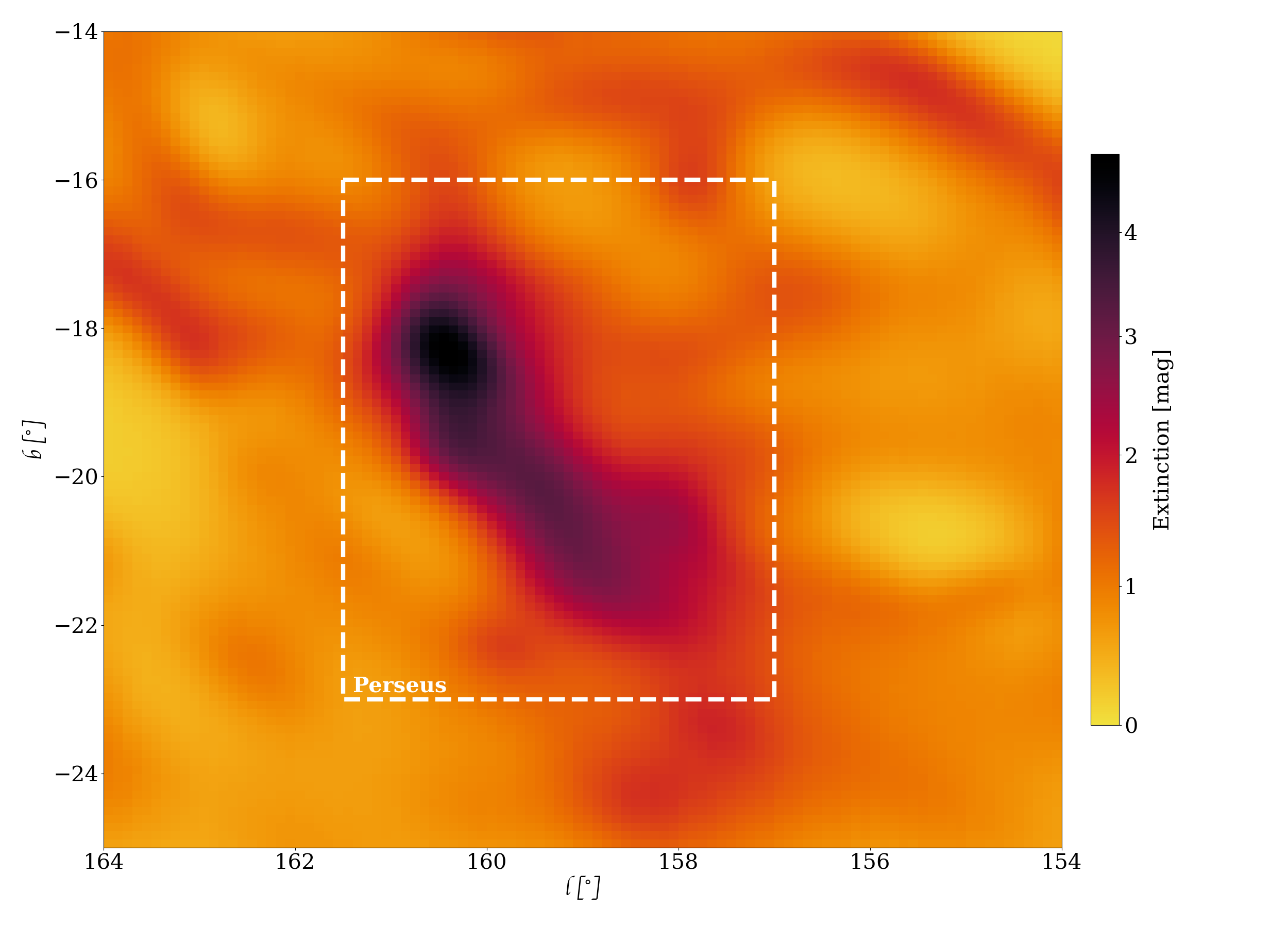}
    \caption{The predicted integrated 2D extinction of the Perseus molecular cloud complex as seen from the Sun (i.e: on the plane of the sky).}
    \label{fig:Per_2D} 
\end{figure}

\begin{figure}
    \centering
    \includegraphics[width=0.55\textwidth]{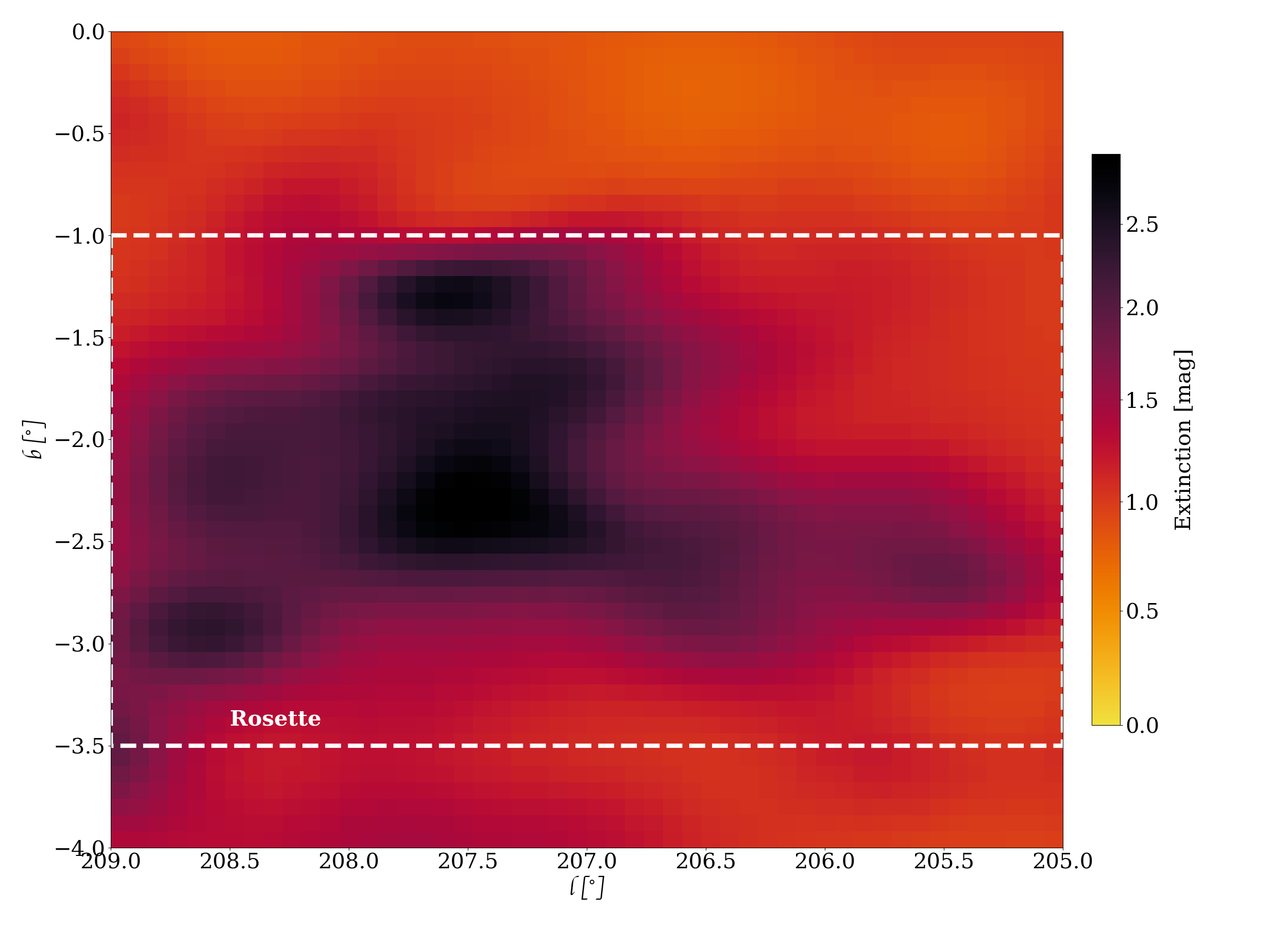}
    \caption{The predicted integrated 2D extinction of the Rosette molecular cloud complex as seen from the Sun (i.e: on the plane of the sky).}
    \label{fig:Rose_2D} 
\end{figure}

\begin{figure}
    \centering
    \includegraphics[width=0.55\textwidth]{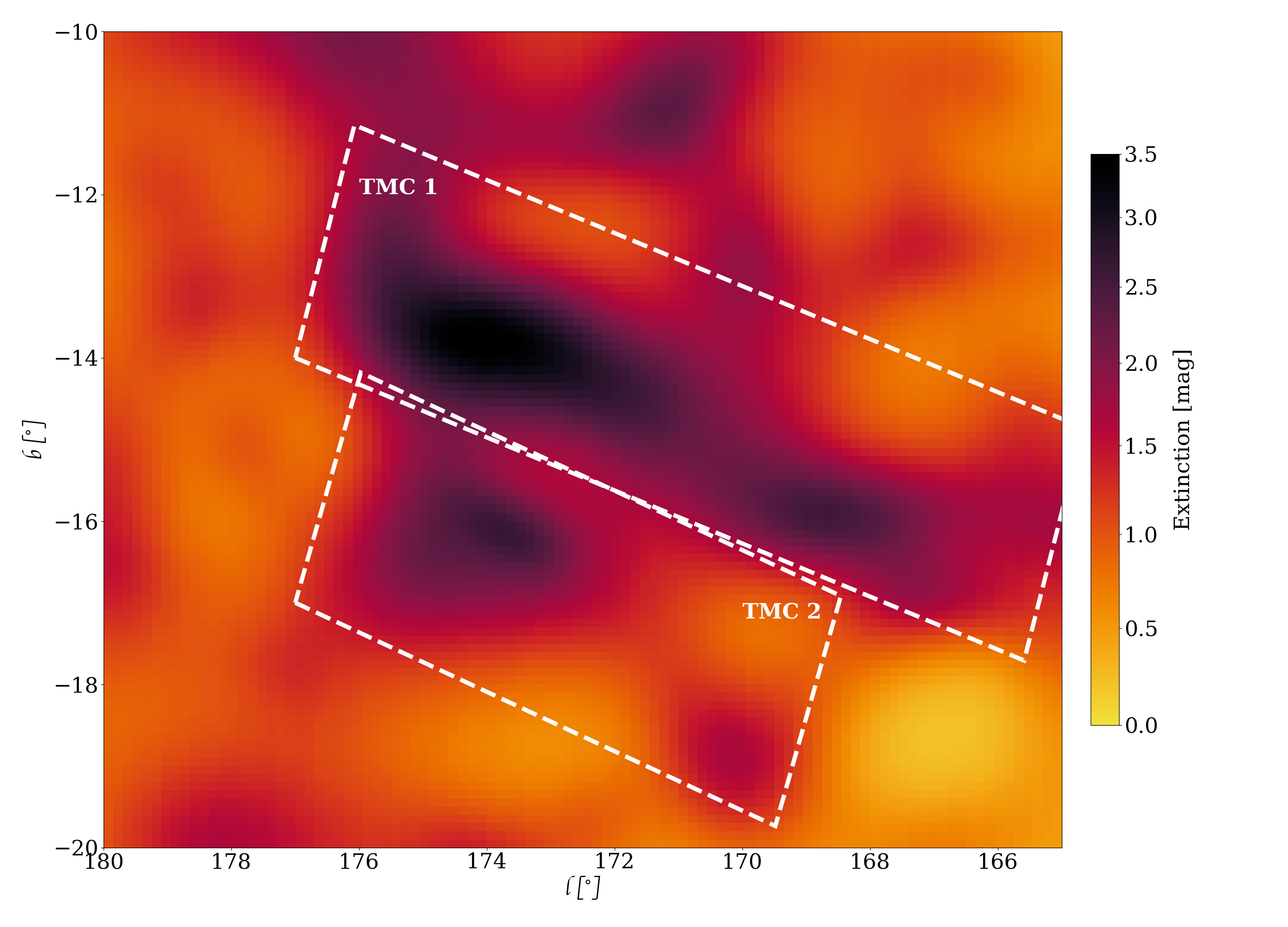}
    \caption{ The predicted integrated 2D extinction of the Taurus molecular cloud complex as seen from the Sun (i.e: on the plane of the sky). TMC 1 and 2 are highlighted.}
    \label{fig:Tau_2D} 
\end{figure}

\subsection{Predicted 3D density plots of the molecular cloud complexes}

\label{sec:Ap:3Ddense}

\begin{landscape}
\begin{figure*}
\begin{adjustwidth}{-7.5cm}{0cm}
    \centering
    \begin{multicols}{2}
    \includegraphics[width=0.75\textwidth]{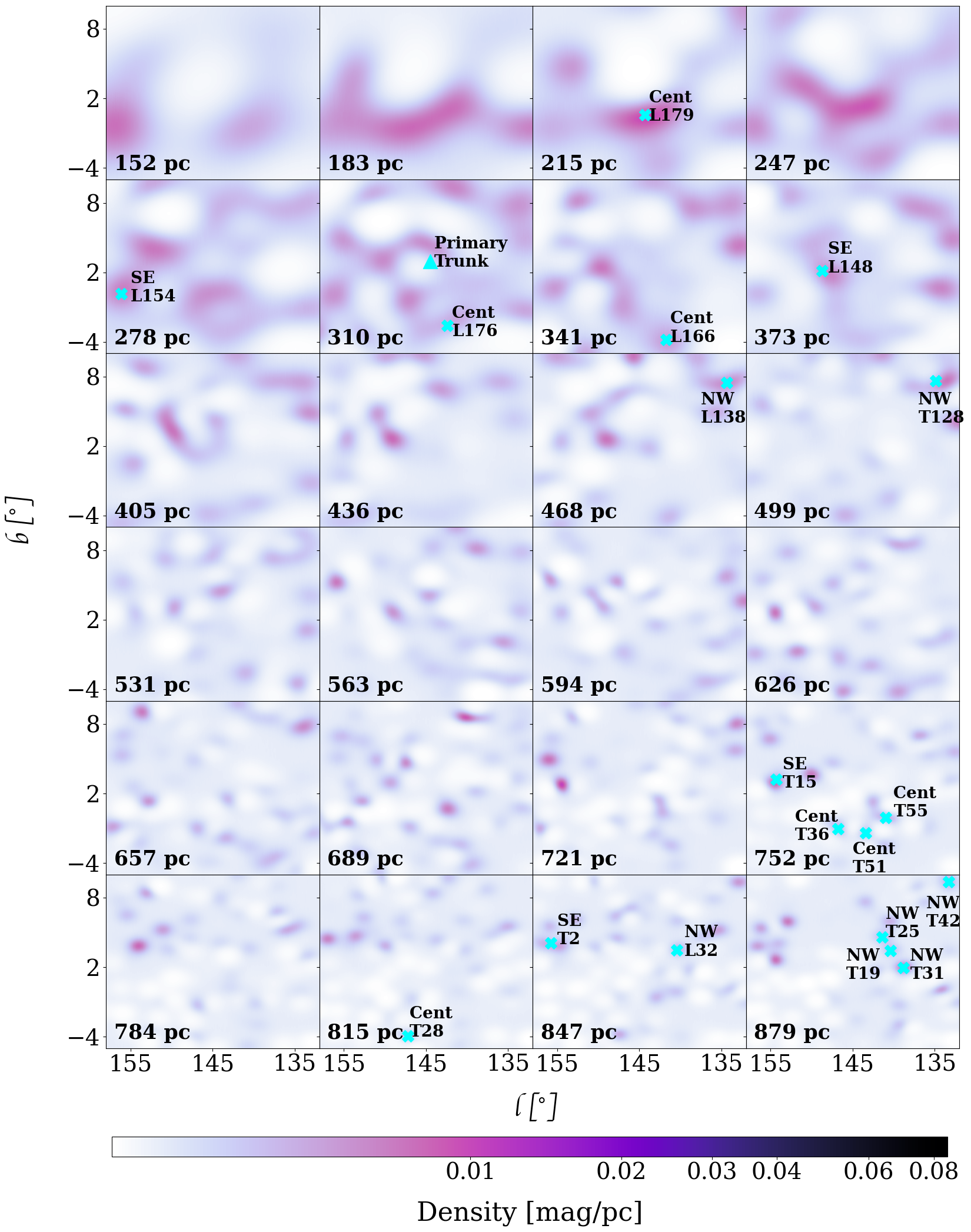}
    
    \hfill
    \includegraphics[width=0.6\textwidth]{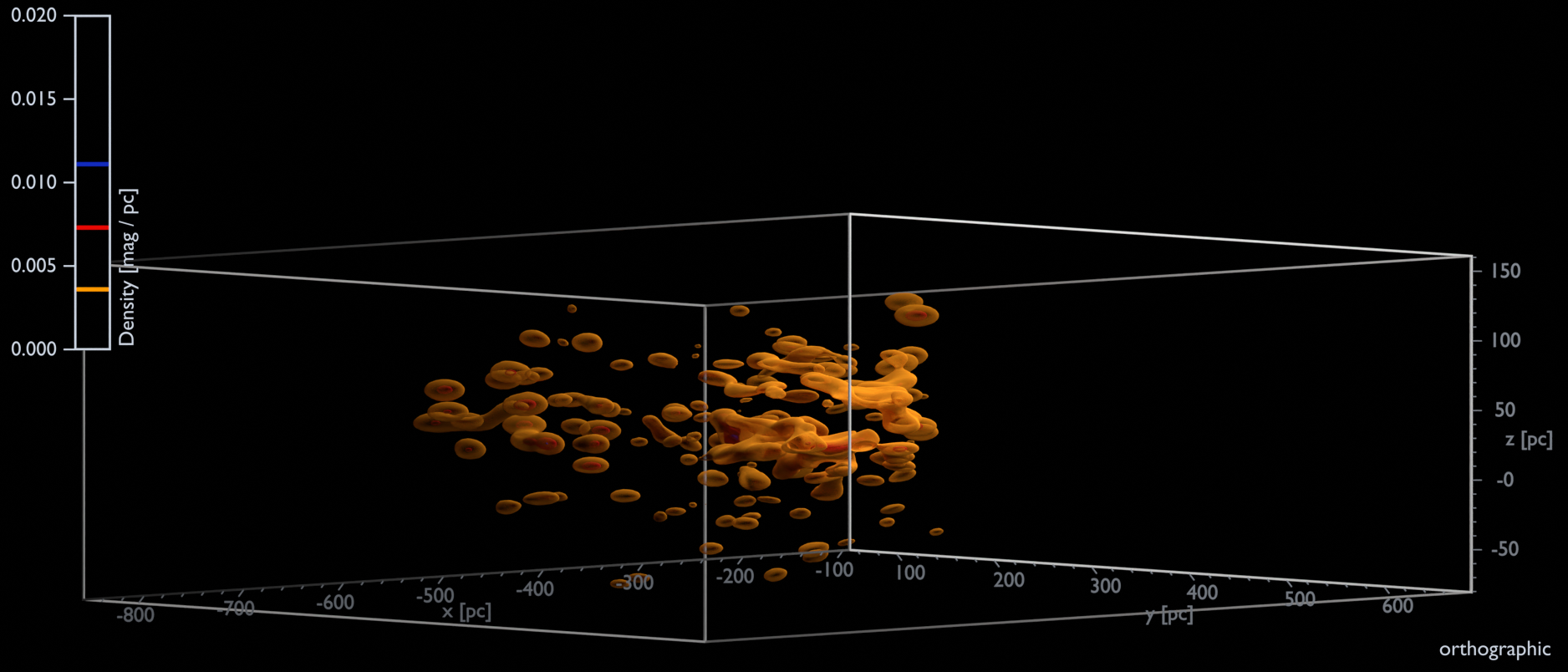}
    
    \hfill
    \includegraphics[width=0.6\textwidth]{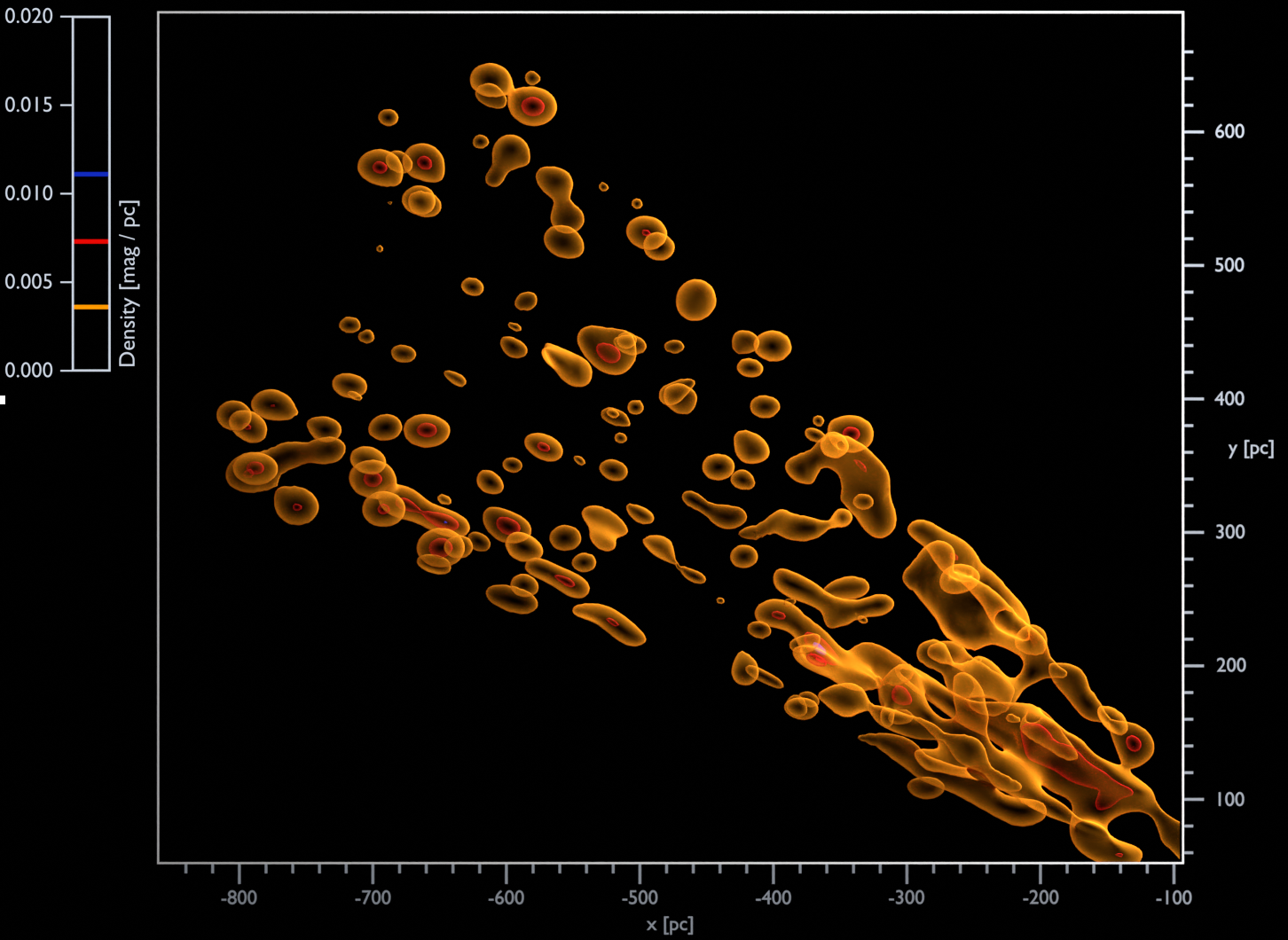}
    
    \end{multicols}
    \vspace{-0.8cm}
    \caption{ Predicted 3D density structure of the Camelopardalis molecular cloud complex. Left: Slices along the line-of-sight of the predicted 3D density structure. With the Cyan triangle we have marked the mass weighted centroid of the primary trunk (as given in Table~\ref{tab:maintrunk_params} placed on the closest distance slice included in the plot. The Cyan $\times$s mark the mass weighted centroids of the interesting features discussed in Sec.\ref{sec:IndGMCs} and highlighted in Table~\ref{tab:leaf_params} placed on the closest distance slice included in the plot.; Right top: Video of a volume rendering of the predicted 3D density structure which begins from the view as seen from the Sun. It then rotates anti-clockwise about an axis perpendicular to the initial viewing angle. The semi-transparent iso-surfaces mark three different density levels with orange being the least dense to blue being the most dense as shown by the colour bar.; Right bottom: Still image showing the top down view of the predicted 3D density structure of the molecular cloud region using identical rendering to the preceding video.}
    
\label{fig:Camel_3Ddens}
\end{adjustwidth}   
\end{figure*}
\end{landscape}

\begin{landscape}
\begin{figure*}
\begin{adjustwidth}{-7.5cm}{0cm}
    \centering
    \begin{multicols}{2}
    \includegraphics[width=0.75\textwidth]{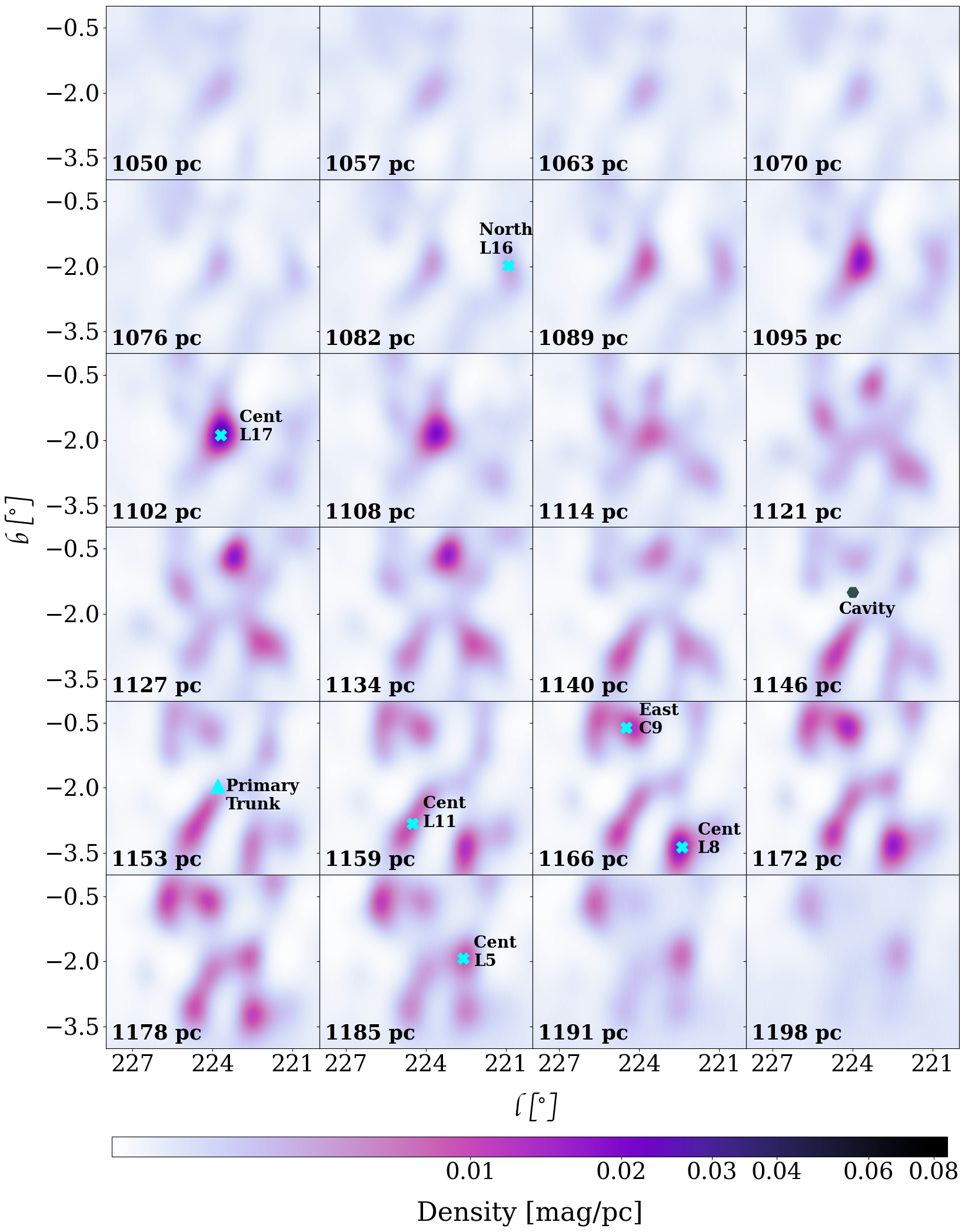}
    
    \hfill
    \includegraphics[width=0.6\textwidth]{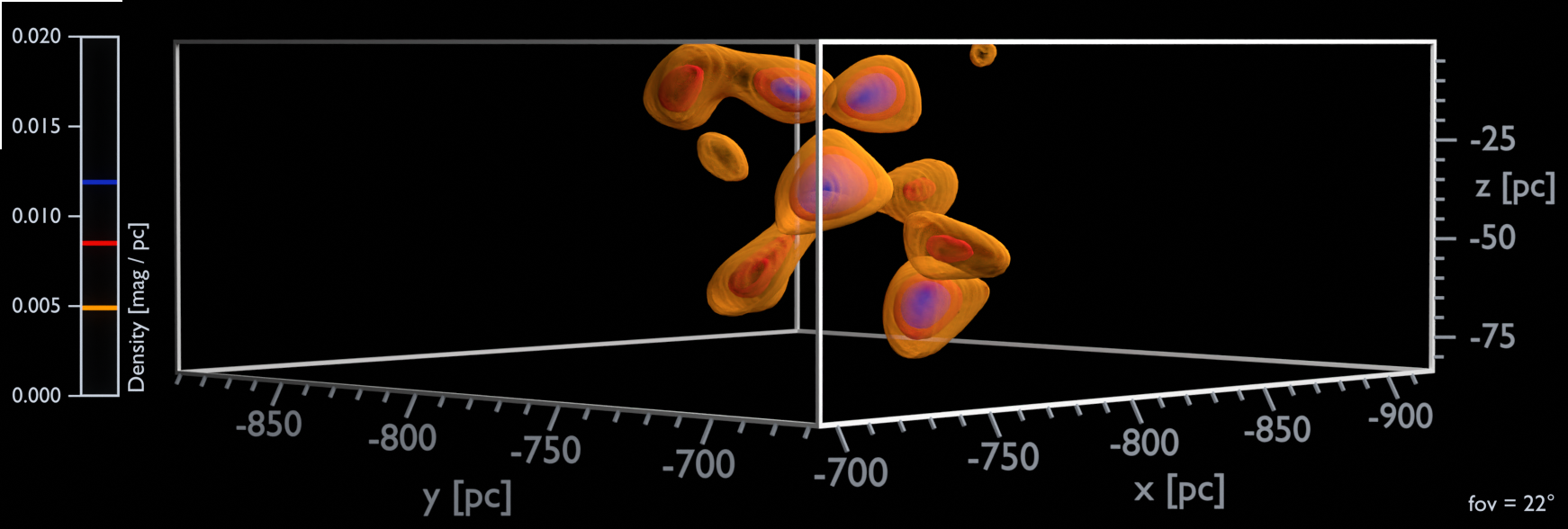}
    
    \hfill
    \includegraphics[width=0.6\textwidth]{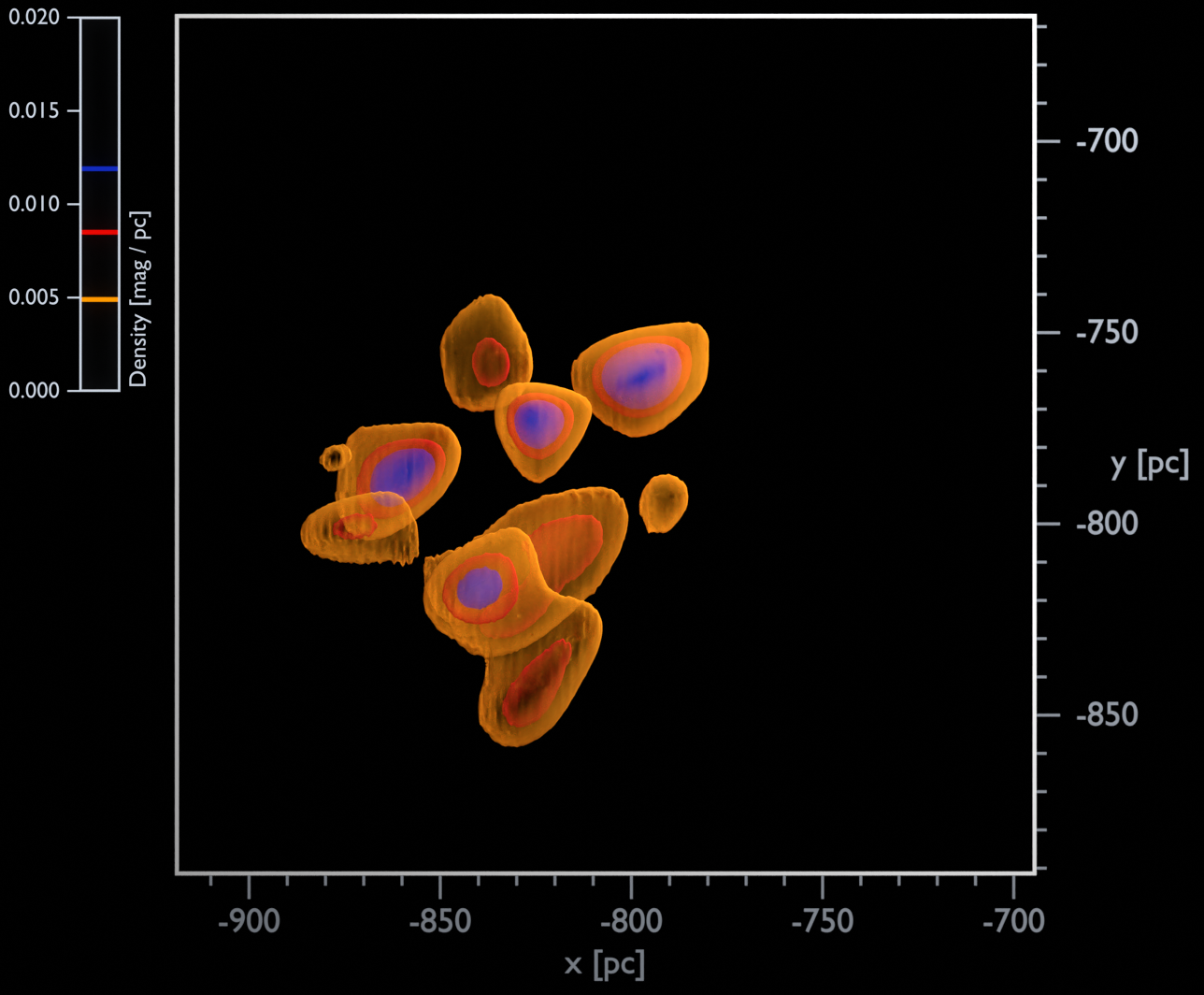}
    
    \end{multicols}
    \vspace{-0.8cm}
    \caption{ Predicted 3D density structure of the Canis Major molecular cloud complex. Left: Slices along the line-of-sight of the predicted 3D density structure. With the Cyan triangle we have marked the mass weighted centroid of the primary trunk (as given in Table~\ref{tab:maintrunk_params} placed on the closest distance slice included in the plot. The Cyan $\times$s mark the mass weighted centroids of the interesting features discussed in Sec.\ref{sec:IndGMCs} and highlighted in Table~\ref{tab:leaf_params} placed on the closest distance slice included in the plot. The grey hexagon marks the centre of the Canis Major cavity; Right top: Video of a volume rendering of the predicted 3D density structure which begins from the view as seen from the Sun. It then rotates anti-clockwise about an axis perpendicular to the initial viewing angle. The semi-transparent iso-surfaces mark three different density levels with orange being the least dense to blue being the most dense as shown by the colour bar.; Right bottom: Still image showing the top down view of the predicted 3D density structure of the molecular cloud region using identical rendering to the preceding video.}
    
\label{fig:CanMaj_3Ddens}
\end{adjustwidth}   
\end{figure*}
\end{landscape}

\begin{landscape}
\begin{figure*}
\begin{adjustwidth}{-7.5cm}{0cm}
    \centering
    \begin{multicols}{2}
    \includegraphics[width=0.75\textwidth]{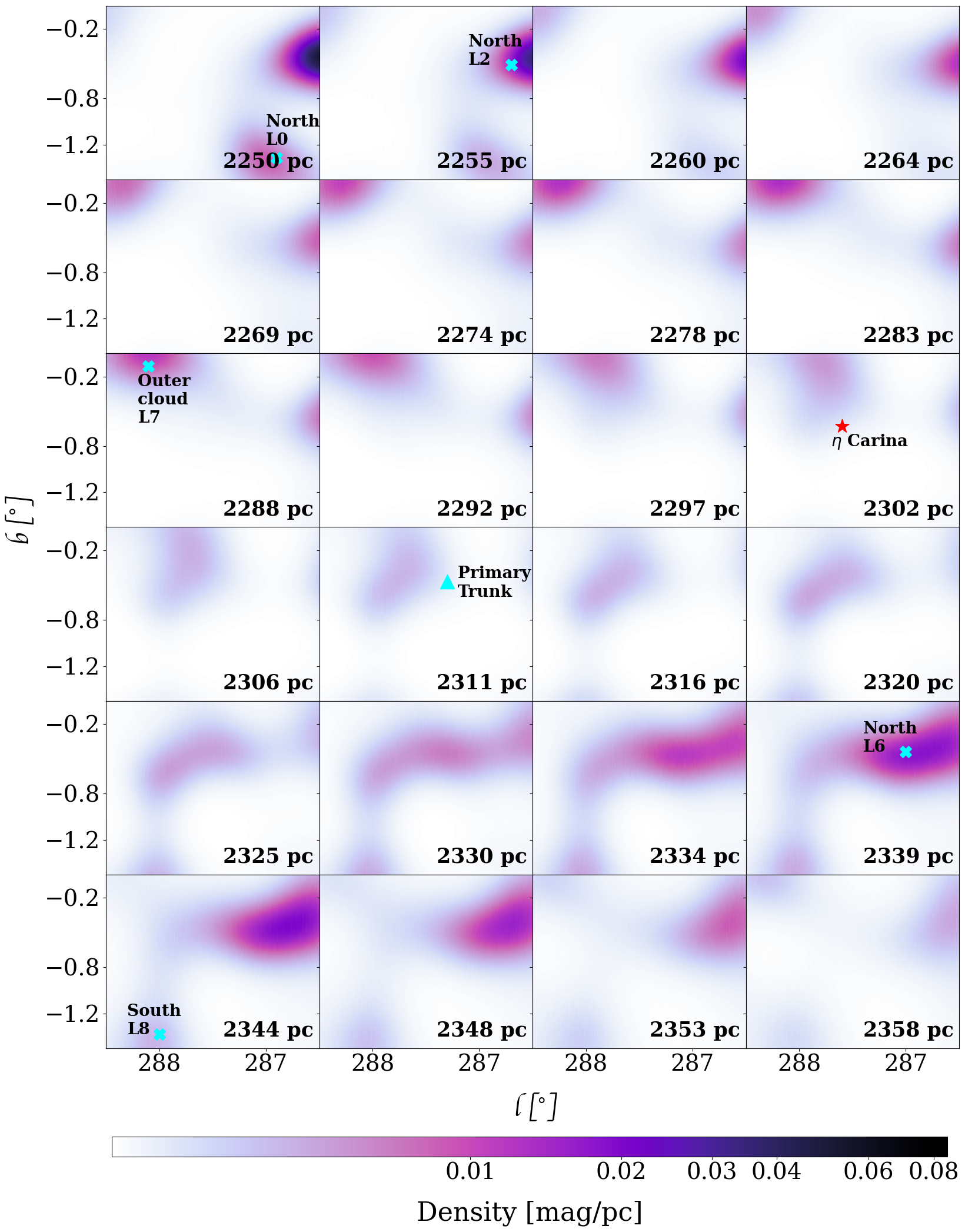}
    
    \hfill
    \includegraphics[width=0.6\textwidth]{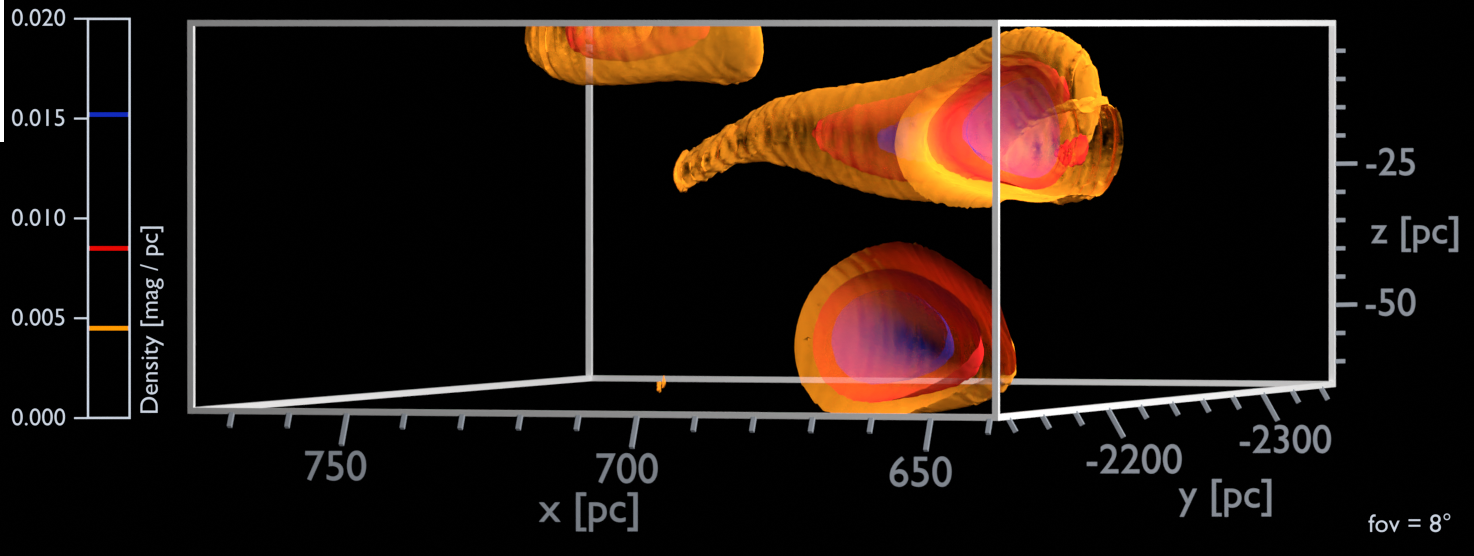}
    
    \hfill
    \includegraphics[width=0.6\textwidth]{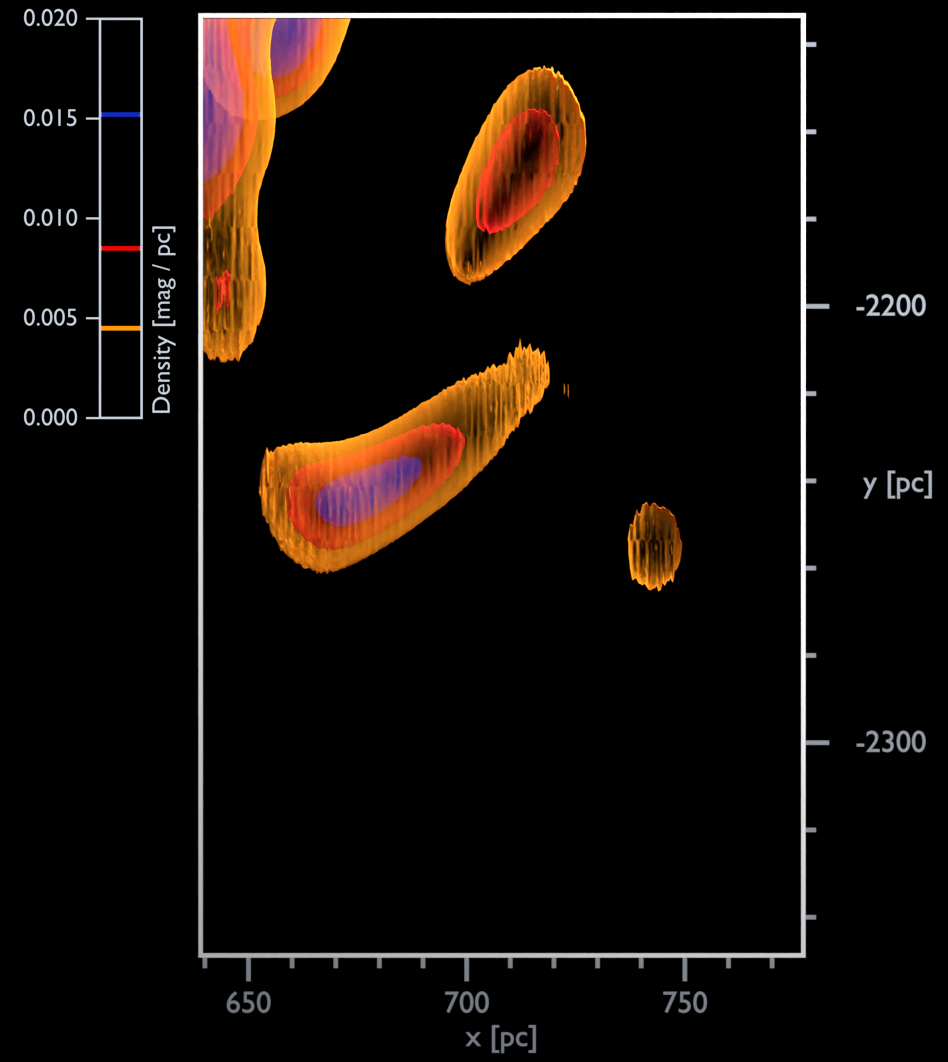}
    
    \end{multicols}
    \vspace{-0.8cm}
    \caption{ Predicted 3D density structure of the Carina molecular cloud complex. Left: Slices along the line-of-sight of the predicted 3D density structure. With the Cyan triangle we have marked the mass weighted centroid of the primary trunk (as given in Table~\ref{tab:maintrunk_params} placed on the closest distance slice included in the plot. The Cyan $\times$s mark the mass weighted centroids of the interesting features discussed in Sec.\ref{sec:IndGMCs} and highlighted in Table~\ref{tab:leaf_params} placed on the closest distance slice included in the plot. The red star marks the position of $\eta$ Carina as given in \citet{Smith2008_Carina_SFHB}; Right top: Video of a volume rendering of the predicted 3D density structure which begins from the view as seen from the Sun. It then rotates anti-clockwise about an axis perpendicular to the initial viewing angle. The semi-transparent iso-surfaces mark three different density levels with orange being the least dense to blue being the most dense as shown by the colour bar.; Right bottom: Still image showing the top down view of the predicted 3D density structure of the molecular cloud region using identical rendering to the preceding video.}
    
\label{fig:Carina_3Ddens}
\end{adjustwidth}   
\end{figure*}
\end{landscape}

\begin{landscape}
\begin{figure*}
\begin{adjustwidth}{-7.5cm}{0cm}
    \centering
    \begin{multicols}{2}
    \includegraphics[width=0.75\textwidth]{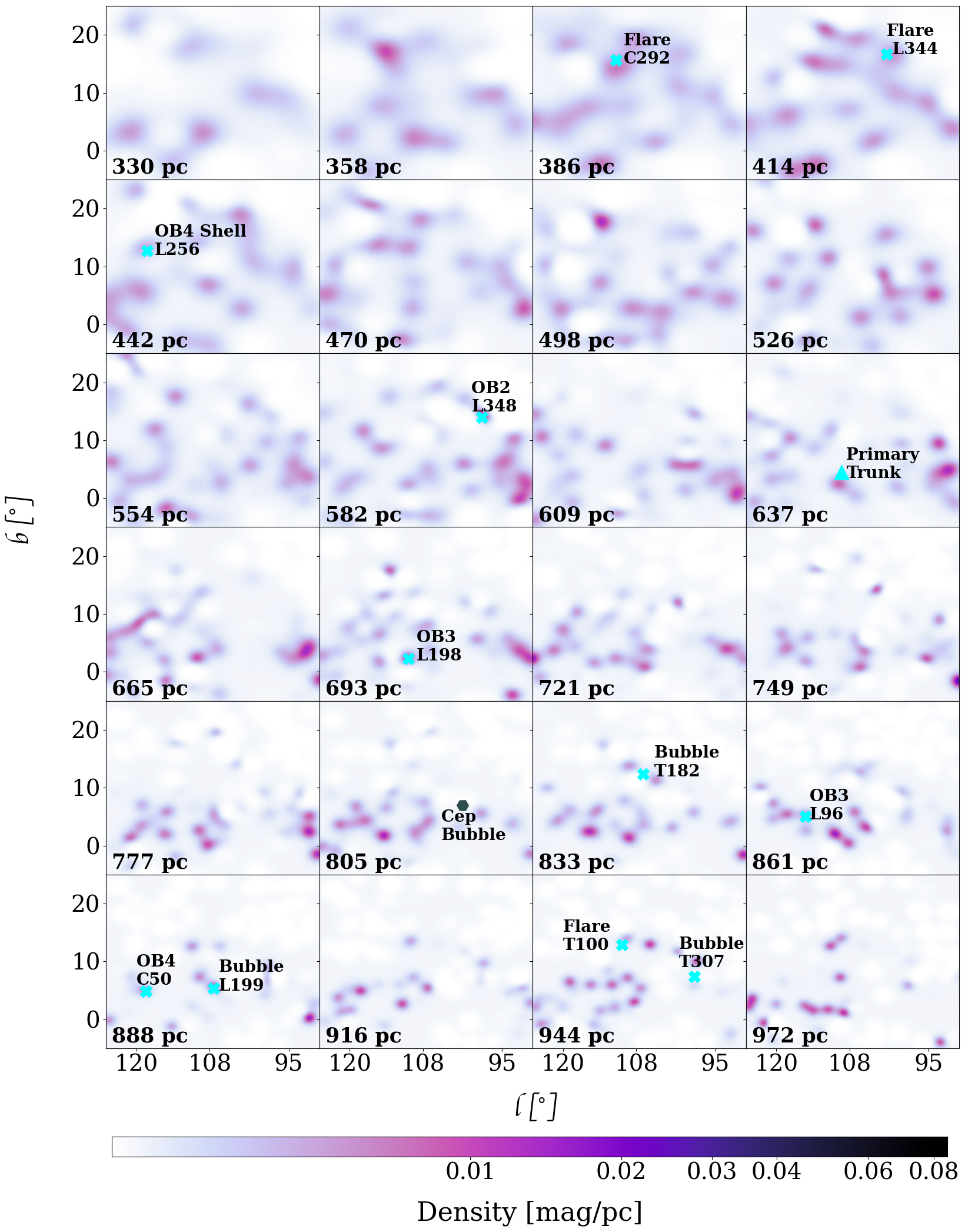}
    
    \hfill
    \includegraphics[width=0.6\textwidth]{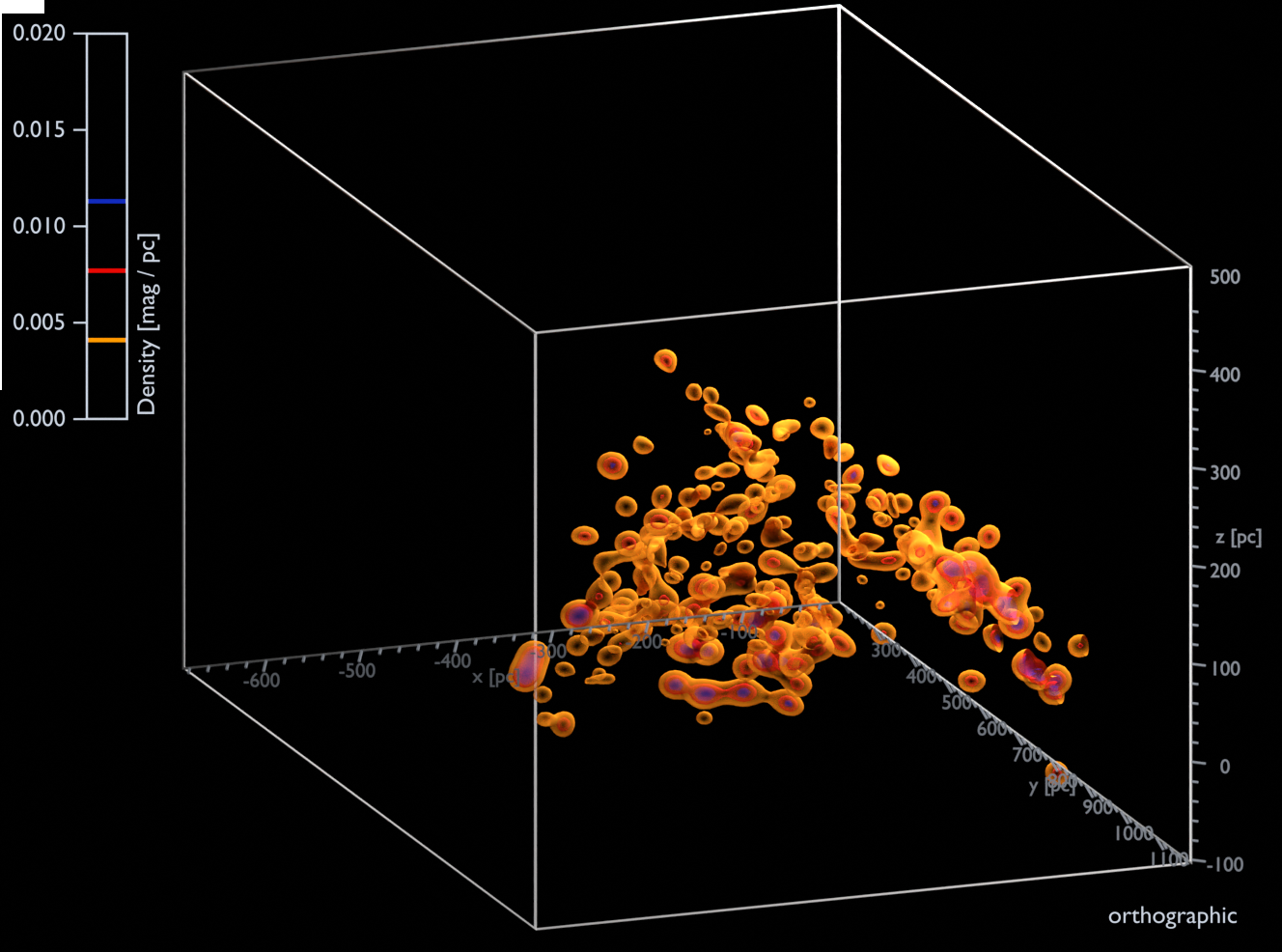}
    
    \hfill
    \includegraphics[width=0.6\textwidth]{Figures/ThreeD_renders/mn_22_2305_mj_cepheus_sun.png}
    
    \end{multicols}
    \vspace{-0.8cm}
    \caption{ Predicted 3D density structure of the Cepheus molecular cloud complex. Left: Slices along the line-of-sight of the predicted 3D density structure. With the Cyan triangle we have marked the mass weighted centroid of the primary trunk (as given in Table~\ref{tab:maintrunk_params} placed on the closest distance slice included in the plot. The Cyan $\times$s mark the mass weighted centroids of the interesting features discussed in Sec.\ref{sec:IndGMCs} and highlighted in Table~\ref{tab:leaf_params} placed on the closest distance slice included in the plot. The Cepheus bubble centroid is marked by the grey hexagon; Right top: Video of a volume rendering of the predicted 3D density structure which begins from the view as seen from the Sun. It then rotates anti-clockwise about an axis perpendicular to the initial viewing angle. The semi-transparent iso-surfaces mark three different density levels with orange being the least dense to blue being the most dense as shown by the colour bar.; Right bottom: Still image showing the top down view of the predicted 3D density structure of the molecular cloud region using identical rendering to the preceding video.}
    
\label{fig:Cep_3Ddens}
\end{adjustwidth}   
\end{figure*}
\end{landscape}

\begin{landscape}
\begin{figure*}
\begin{adjustwidth}{-7.5cm}{0cm}
    \centering
    \begin{multicols}{2}
    \includegraphics[width=0.75\textwidth]{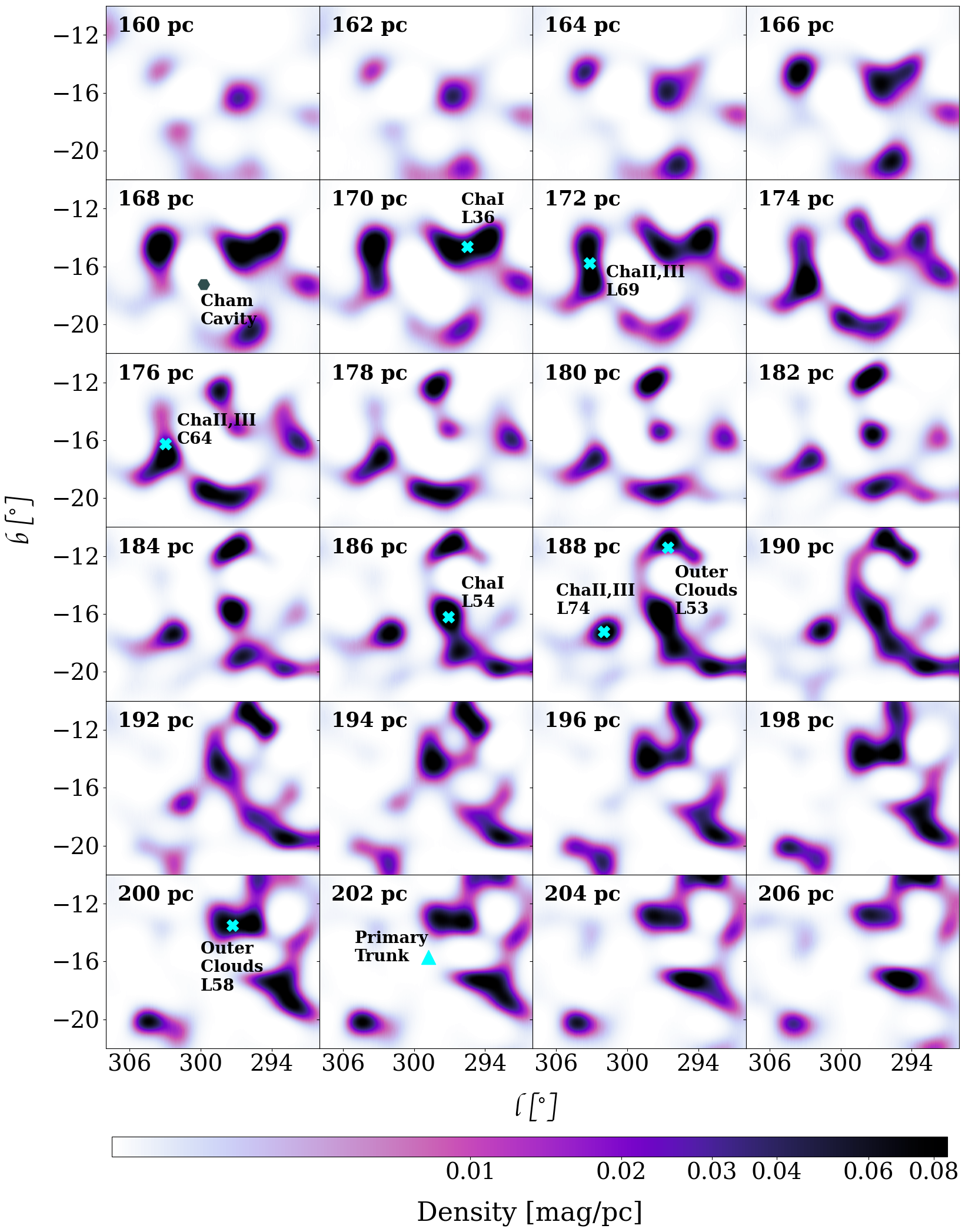}
    
    \hfill
    \includegraphics[width=0.6\textwidth]{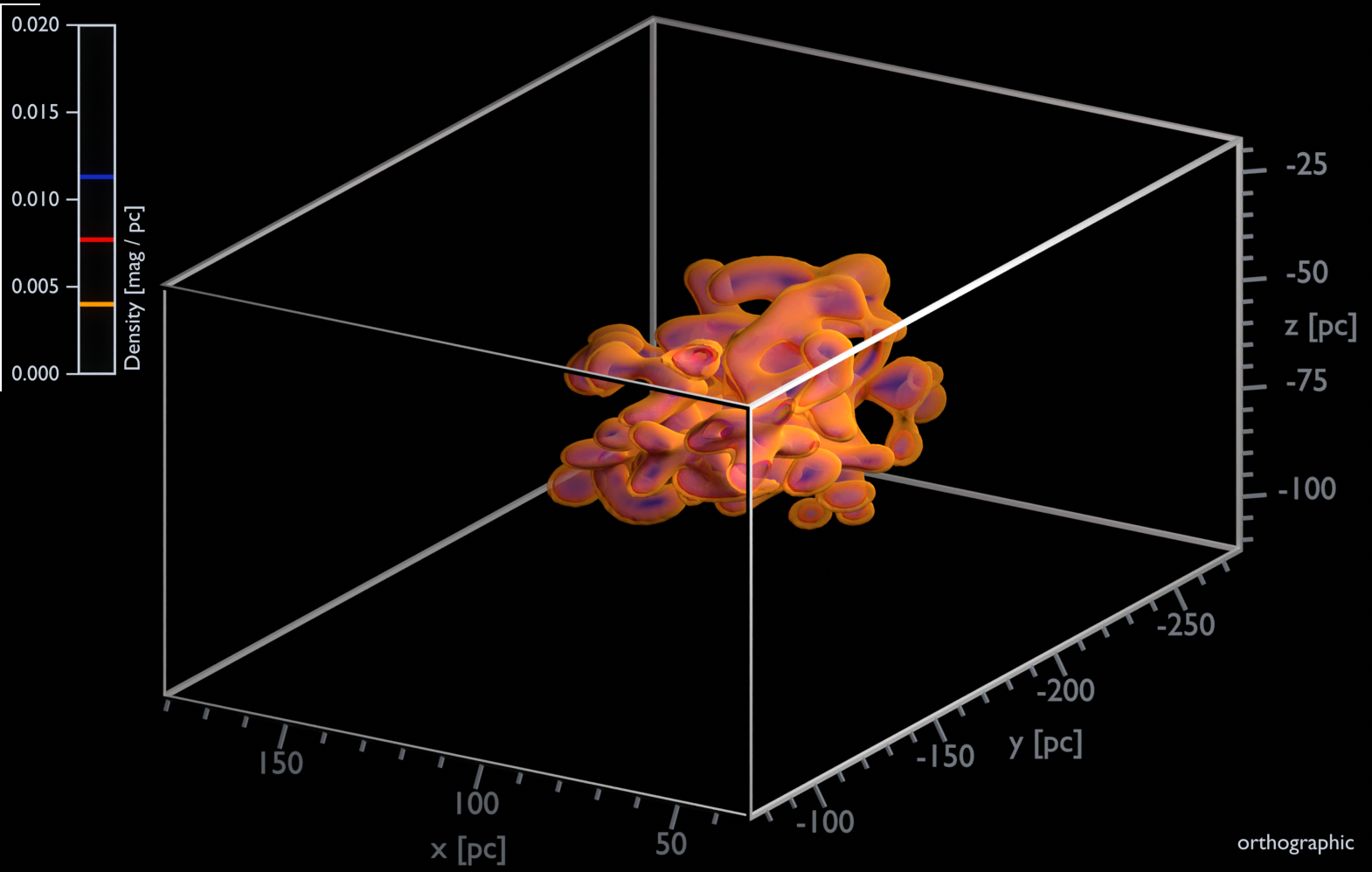}
    
    \hfill
    \includegraphics[width=0.6\textwidth]{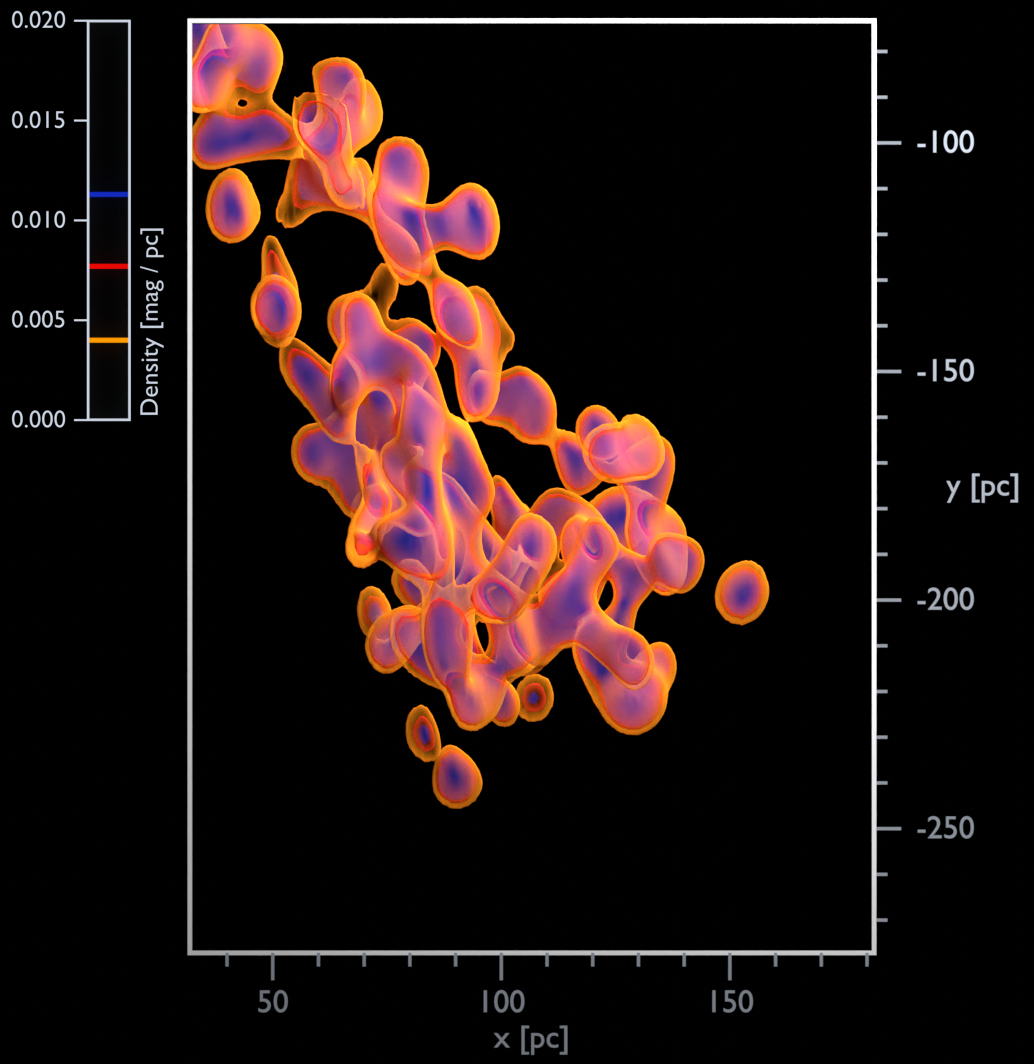}
    
    \end{multicols}
    \vspace{-0.8cm}
    \caption{ Predicted 3D density structure of the Chamaeleon molecular cloud complex. Left: Slices along the line-of-sight of the predicted 3D density structure. With the Cyan triangle we have marked the mass weighted centroid of the primary trunk (as given in Table~\ref{tab:maintrunk_params} placed on the closest distance slice included in the plot. The Cyan $\times$s mark the mass weighted centroids of the interesting features discussed in Sec.\ref{sec:IndGMCs} and highlighted in Table~\ref{tab:leaf_params} placed on the closest distance slice included in the plot. The centroid of the Chamaeleon cavity is marked by the grey hexagon; Right top: Video of a volume rendering of the predicted 3D density structure which begins from the view as seen from the Sun. It then rotates anti-clockwise about an axis perpendicular to the initial viewing angle. The semi-transparent iso-surfaces mark three different density levels with orange being the least dense to blue being the most dense as shown by the colour bar.; Right bottom: Still image showing the top down view of the predicted 3D density structure of the molecular cloud region using identical rendering to the preceding video.}
    
\label{fig:Cham_3Ddens}
\end{adjustwidth}   
\end{figure*}
\end{landscape}

\begin{landscape}
\begin{figure*}
\begin{adjustwidth}{-7.5cm}{0cm}
    \centering
    \begin{multicols}{2}
    \includegraphics[width=0.75\textwidth]{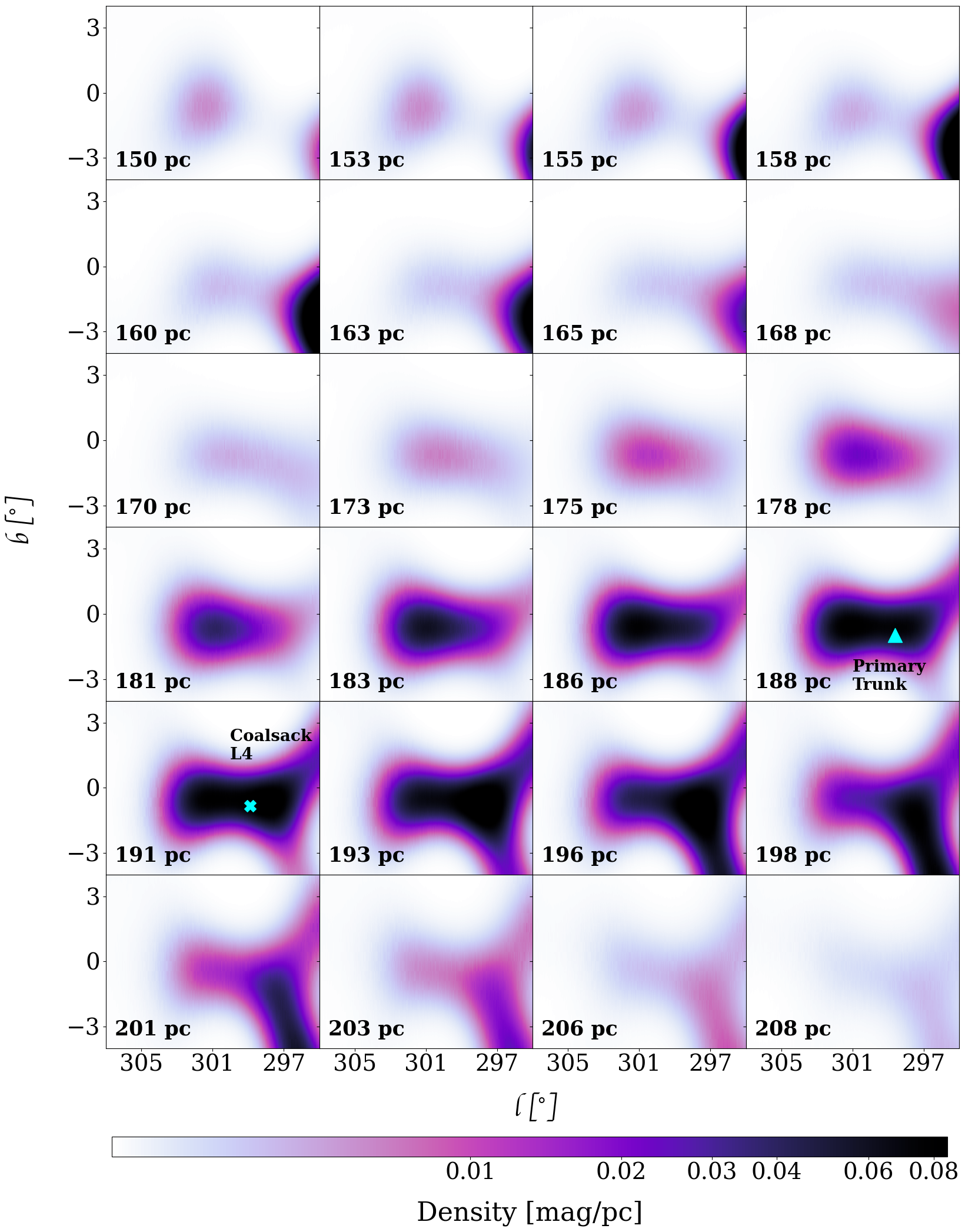}
    
    \hfill
    \includegraphics[width=0.6\textwidth]{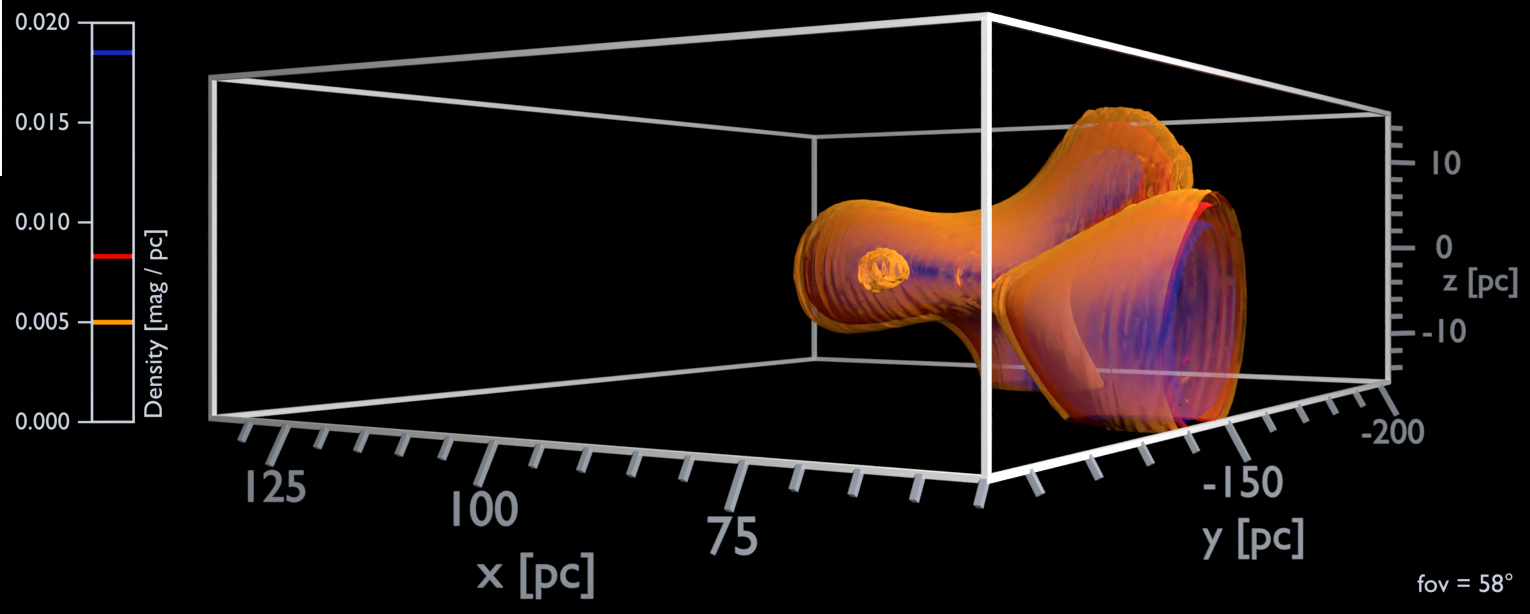}
    
    \hfill
    \includegraphics[width=0.6\textwidth]{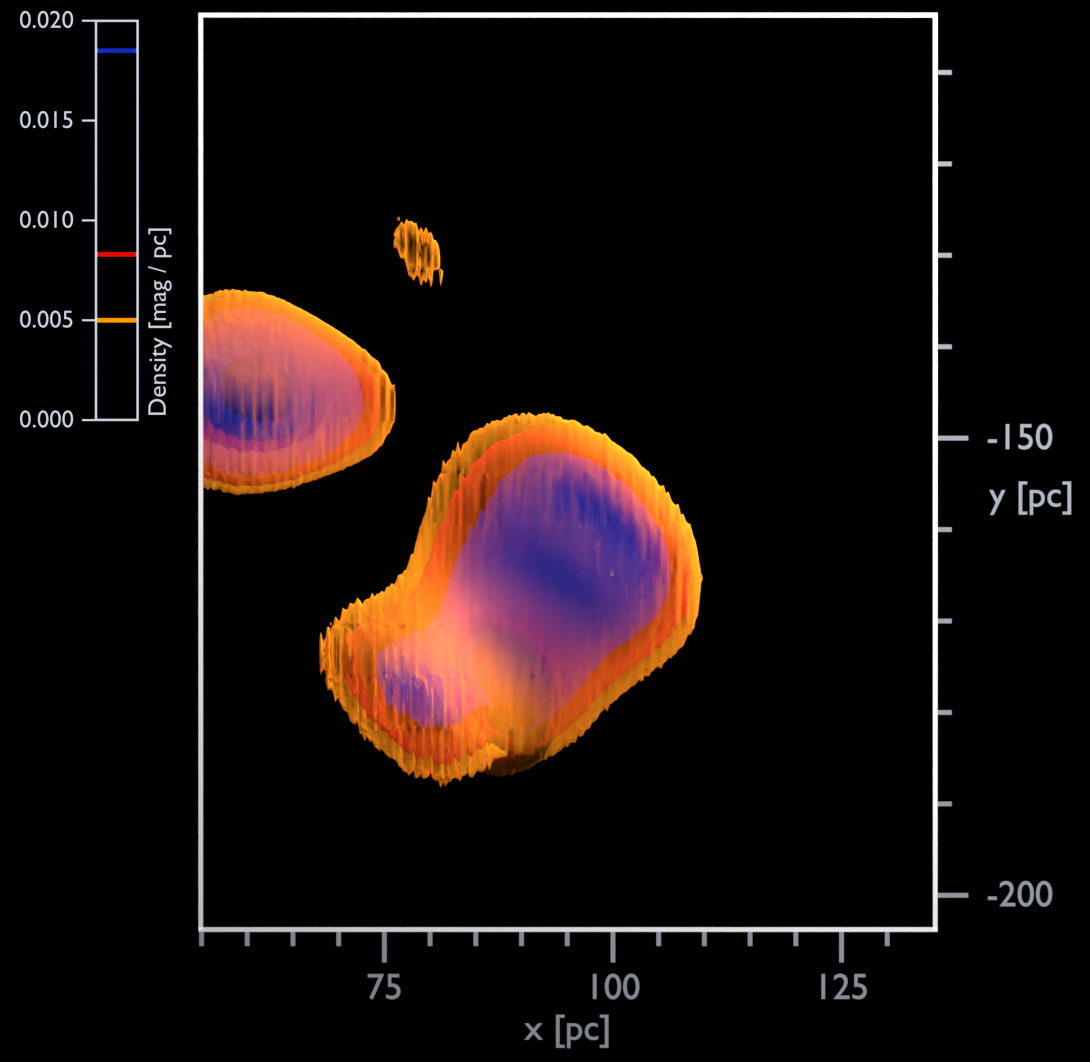}
    
    \end{multicols}
    \vspace{-0.8cm}
    \caption{ Predicted 3D density structure of the Southern Coalsack molecular cloud. Left: Slices along the line-of-sight of the predicted 3D density structure. With the Cyan triangle we have marked the mass weighted centroid of the primary trunk (as given in Table~\ref{tab:maintrunk_params} placed on the closest distance slice included in the plot. The Cyan $\times$s mark the mass weighted centroids of the interesting features discussed in Sec.\ref{sec:IndGMCs} and highlighted in Table~\ref{tab:leaf_params} placed on the closest distance slice included in the plot.; Right top: Video of a volume rendering of the predicted 3D density structure which begins from the view as seen from the Sun. It then rotates anti-clockwise about an axis perpendicular to the initial viewing angle. The semi-transparent iso-surfaces mark three different density levels with orange being the least dense to blue being the most dense as shown by the colour bar.; Right bottom: Still image showing the top down view of the predicted 3D density structure of the molecular cloud region using identical rendering to the preceding video.}
    
\label{fig:Coalsack_3Ddens}
\end{adjustwidth}   
\end{figure*}
\end{landscape}

\begin{landscape}
\begin{figure*}
\begin{adjustwidth}{-7.5cm}{0cm}
    \centering
    \begin{multicols}{2}
    \includegraphics[width=0.75\textwidth]{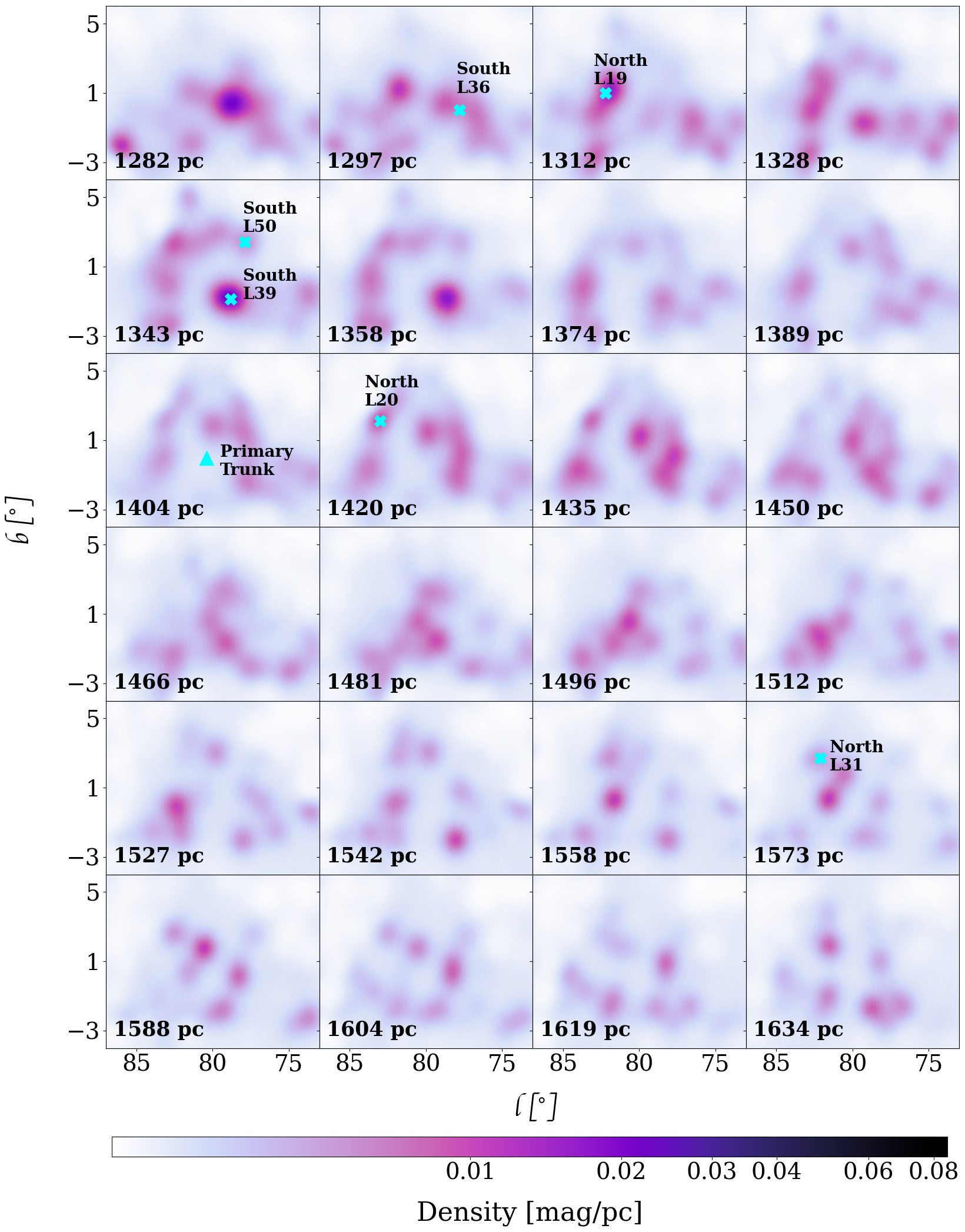}
    
    \hfill
    \includegraphics[width=0.6\textwidth]{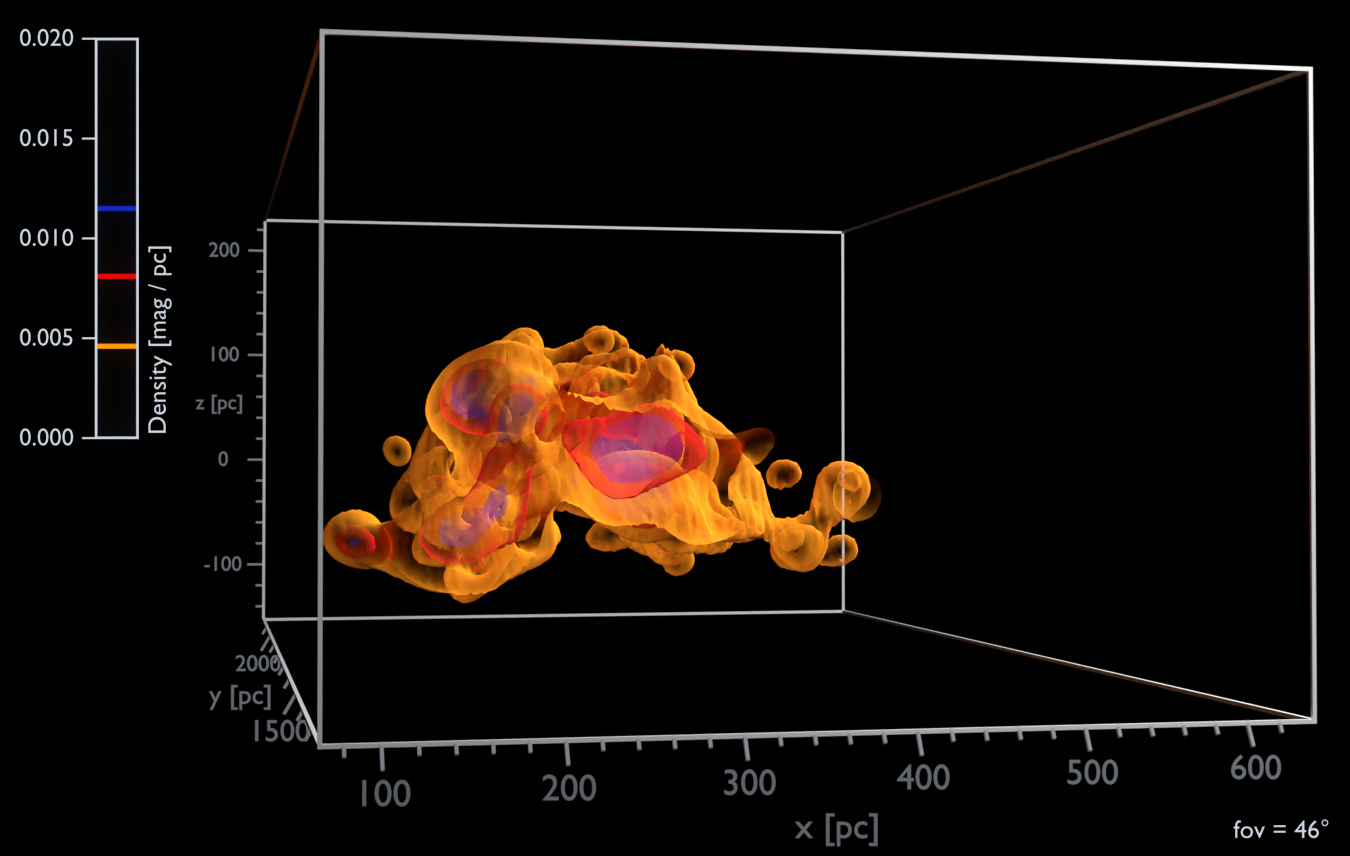}
    
    \hfill
    \includegraphics[width=0.6\textwidth]{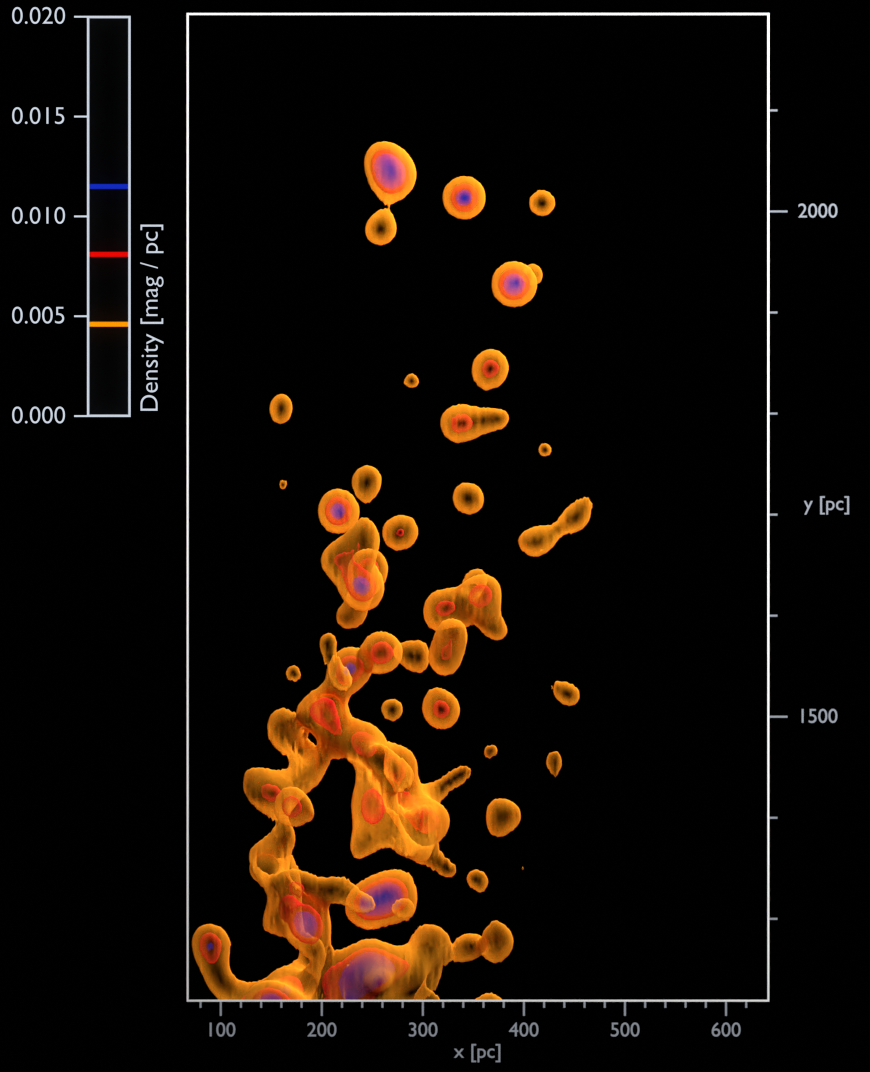}
    
    \end{multicols}
    \vspace{-0.8cm}
    \caption{ Predicted 3D density structure of the Cygnus X molecular cloud complex. Left: Slices along the line-of-sight of the predicted 3D density structure. With the Cyan triangle we have marked the mass weighted centroid of the primary trunk (as given in Table~\ref{tab:maintrunk_params} placed on the closest distance slice included in the plot. The Cyan $\times$s mark the mass weighted centroids of the interesting features discussed in Sec.\ref{sec:IndGMCs} and highlighted in Table~\ref{tab:leaf_params} placed on the closest distance slice included in the plot.; Right top: Video of a volume rendering of the predicted 3D density structure which begins from the view as seen from the Sun. It then rotates anti-clockwise about an axis perpendicular to the initial viewing angle. The semi-transparent iso-surfaces mark three different density levels with orange being the least dense to blue being the most dense as shown by the colour bar.; Right bottom: Still image showing the top down view of the predicted 3D density structure of the molecular cloud region using identical rendering to the preceding video.}
    
\label{fig:Cyg_3Ddens}
\end{adjustwidth}   
\end{figure*}
\end{landscape}

\begin{landscape}
\begin{figure*}
\begin{adjustwidth}{-7.5cm}{0cm}
    \centering
    \begin{multicols}{2}
    \includegraphics[width=0.75\textwidth]{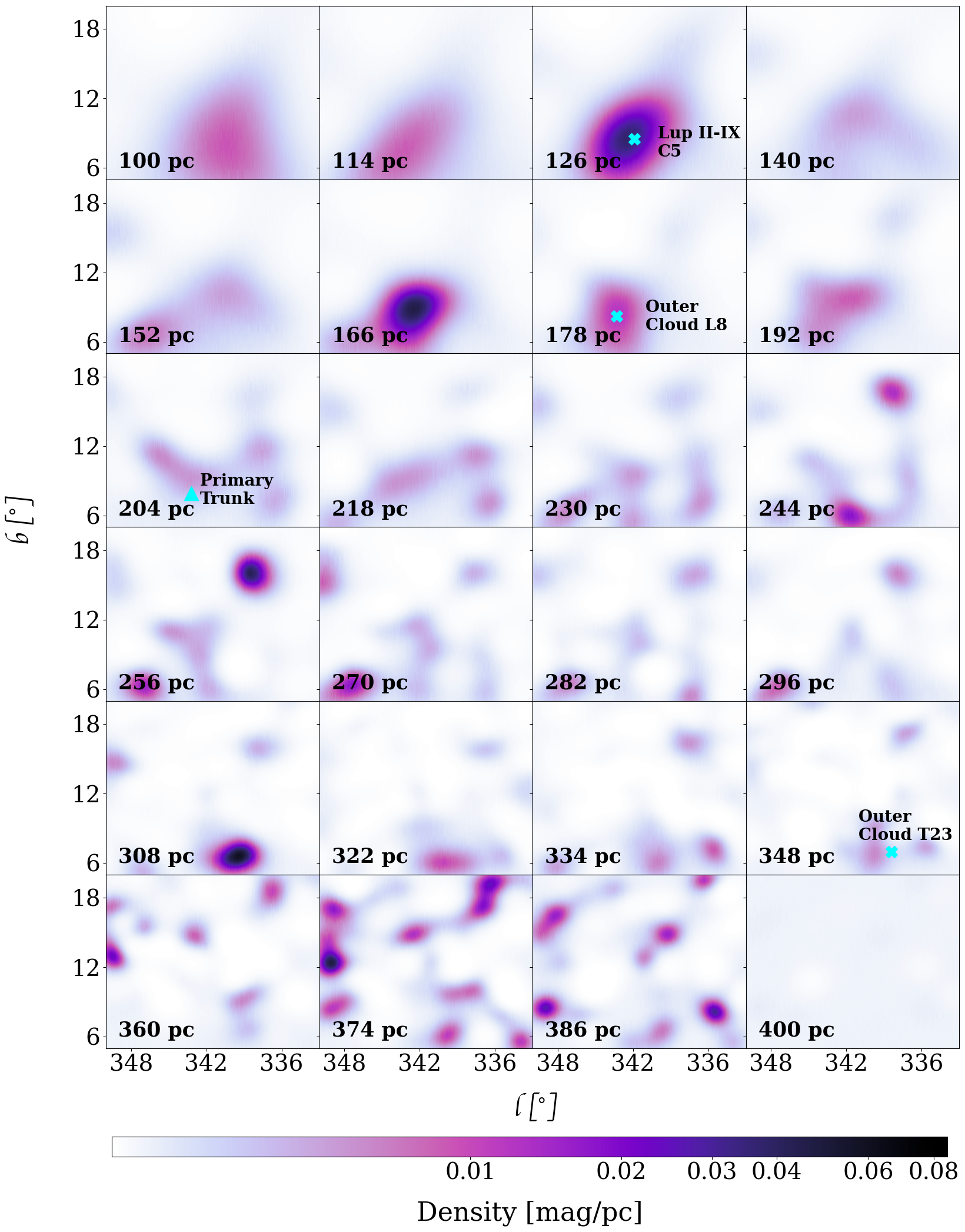}
    
    \hfill
    \includegraphics[width=0.6\textwidth]{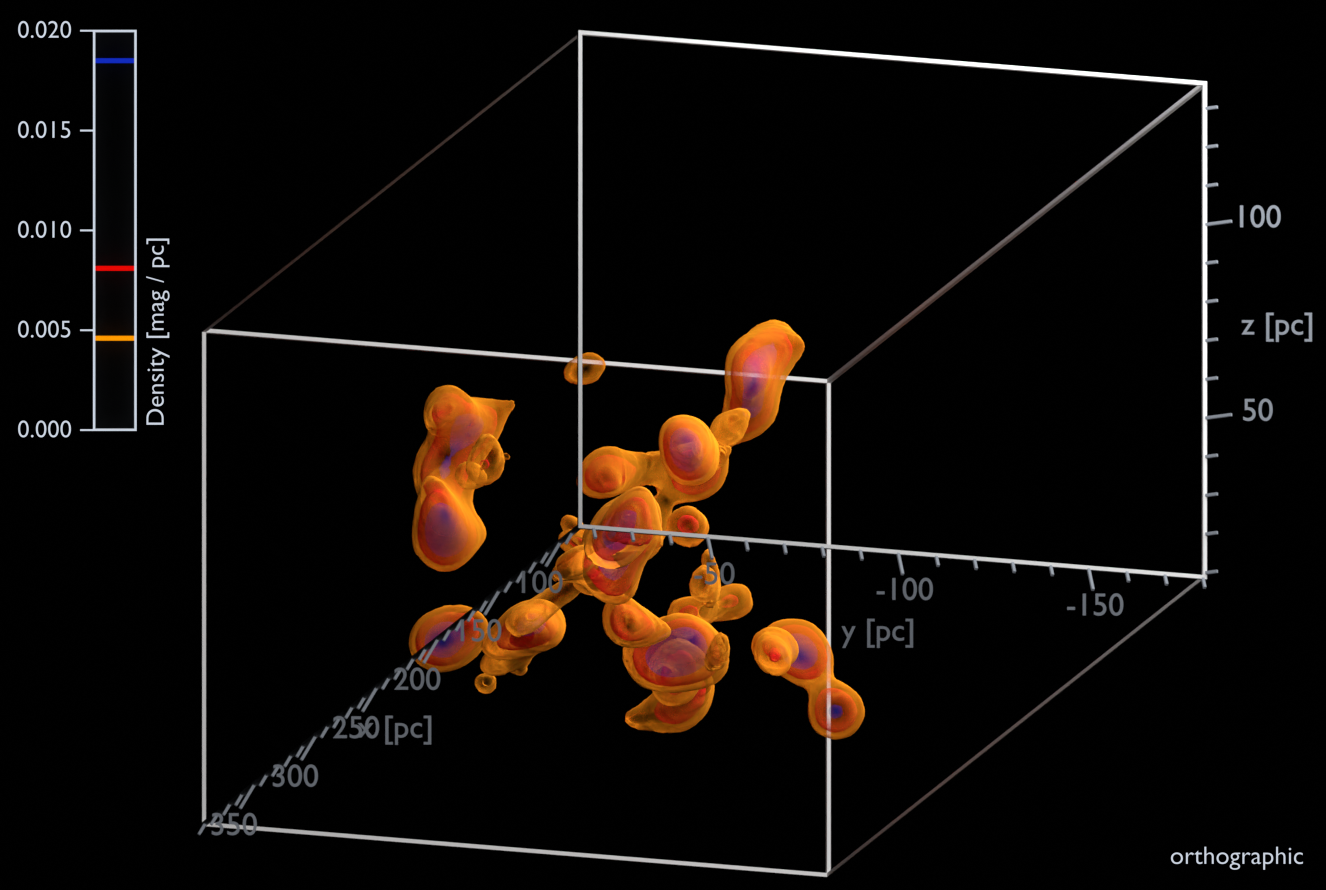}
    
    \hfill
    \includegraphics[width=0.6\textwidth]{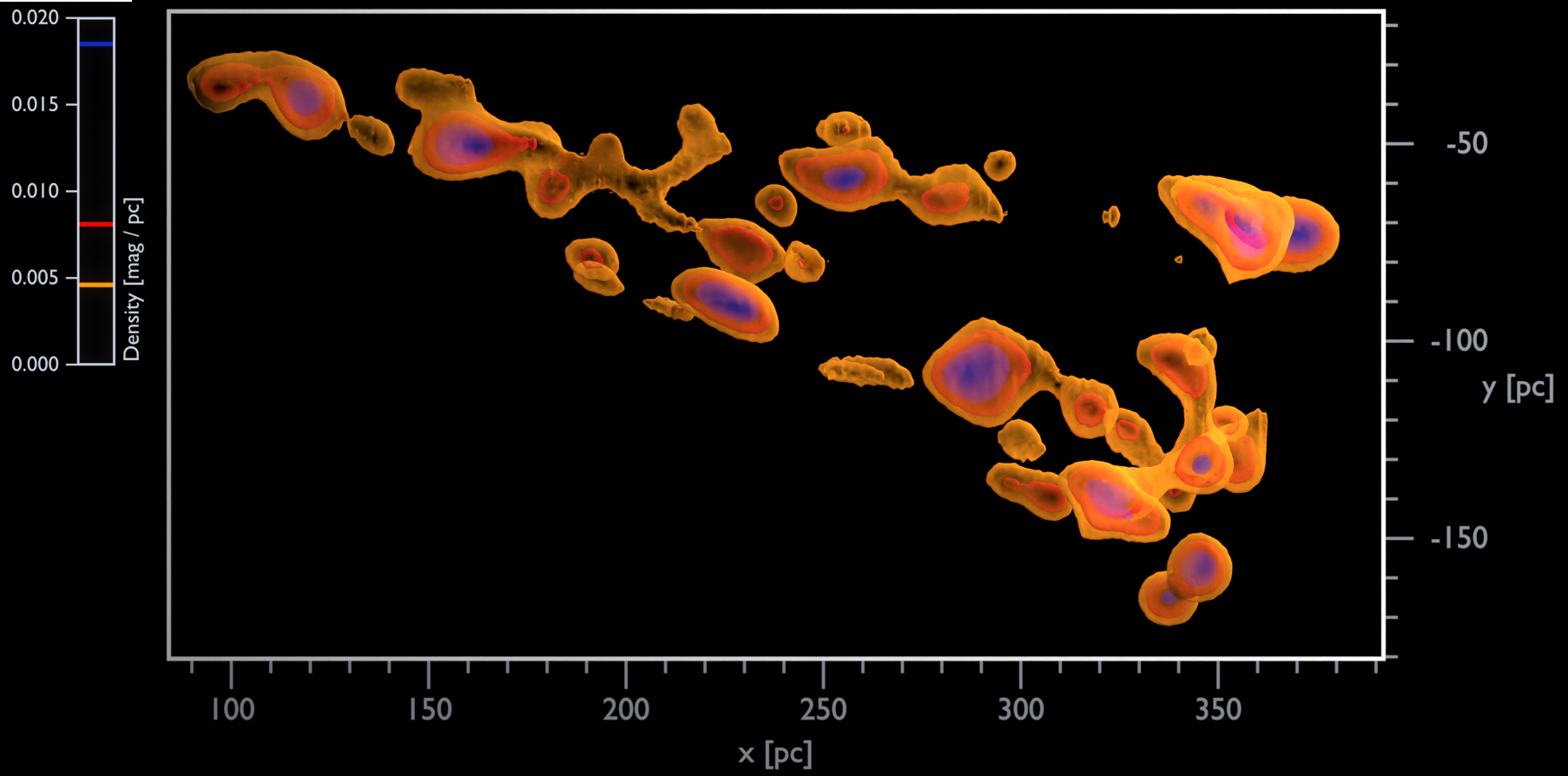}
    
    \end{multicols}
    \vspace{-0.8cm}
    \caption{ Predicted 3D density structure of the Lupus molecular cloud complex. Left: Slices along the line-of-sight of the predicted 3D density structure. With the Cyan triangle we have marked the mass weighted centroid of the primary trunk (as given in Table~\ref{tab:maintrunk_params} placed on the closest distance slice included in the plot. The Cyan $\times$s mark the mass weighted centroids of the interesting features discussed in Sec.\ref{sec:IndGMCs} and highlighted in Table~\ref{tab:leaf_params} placed on the closest distance slice included in the plot.; Right top: Video of a volume rendering of the predicted 3D density structure which begins from the view as seen from the Sun. It then rotates anti-clockwise about an axis perpendicular to the initial viewing angle. The semi-transparent iso-surfaces mark three different density levels with orange being the least dense to blue being the most dense as shown by the colour bar.; Right bottom: Still image showing the top down view of the predicted 3D density structure of the molecular cloud region using identical rendering to the preceding video.}
    
\label{fig:Lup_3Ddens}
\end{adjustwidth}   
\end{figure*}
\end{landscape}

\begin{landscape}
\begin{figure*}
\begin{adjustwidth}{-7.5cm}{0cm}
    \centering
    \begin{multicols}{2}
    \includegraphics[width=0.75\textwidth]{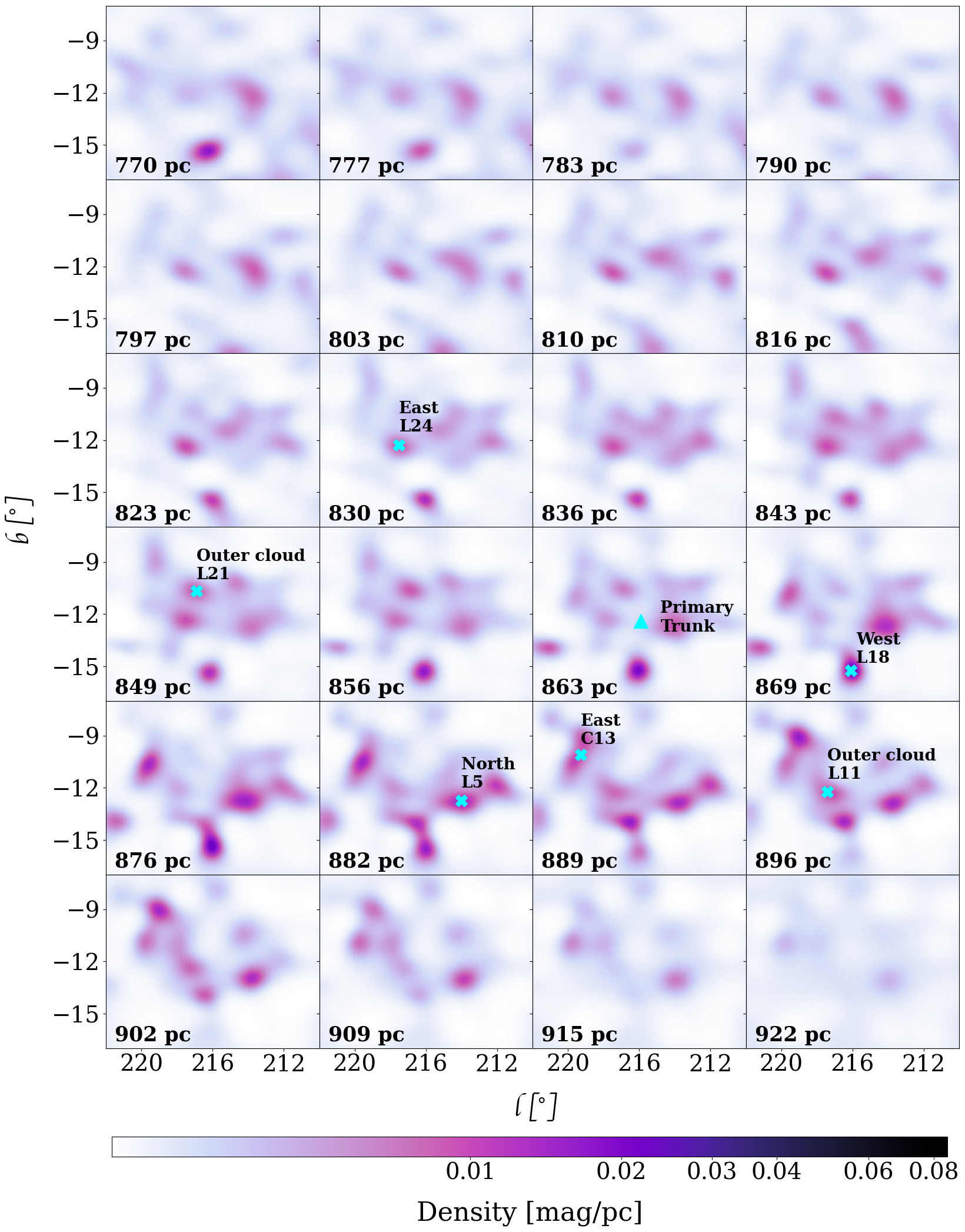}
    
    \hfill
    \includegraphics[width=0.6\textwidth]{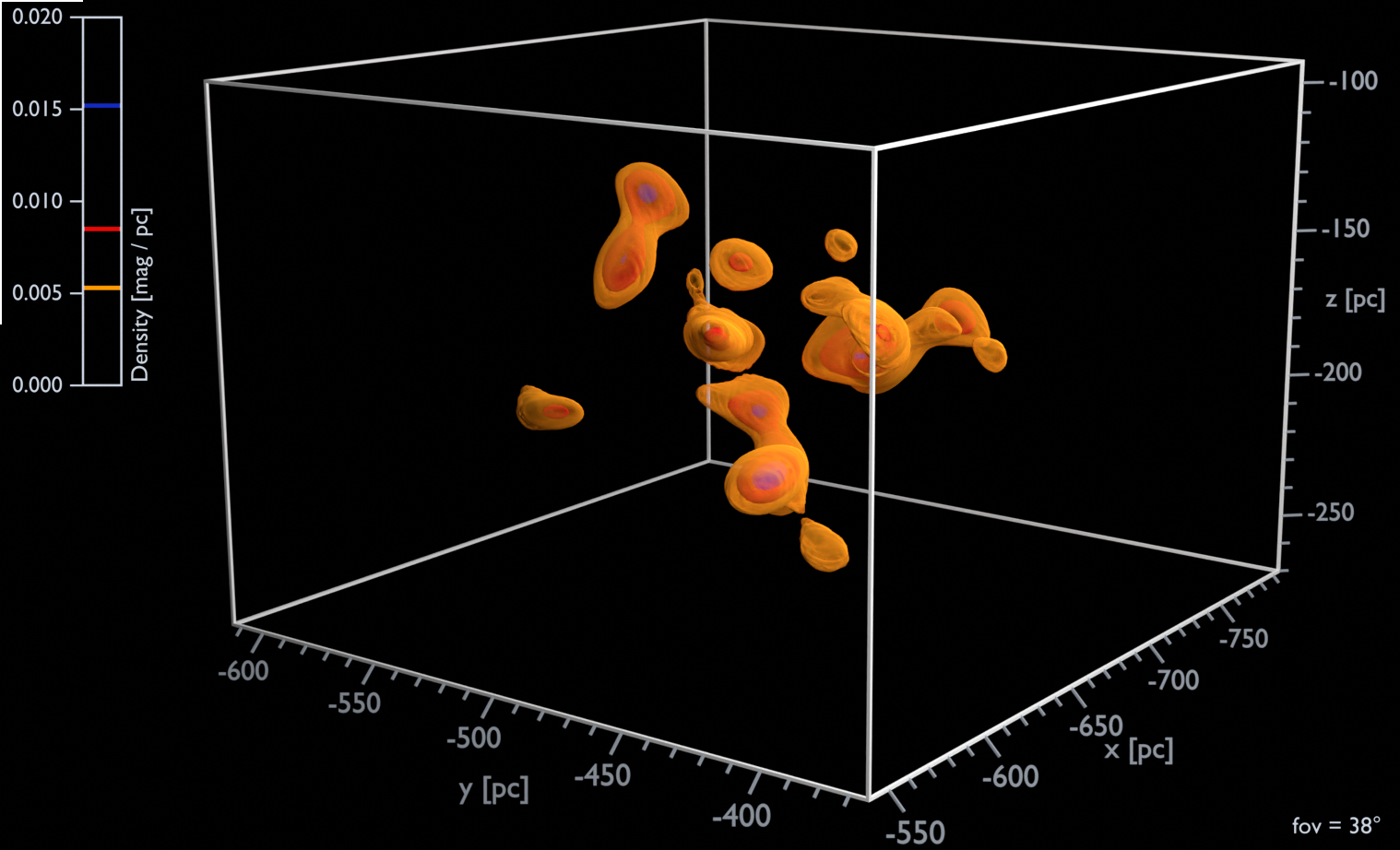}
    
    \hfill
    \includegraphics[width=0.6\textwidth]{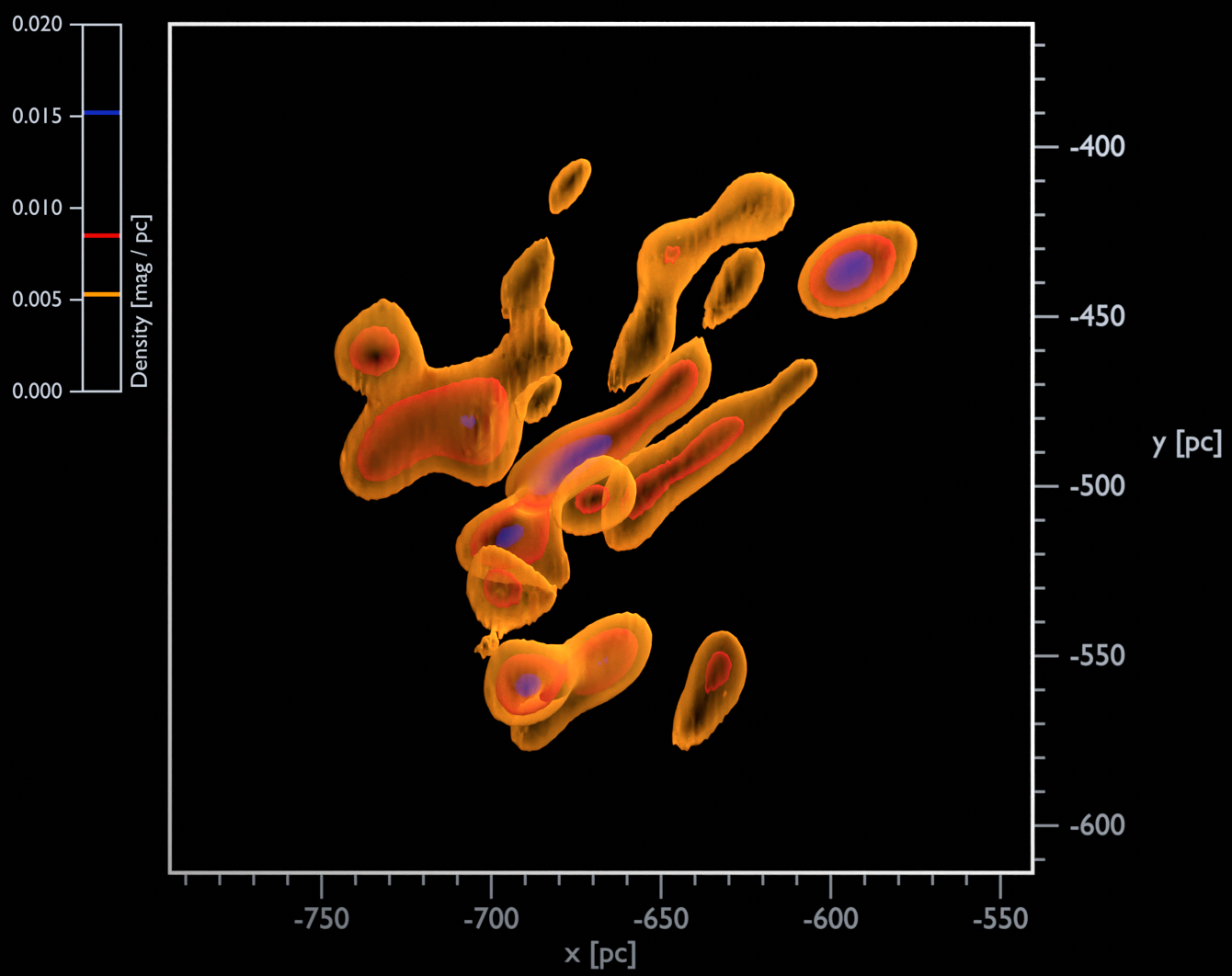}
    
    \end{multicols}
    \vspace{-0.8cm}
    \caption{ Predicted 3D density structure of the Mon R2 molecular cloud complex. Left: Slices along the line-of-sight of the predicted 3D density structure. With the Cyan triangle we have marked the mass weighted centroid of the primary trunk (as given in Table~\ref{tab:maintrunk_params} placed on the closest distance slice included in the plot. The Cyan $\times$s mark the mass weighted centroids of the interesting features discussed in Sec.\ref{sec:IndGMCs} and highlighted in Table~\ref{tab:leaf_params} placed on the closest distance slice included in the plot.; Right top: Video of a volume rendering of the predicted 3D density structure which begins from the view as seen from the Sun. It then rotates anti-clockwise about an axis perpendicular to the initial viewing angle. The semi-transparent iso-surfaces mark three different density levels with orange being the least dense to blue being the most dense as shown by the colour bar.; Right bottom: Still image showing the top down view of the predicted 3D density structure of the molecular cloud region using identical rendering to the preceding video.}
    
\label{fig:MonR2_3Ddens}
\end{adjustwidth}   
\end{figure*}
\end{landscape}

\begin{landscape}
\begin{figure*}
\begin{adjustwidth}{-7.5cm}{0cm}
    \centering
    \begin{multicols}{2}
    \includegraphics[width=0.75\textwidth]{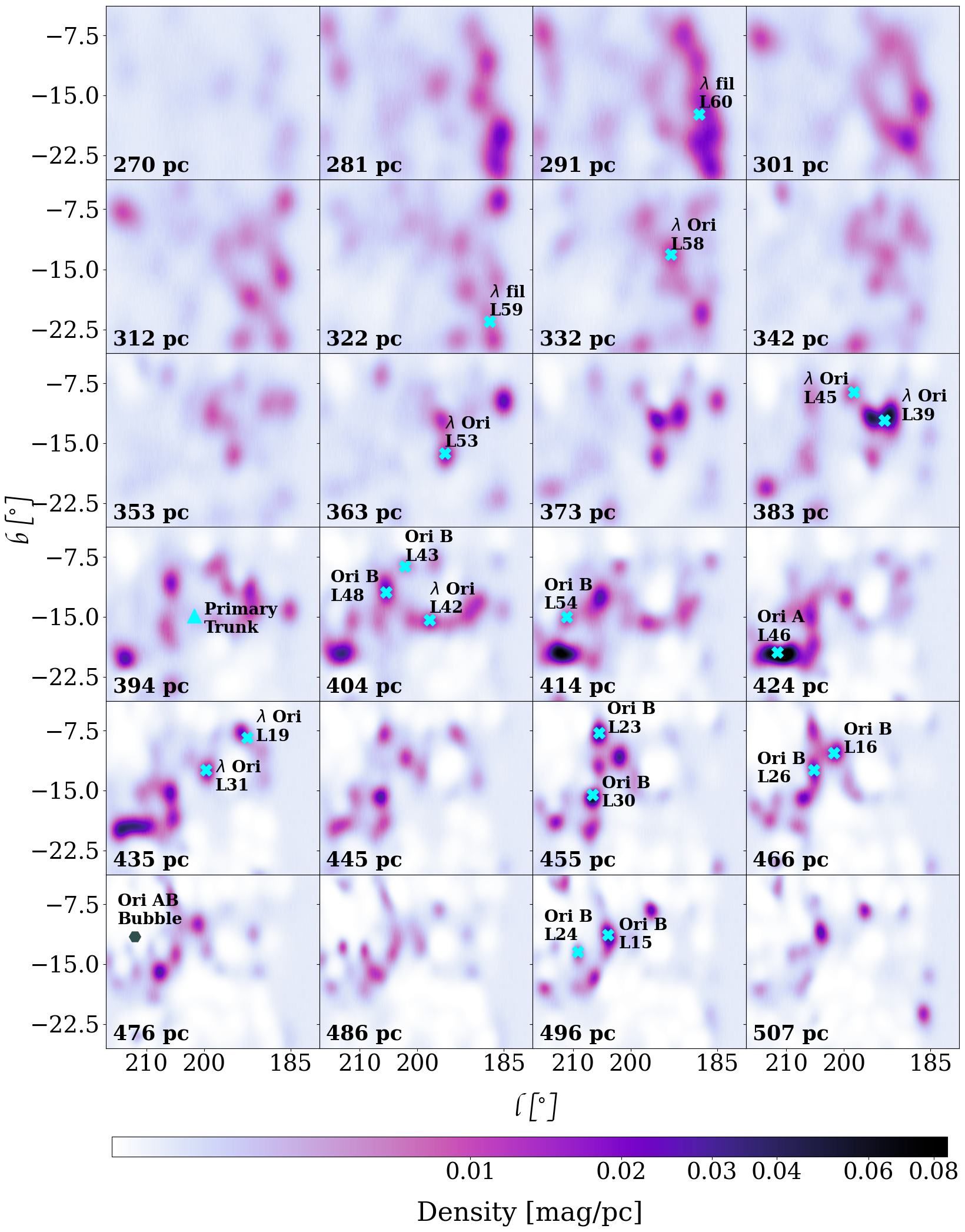}
    
    \hfill
    \includegraphics[width=0.6\textwidth]{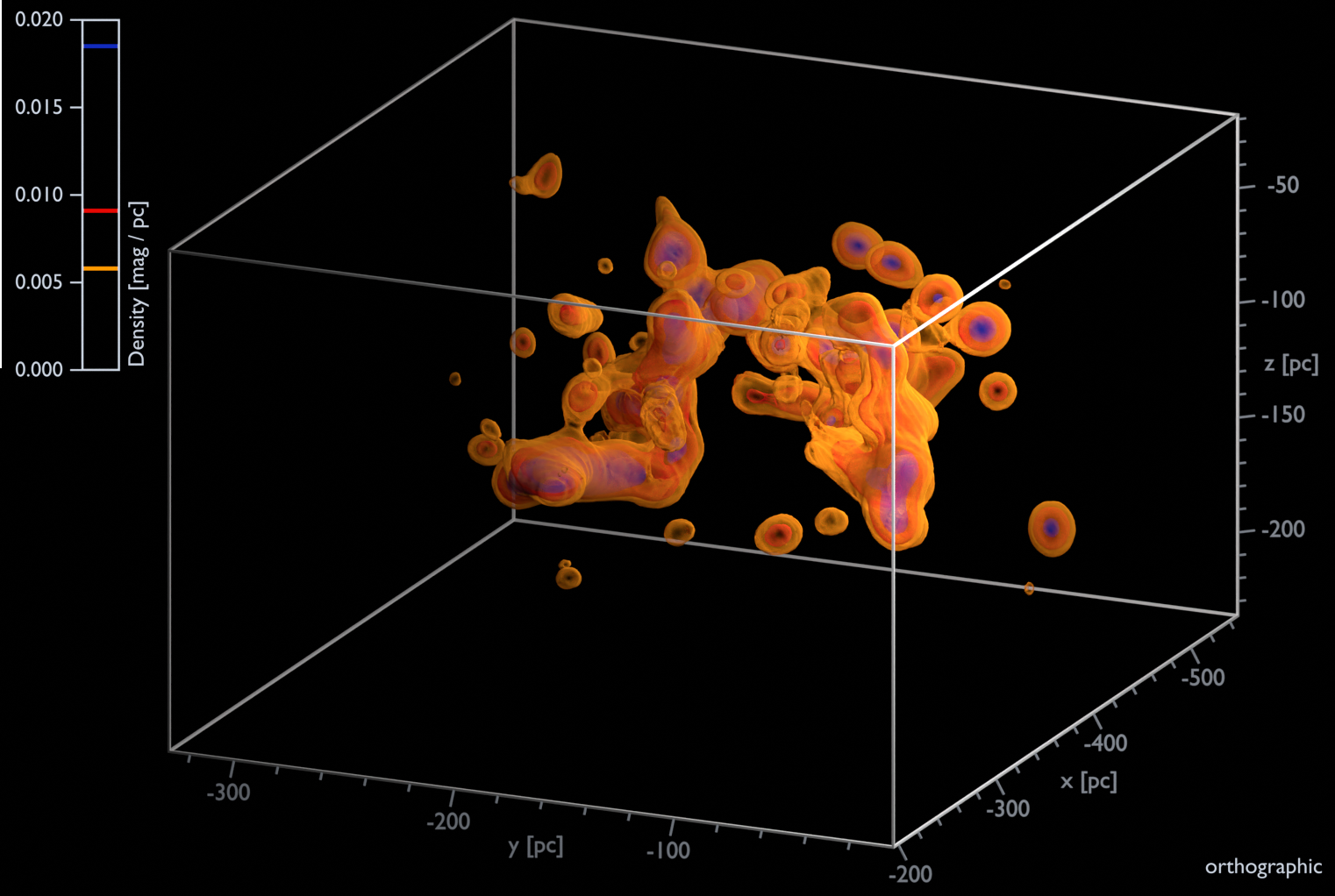}
    
    \hfill
    \includegraphics[width=0.6\textwidth]{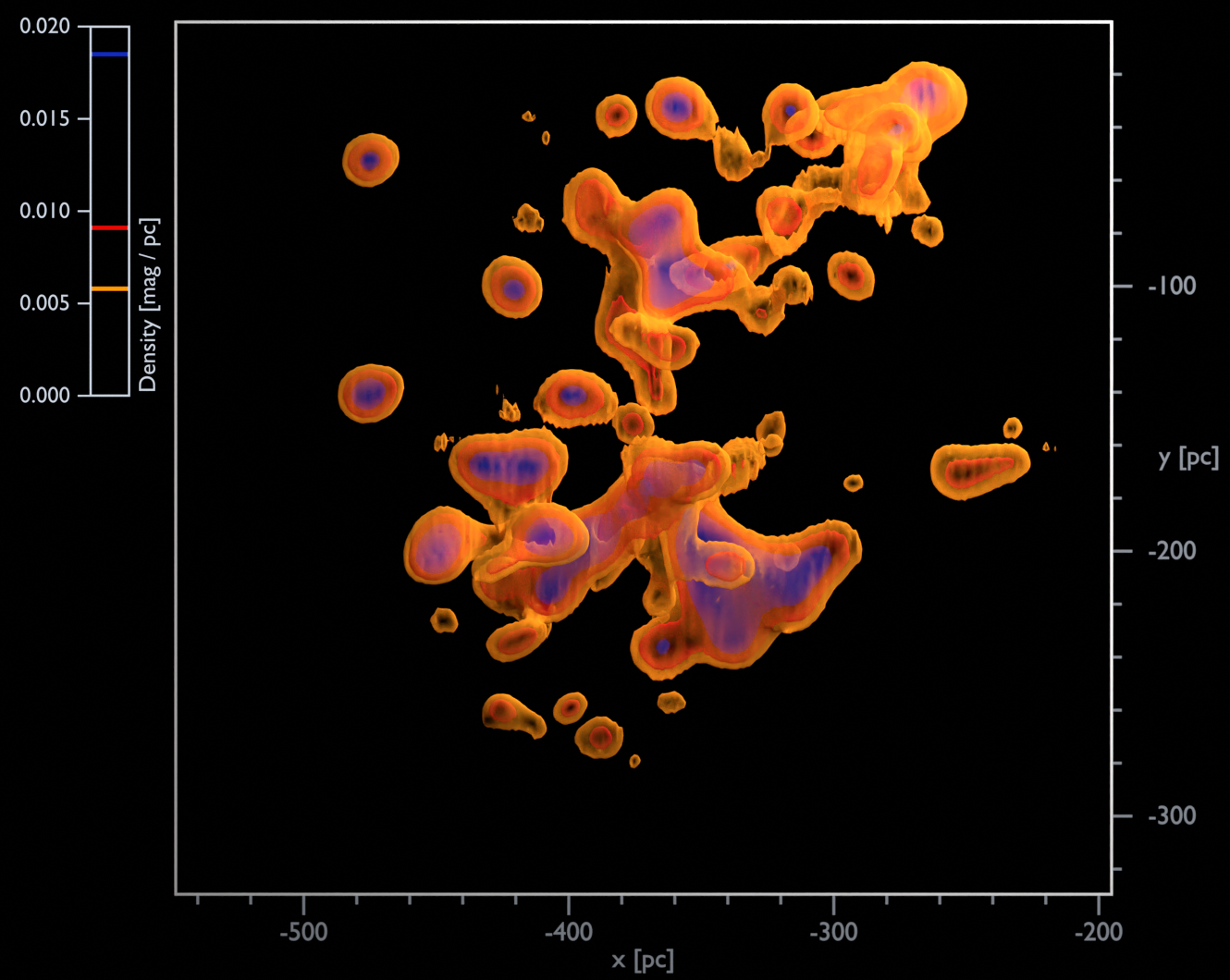}
    
    \end{multicols}
    \vspace{-0.8cm}
    \caption{ Predicted 3D density structure of the Orion molecular cloud complex. Left: Slices along the line-of-sight of the predicted 3D density structure. With the Cyan triangle we have marked the mass weighted centroid of the primary trunk (as given in Table~\ref{tab:maintrunk_params} placed on the closest distance slice included in the plot. The Cyan $\times$s mark the mass weighted centroids of the interesting features discussed in Sec.\ref{sec:IndGMCs} and highlighted in Table~\ref{tab:leaf_params} placed on the closest distance slice included in the plot.; Right top: Video of a volume rendering of the predicted 3D density structure which begins from the view as seen from the Sun. It then rotates anti-clockwise about an axis perpendicular to the initial viewing angle. The semi-transparent iso-surfaces mark three different density levels with orange being the least dense to blue being the most dense as shown by the colour bar.; Right bottom: Still image showing the top down view of the predicted 3D density structure of the molecular cloud region using identical rendering to the preceding video.}
    
\label{fig:Ori_3Ddens}
\end{adjustwidth}   
\end{figure*}
\end{landscape}

\begin{landscape}
\begin{figure*}
\begin{adjustwidth}{-7.5cm}{0cm}
    \centering
    \begin{multicols}{2}
    \includegraphics[width=0.75\textwidth]{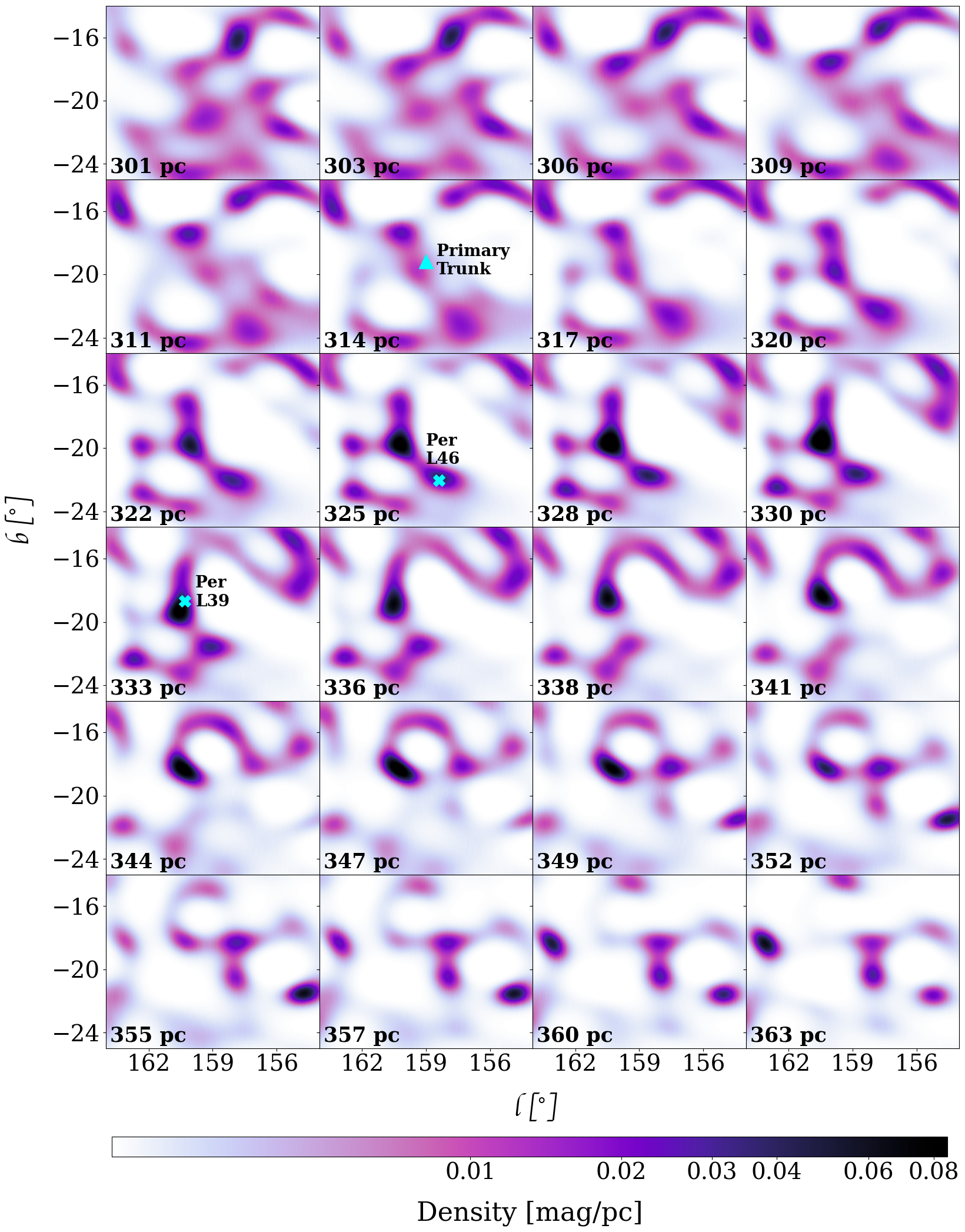}
    
    \hfill
    \includegraphics[width=0.6\textwidth]{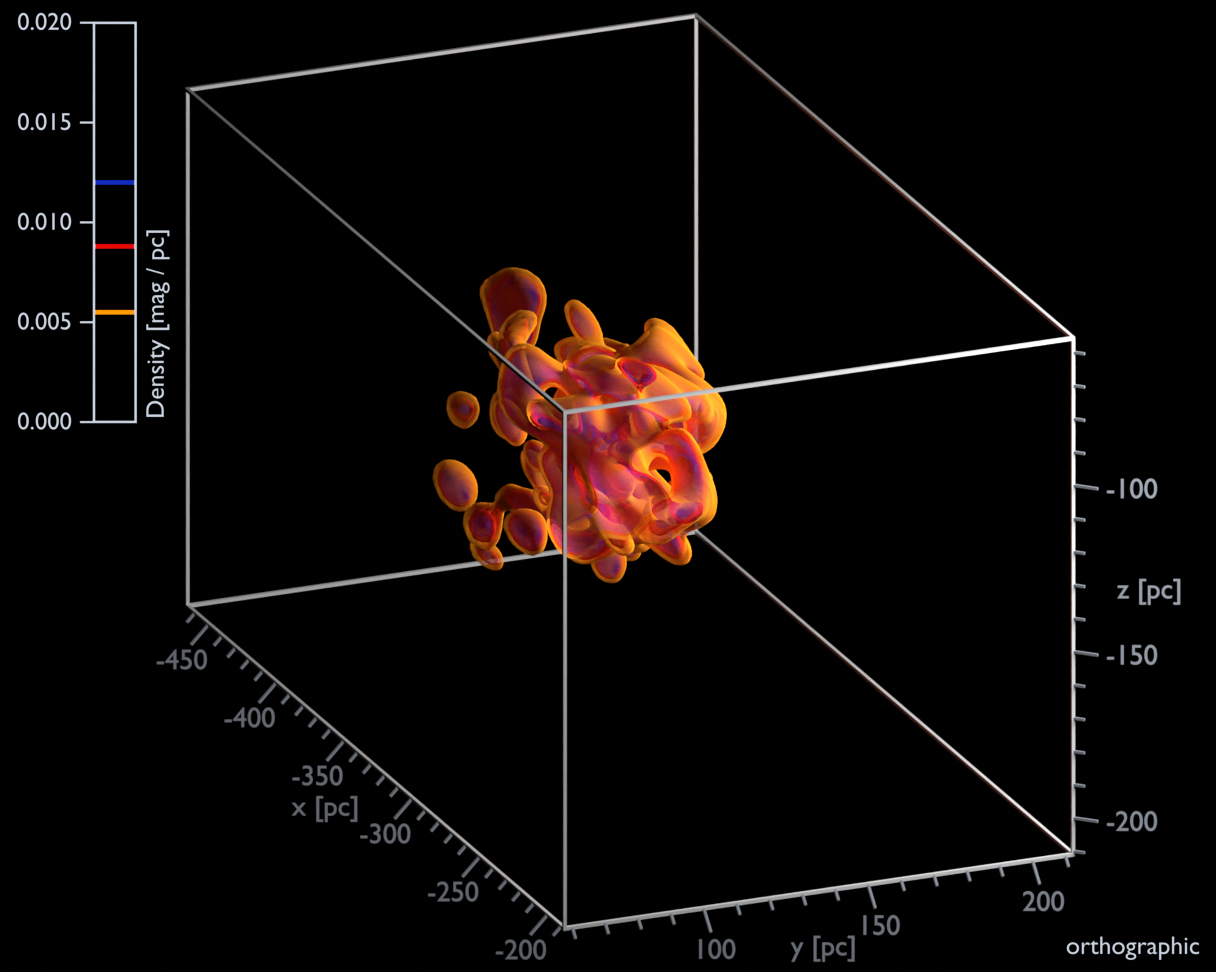}
    
    \hfill
    \includegraphics[width=0.6\textwidth]{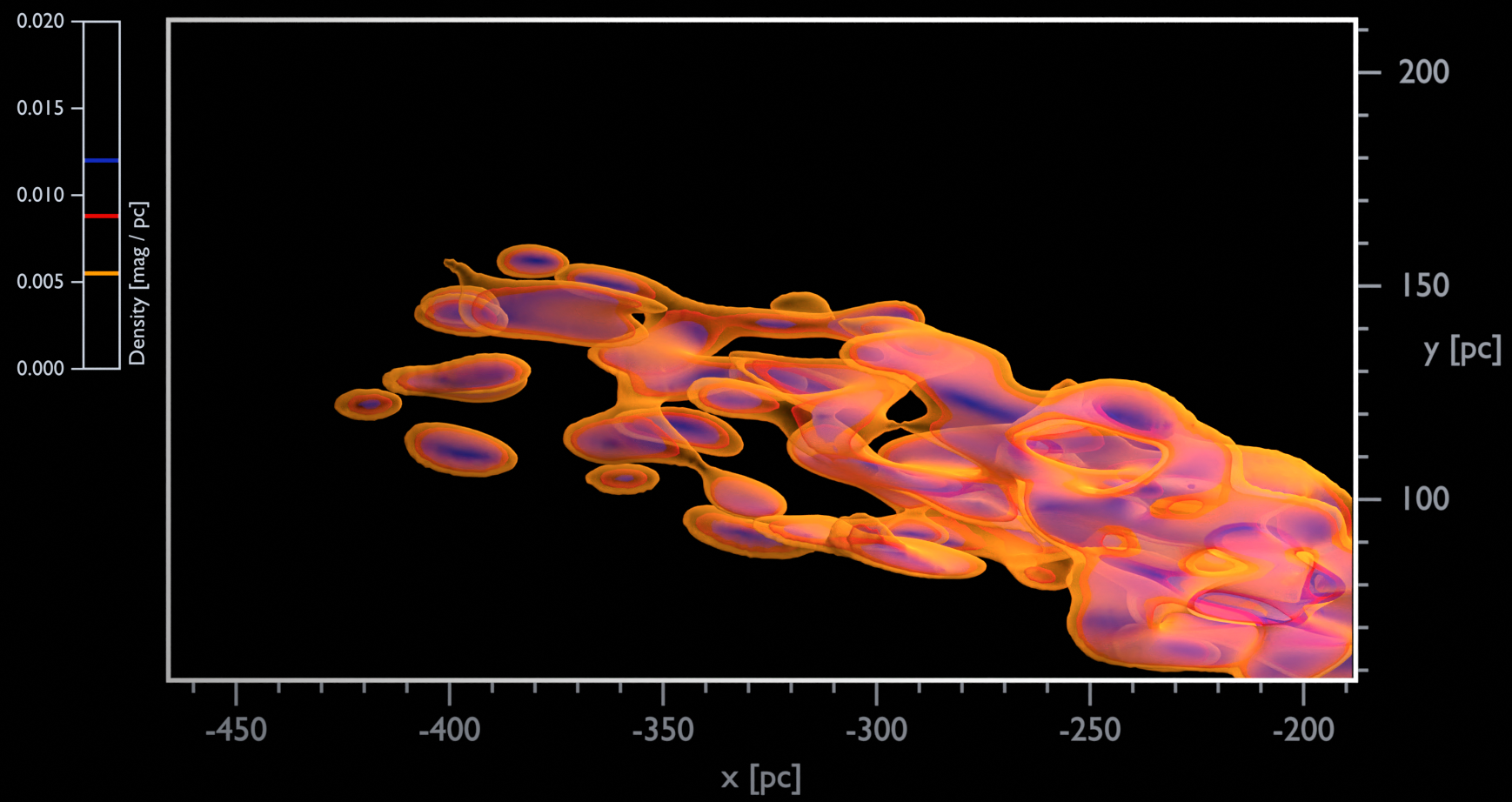}
    
    \end{multicols}
    \vspace{-0.8cm}
    \caption{ Predicted 3D density structure of the Perseus molecular cloud complex. Left: Slices along the line-of-sight of the predicted 3D density structure. With the Cyan triangle we have marked the mass weighted centroid of the primary trunk (as given in Table~\ref{tab:maintrunk_params} placed on the closest distance slice included in the plot. The Cyan $\times$s mark the mass weighted centroids of the interesting features discussed in Sec.\ref{sec:IndGMCs} and highlighted in Table~\ref{tab:leaf_params} placed on the closest distance slice included in the plot.; Right top: Video of a volume rendering of the predicted 3D density structure which begins from the view as seen from the Sun. It then rotates anti-clockwise about an axis perpendicular to the initial viewing angle. The semi-transparent iso-surfaces mark three different density levels with orange being the least dense to blue being the most dense as shown by the colour bar.; Right bottom: Still image showing the top down view of the predicted 3D density structure of the molecular cloud region using identical rendering to the preceding video.}
    
\label{fig:Per_3Ddens}
\end{adjustwidth}   
\end{figure*}
\end{landscape}

\begin{landscape}
\begin{figure*}
\begin{adjustwidth}{-7.5cm}{0cm}
    \centering
    \begin{multicols}{2}
    \includegraphics[width=0.75\textwidth]{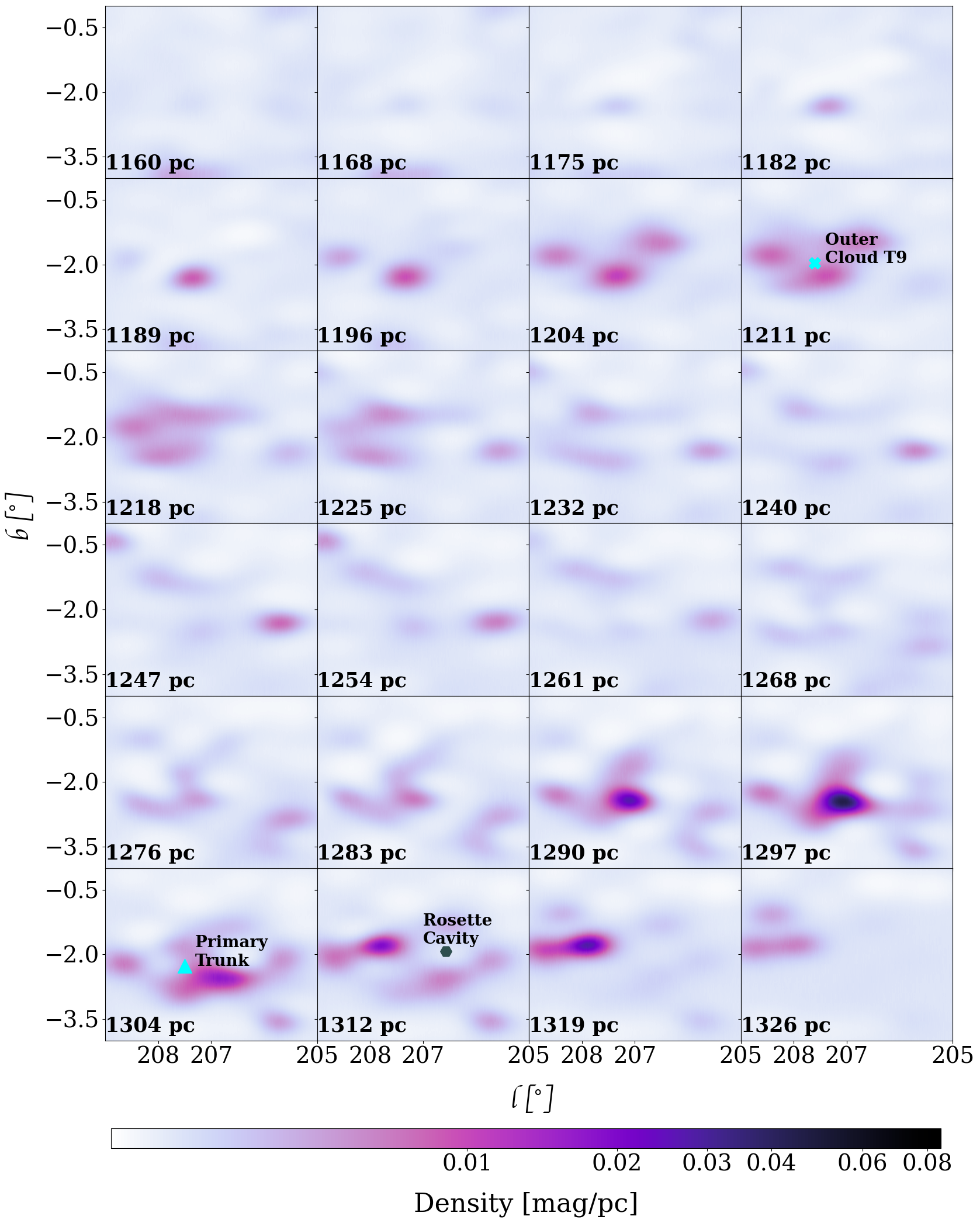}
    
    \hfill
    \includegraphics[width=0.6\textwidth]{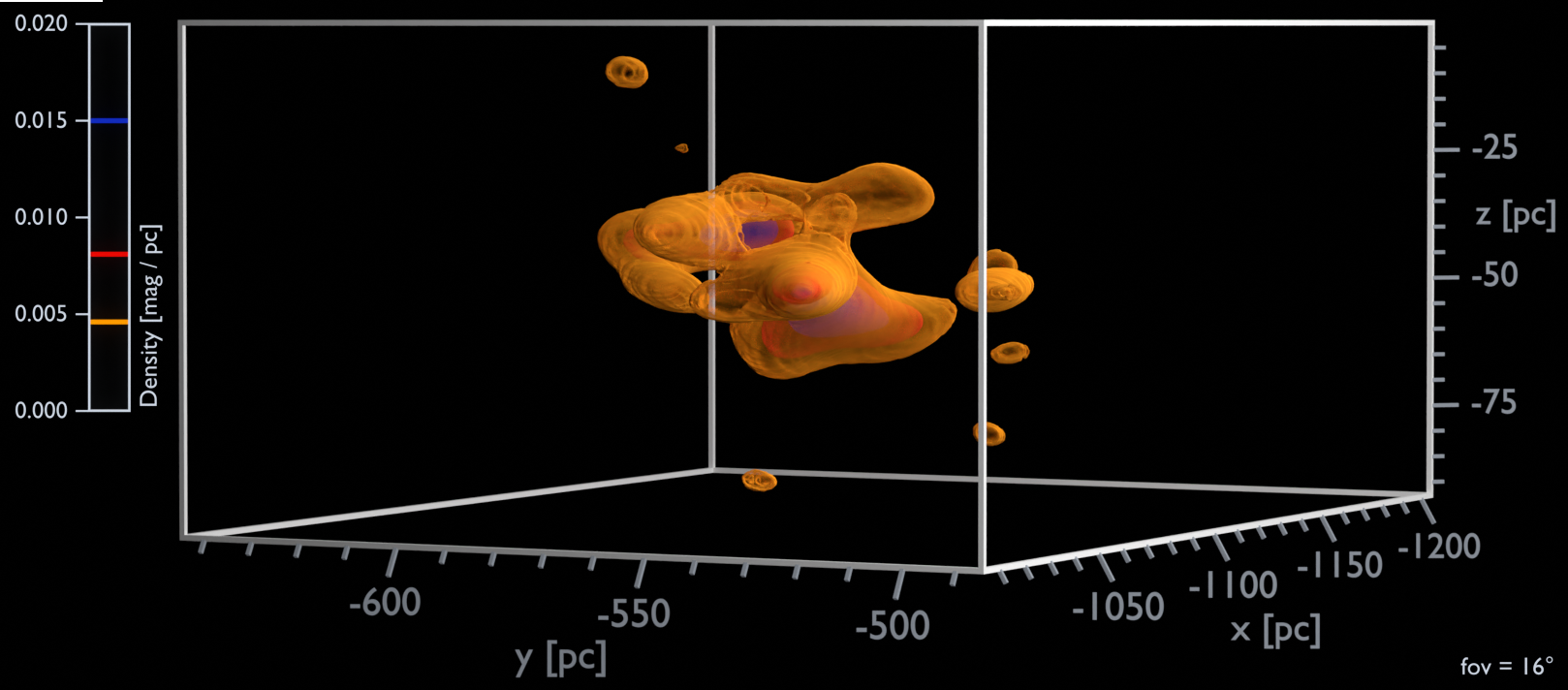}
    
    \hfill
    \includegraphics[width=0.6\textwidth]{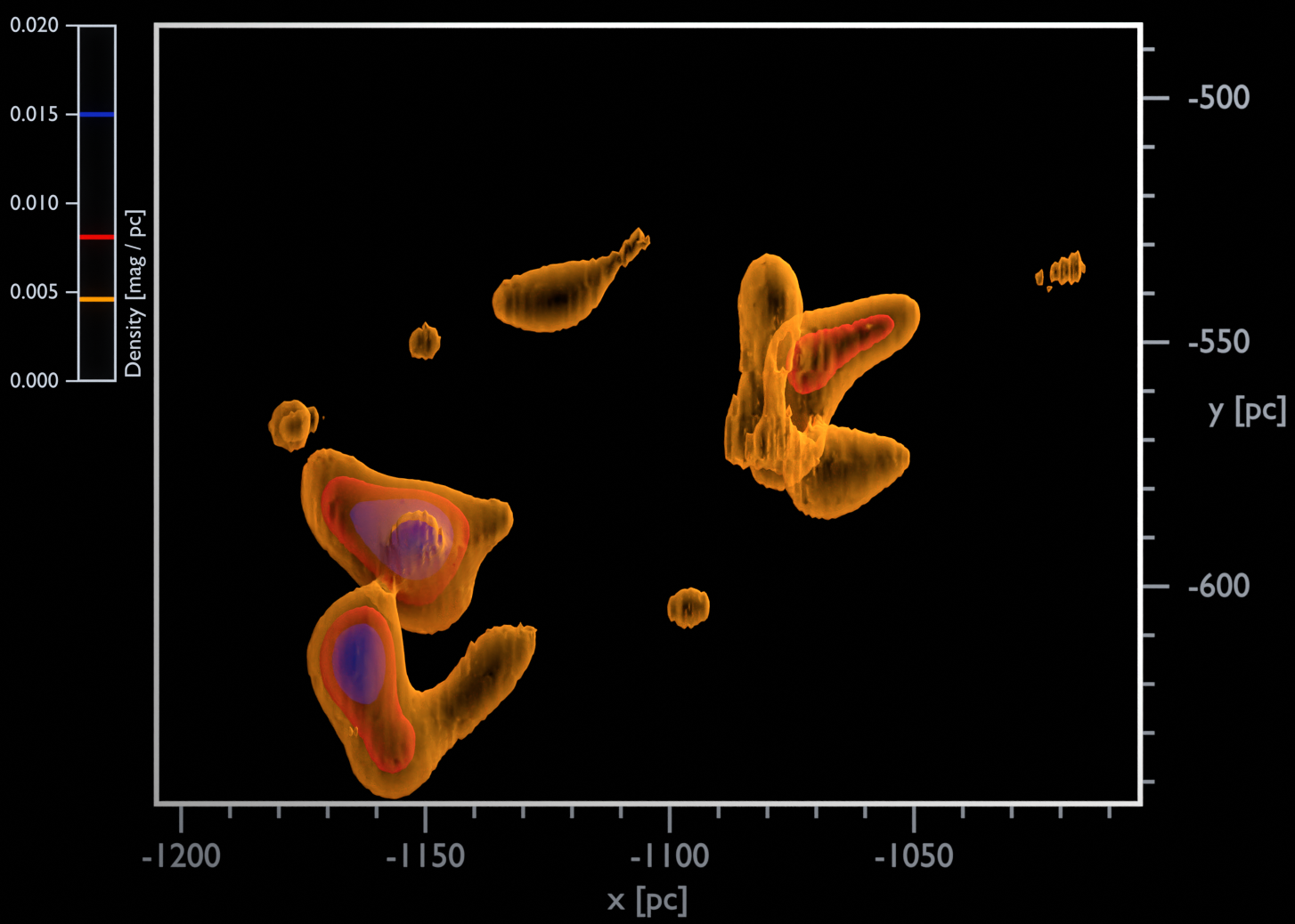}
    
    \end{multicols}
    \vspace{-0.8cm}
    \caption{ Predicted 3D density structure of the Rosette molecular cloud complex. Left: Slices along the line-of-sight of the predicted 3D density structure. With the Cyan triangle we have marked the mass weighted centroid of the primary trunk (as given in Table~\ref{tab:maintrunk_params} placed on the closest distance slice included in the plot. The Cyan $\times$s mark the mass weighted centroids of the interesting features discussed in Sec.\ref{sec:IndGMCs} and highlighted in Table~\ref{tab:leaf_params} placed on the closest distance slice included in the plot.; Right top: Video of a volume rendering of the predicted 3D density structure which begins from the view as seen from the Sun. It then rotates anti-clockwise about an axis perpendicular to the initial viewing angle. The semi-transparent iso-surfaces mark three different density levels with orange being the least dense to blue being the most dense as shown by the colour bar.; Right bottom: Still image showing the top down view of the predicted 3D density structure of the molecular cloud region using identical rendering to the preceding video.}
    
\label{fig:Rose_3Ddens}
\end{adjustwidth}   
\end{figure*}
\end{landscape}

\begin{landscape}
\begin{figure*}
\begin{adjustwidth}{-7.5cm}{0cm}
    \centering
    \begin{multicols}{2}
    \includegraphics[width=0.75\textwidth]{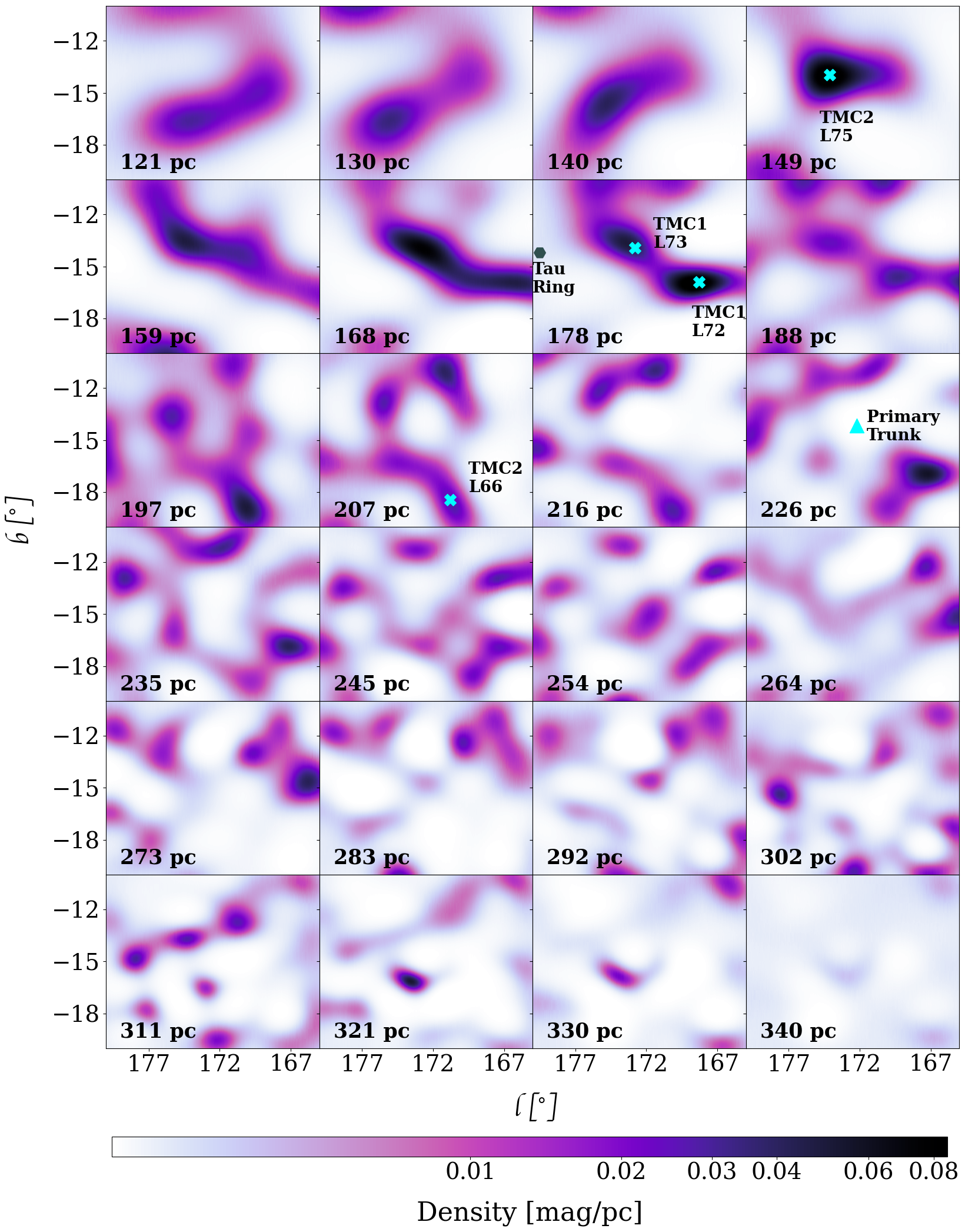}
    
    \hfill
    \includegraphics[width=0.6\textwidth]{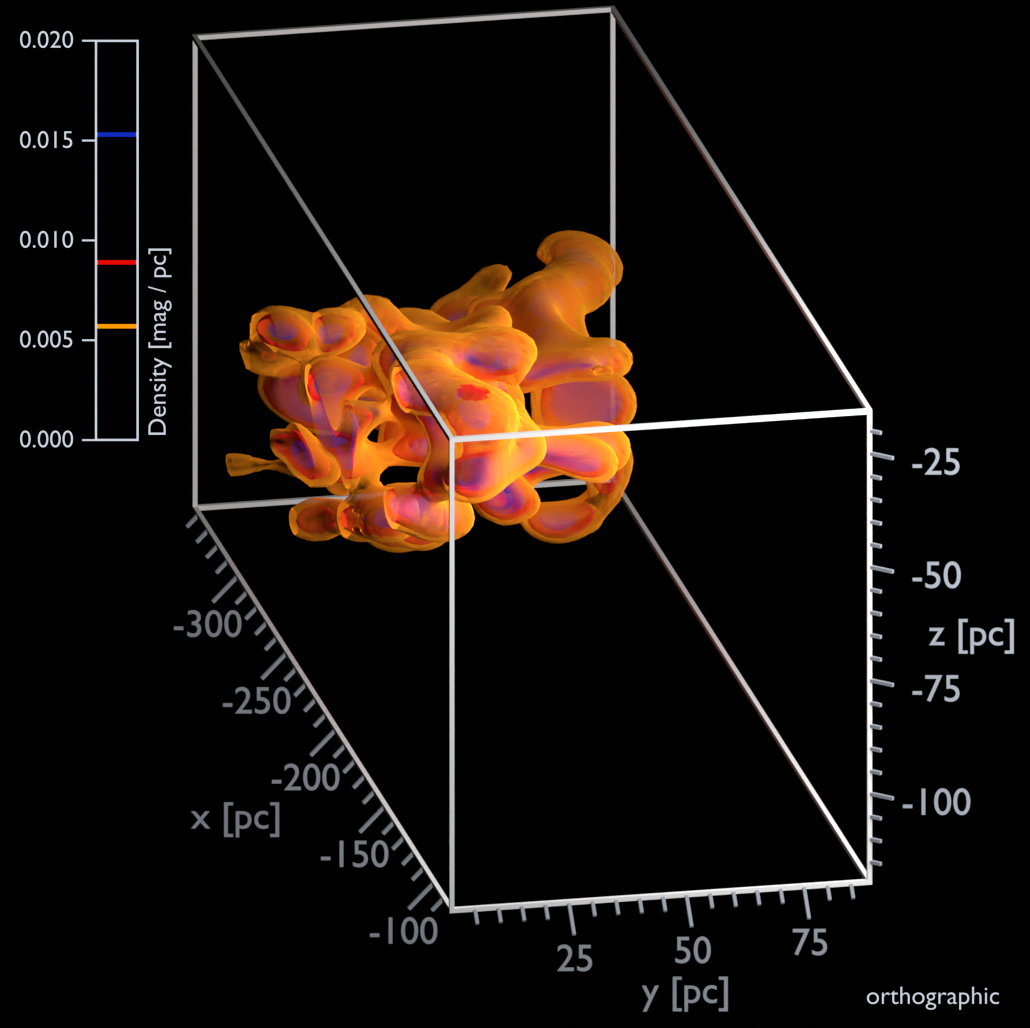}
    
    \hfill
    \includegraphics[width=0.6\textwidth]{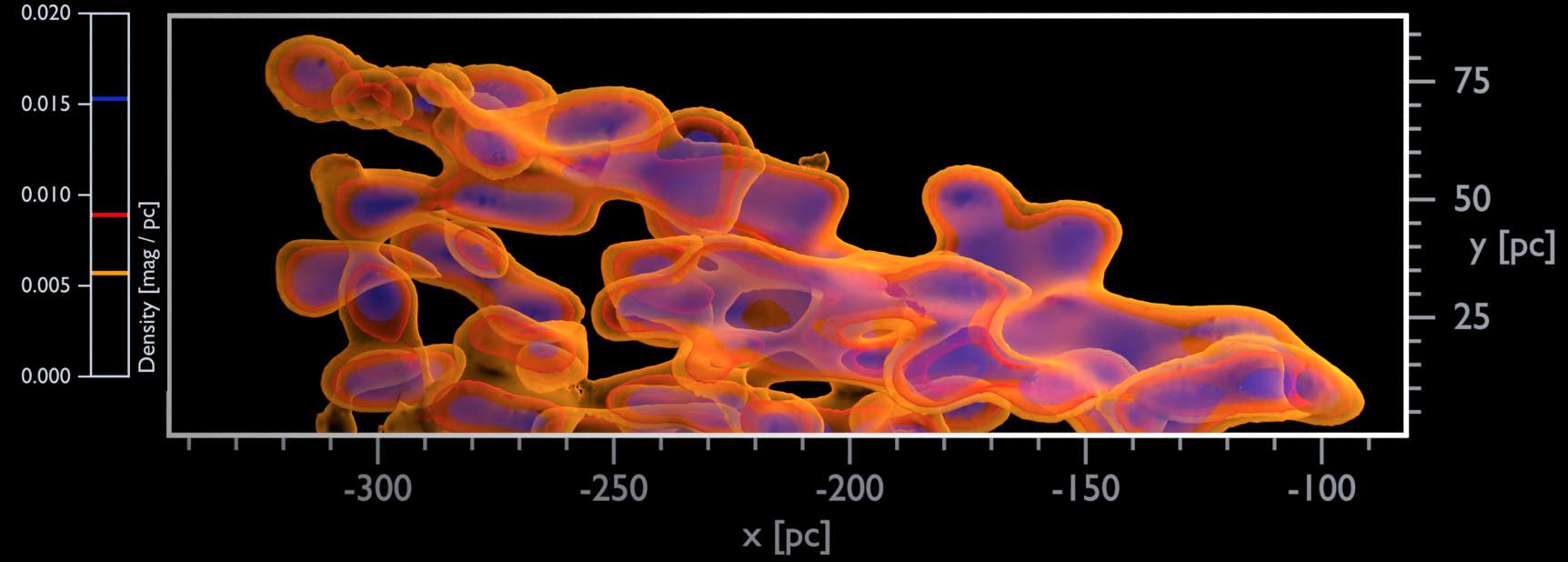}
    
    \end{multicols}
    \vspace{-0.8cm}
    \caption{ Predicted 3D density structure of the Taurus molecular cloud complex. Left: Slices along the line-of-sight of the predicted 3D density structure. With the Cyan triangle we have marked the mass weighted centroid of the primary trunk (as given in Table~\ref{tab:maintrunk_params} placed on the closest distance slice included in the plot. The Cyan $\times$s mark the mass weighted centroids of the interesting features discussed in Sec.\ref{sec:IndGMCs} and highlighted in Table~\ref{tab:leaf_params} placed on the closest distance slice included in the plot. The centroid of the Tau ring from \citet{Bialy2021} is marked in the grey hexagon; Right top: Video of a volume rendering of the predicted 3D density structure which begins from the view as seen from the Sun. It then rotates anti-clockwise about an axis perpendicular to the initial viewing angle. The semi-transparent iso-surfaces mark three different density levels with orange being the least dense to blue being the most dense as shown by the colour bar.; Right bottom: Still image showing the top down view of the predicted 3D density structure of the molecular cloud region using identical rendering to the preceding video.}
    
\label{fig:Tau_3Ddens}
\end{adjustwidth}   
\end{figure*}
\end{landscape}


\bsp	
\label{lastpage}
\end{document}